\documentclass[iop]{emulateapj}

\usepackage{subfigure}
\usepackage{floatflt,graphicx}
\usepackage{amsmath}
\usepackage{array}
\usepackage{verbatim}
\usepackage{rotating}
\usepackage[section] {placeins}

\RequirePackage{epsf,graphicx}

\newcommand{\HI}{H\,{\sc i}}
\newcommand{\HII}{H\,{\sc ii}}
\newcommand{\Ht}{H$_2$}

\newcommand{\arcs}{\arcsec }

\shorttitle{HCG MOHEGs}
\shortauthors{Cluver et al.}

\begin{document}

\title{Enhanced Warm \Ht\ emission in the Compact Group Mid-Infrared ``Green Valley"}

\author{M.E. Cluver\altaffilmark{1,2},\email{mcluver@aao.gov.au} P.N. Appleton\altaffilmark{3}, P. Ogle\altaffilmark{1}, T.H. Jarrett\altaffilmark{4},  J. Rasmussen\altaffilmark{5}, U. Lisenfeld\altaffilmark{6}, P. Guillard\altaffilmark{1},  L. Verdes-Montenegro\altaffilmark{7},  R. Antonucci\altaffilmark{8}, T. Bitsakis\altaffilmark{9}, V. Charmandaris\altaffilmark{9,10,11}, F. Boulanger\altaffilmark{12},  E. Egami\altaffilmark{13}, C.K. Xu\altaffilmark{2}, M.S. Yun\altaffilmark{14}}

\slugcomment{Accepted to ApJ: 17 January 2013}

\altaffiltext{1}{Spitzer Science Center, IPAC, California Institute of Technology, Pasadena, CA 91125, USA}
\altaffiltext{2}{ARC Super Science Fellow, Australian Astronomical Observatory, PO Box 915, North Ryde, NSW 1670, Australia}
\altaffiltext{3}{NASA Herschel S Center, California Institute of Technology, Pasadena, CA 91125, USA}
\altaffiltext{4}{Department of Astronomy, University of Cape Town, Private Bag X3, Rondebosch, 7701, South Africa}
\altaffiltext{5}{Dark Cosmology Centre, Niels Bohr Institute,
  University of Copenhagen, Juliane Maries Vej 30, DK-2100 Copenhagen,
  Denmark}
\altaffiltext{6}{Departmento de F\`{i}sica Te\'{o}rica y del Cosmos, Facultad de Ciencias, Universidad de Granada, Spain }
\altaffiltext{7}{Instituto de Astrof\'{i}sica de Andaluc\'{i}a (IAA/CSIC), Apdo. 3004, 18080 Granada, Spain}
\altaffiltext{8}{University of California Santa Barbara, Department of Physics, Santa Barbara, CA 93106, USA}
\altaffiltext{9}{Department of Physics, University of Crete, GR-71003, Heraklion,
Greece}
\altaffiltext{10}{IESL/Foundation for Research \& Technology-Hellas, GR-71110,
Heraklion, Greece}
\altaffiltext{11}{Chercheur Associ\'e, Observatoire de Paris, F-75014,  Paris, France}
\altaffiltext{12}{Institute d'Astrophysique Spatiale, Universite Paris Sud 11, Orsay, France}
\altaffiltext{13}{Steward Observatory, University of Arizona, 933 N. Cherry Avenue, Tucson, AZ 85721, USA}
\altaffiltext{14}{
Department of Astronomy, University of Massachusetts, Amherst, MA 01003, USA}

\begin{abstract}
 
We present results from a {\it Spitzer}, mid-infrared spectroscopy study of a
sample of 74 galaxies located in 23 Hickson Compact Groups, chosen to be at a
dynamically-active stage of \HI\ depletion. We find evidence for
enhanced warm \Ht\ emission (i.e. above that associated with UV
excitation in star-forming regions) in 14 galaxies ($\sim$20\%), with 8 galaxies having
extreme values of L(\Ht\,S(0)-S(3))/L(7.7\micron\ PAH), in excess of 0.07. 
Such emission has been seen previously in the compact group HCG 92 (Stephan's Quintet), and was shown to be associated with the dissipation of mechanical energy associated with a large-scale shock caused when one group member collided, at high velocity, with tidal debris in the intragroup medium. 
Similarly, shock excitation or turbulent heating is likely responsible for the enhanced
\Ht\ emission in the compact group galaxies, since other sources of heating (UV or X-ray
excitation from star formation or AGN) are insufficient to account for the observed emission.
The group galaxies fall predominantly in a region of mid-infrared color-color space identified by previous studies as being connected to rapid transformations in HCG galaxy evolution. Furthermore, the majority of \Ht-enhanced galaxies lie in the optical ``green valley'' between the blue cloud and red-sequence, and are primarily early-type disk systems. 
We suggest that \Ht-enhanced systems may represent a specific phase in the evolution of galaxies in dense environments and provide new insight into mechanisms which transform galaxies onto the optical red sequence.

\end{abstract}


\section{Introduction}

Compact Groups are key laboratories for studying morphological transformations as they represent the highest density enhancements
outside of clusters, and their relatively low velocity dispersions prolong
gravitational interactions \citep{Hick92}. Disentangling the mechanisms that influence galaxy evolution are made more challenging with
growing evidence that cluster galaxies have been ``pre-processed" in
groups, and then subsequently assimilated into larger systems
\citep[e.g.][]{Cor06}. Simulations show that spirals in a group
environment are strongly influenced by repetitive slow
encounters, building bulge mass as gas is funnelled into the
central regions, transforming them into S0 galaxies with young, metal-rich stellar populations in their inner bulges \citep{Bek11}. In addition, galaxy interactions in groups may differ from isolated binary
interactions because they can exhibit a broader range of behaviours,
including tidal stripping and
interactions with the intra-group medium \citep[IGM; see e.g. ][]{All80}. 

\citet{Hick82} identified a uniform sample of 100 nearby
compact groups (Hickson Compact Groups; hereafter HCG)
using the Palomar Sky Survey, and applied the criterion
of 4 or more members within a 3 magnitude range ($\delta$m$_B$) that
also satisfied an isolation constraint. These groups have been the
subject of extensive follow-up and study. Radial velocities \citep{Hick92} have led to the identification of true associations of
galaxies, with 92 groups consisting of at least three accordant
members.  That many of the groups are real physical associations is
further attested by the presence of hot intragroup gas in many of them \citep{Pon96, Des12}. Signs of interactions within these groups include
peculiar rotation curves and disturbed
morphologies of group members \citep{Rub91,Mend94}, as well as the presence of
intragroup light \citep{Roch08}.

\HI\ deficiency in compact group galaxies has long been suspected \citep{Will87,Huch97} and the \HI\ study of 72 HCGs led \citet{VM01} to propose an evolutionary sequence
in which compact group galaxies become increasingly deficient in neutral
hydrogen. Multiple tidal interactions, and possible gas stripping (via
interaction with the IGM) may be the cause of the observed \HI\ depletion. More
recent observations using the Green Bank Telescope, sensitive to
extended, faint \HI\ emission, have revealed a diffuse \HI\ component in all the groups studied \citep{Bor10}, explaining in part the ``missing \HI" reported by \citet{VM01}.
Groups containing galaxies with the largest
\HI\ deficiencies are found to have a more massive
diffuse-\HI\ component \citep{Bor10}. These observations
have led to the conclusion that \HI-deficient group galaxies lose
\HI\ into the IGM, primarily through tidal interactions.  In the
most evolved groups, gravitational heating may eventually create a hot
X-ray emitting medium \citep[see ][]{Pon96,Des12}.

\citet{Leon98} observed that galaxies located in the most compact groups within their sample of 45 HCGs have more
molecular gas concentrated in their nuclei, as expected from the effects of tidal forcing within the disk.  However, the link between molecular gas and star formation properties in these interacting systems is less obvious. Studies of far-infrared and CO emission
of galaxies in HCGs indicate no enhanced star formation, but
20\% of spiral galaxies are tentatively found to be deficient in CO
emission compared to isolated and weakly interaction systems
\citep{VM98}. A recent, more extensive CO study of
HCGs finds that the specific star formation rate (SFR per unit
stellar mass or sSFR) is lower in \HI- and CO-deficient (as compared to isolated systems) HCG galaxies, but that the star formation efficiency (SFR per unit cold \Ht\ mass)
in these galaxies appears unaffected \citep{Mart12}. 

The question of whether galaxies are stripped by interaction with a
dense medium, or by tidal forces, still remains unclear. X-ray observations
of 8 HCGs by \citet{Ras08} showed no obvious correlation
between the presence of detectable hot intragroup gas and \HI\ deficiency. Furthermore, in groups
where X-ray emitting gas was strongly detected, it was shown that it
was not of sufficient density to significantly strip \HI\ from the group
members, thus calling into question whether gas stripping by a hot X-ray medium is a viable stripping mechanism within compact groups.

Observations in the infrared with the {\it Spitzer} Space Telescope have led to potentially new 
evidence of evolutionary effects in HCGs. 
\citet{Joh07} observe a correlation in the
IRAC color-color diagram of HCG galaxies, and argue for an evolutionary
sequence -- from groups dominated by dusty spirals with ``red" IRAC colors,
to groups containing evolved stellar-dominated galaxies with
``blue" IRAC colors. These colors appear to correlate with the degree of \HI\
depletion, supporting the idea of group evolution.  \citet{Joh07} discuss a ``gap'' in the IRAC colour space between dusty/gas-rich
and gas-poor galaxies; this apparent absence of intermediate mid-infrared
colours suggests a rapid evolution from gas-rich to \HI-depleted
systems \citep{Walk10}. The role of environment is further investigated in \citet{Walk12} using a sample of 49 compact groups. They find a statistically significant deficit of galaxies in the gap region with similarities to that found for the Coma Infall region. Accelerated transformation, possibly preceded by enhanced star formation in some galaxies, is also
suggested by the significant bimodality in specific star formation
rate (sSFR) seen in the HCG sample of \citet{Tzan10}.

\citet{Bit11} performed a UV to mid-infrared analysis of a sample of 32
Hickson Compact Groups and find that dynamically ``old" groups (containing $>$25\% early-type galaxies) are more compact and display higher velocity dispersions compared to dynamically ``young" systems (containing $>$75\% late-type galaxies). Late-type
galaxies in dynamically old groups are found, on average, to have higher stellar mass and lower sSFR, attributed to a faster build in stellar mass due to past interactions compared to the dynamically young groups. Their study also finds that the majority (73\%) of compact group galaxies lie in either the optical ``green valley" or the ``red sequence", as defined by their NUV$-$r colors. More than half of the early-type galaxies in dynamically ``old" groups were found to be located in the ``green valley" and these are predominantly ($>$70\%) S0/SB0's.

The AGN (Active Galactic Nuclei) activity of galaxies in HCGs are a key consideration; although 46\% of the sample of \citet{Bit11} have optically identified AGN, from nuclear spectra, they find no evidence of enhanced AGN activity at any stage of group evolution. This is consistent with the findings of \citet{Mar10} where the median HCG AGN luminosity corresponds to a low luminosity AGN (LLAGN), likely caused by gas depletion resulting in relatively low accretion rates, and also \citet{Ras08}, where the frequency and strength of nuclear X-ray activity in 8 groups showed no clear correlation with the dynamical state of the group, as measured by either diffuse X-ray emission or \HI-deficiency.


Our current paper is motivated by a possible new diagnostic of HCG
evolution which uses mid-infrared spectroscopy from the {\it Spitzer} Space Telescope as a
probe of the warm molecular gas in galaxies. Our team \citep{App06, Clu10} discovered 
powerful (L$_{{\rm H}_2}$ $>$ 10$^{35}$
W) mid-infrared molecular hydrogem (\Ht) line emission from an intergalactic shock wave in
Stephan's Quintet (SQ; HCG 92). The emission was found to be spatially
associated with a 40\,kpc-long X-ray and radio-continuum filament
believed to be formed as a result of a high-speed collision ($\sim$ 1000 km\,s$^{-1}$) between a
group member and tidal debris from a previous encounter within the
group.  In this case, the high power of the \Ht\ relative to both the infrared continuum and very faint PAH (polycyclic aromatic hydrocarbon) emission, and the close association with a known group-wide shock wave, makes a strong case for shock-heating as a viable mechanism in that compact group \citep{App06, Clu10}. Models demonstrate that driving a shock into a multi-phase medium, such as a pre-existing \HI\ tidal arm are capable of explaining many of the observed properties of the warm \Ht\ emission in SQ \citep{Guil09}.
The group members are known to be \HI\ and CO depleted \citep[see][]{Yun97,Gao00}, and may be one of the best
candidates for hydrodynamic stripping effects as much of the molecular
gas appears to reside in the IGM \citep{Guil12a} based on
deep IRAM CO observations. The discovery of \HI\ and molecular
gas between galaxies in HCG 92, as well as in the tidal bridge between the
Taffy galaxies \citep{Pet12} -- a system which has
recently experienced a head-on collision similar to those expected in
dense environments -- led to our present study of a much larger sample
of 23 HCGs with {\it Spitzer}.

The discovery of a class of powerful H$_2$-emitting radio galaxies, with
similar {\it Spitzer} IRS spectra to SQ, led to the coining of the term MOHEG (MOlecular
Hydrogen Emission-line Galaxies) and are defined as having large \Ht\ to 7.7\micron\ PAH emission ratios, $\ge0.04$, indicating excitation above that expected from UV heating \citep{Og10}.

In this paper we focus on the excited \Ht\ properties of a sample of 74 HCG galaxies, in particular those that show \Ht-enhancement as defined for a MOHEG. A subsequent paper will focus on the cold molecular gas properties of \Ht-enhanced systems through IRAM CO observations and comparisons of the cold versus warm \Ht\ masses (and temperatures of the excited \Ht).
The paper is organised as follows: in Section 2 we outline the sample chosen for this study, in Section 3 we summarise the observations and data reduction procedures, and in Section 4 present results of the excited \Ht\ line emission survey.  Section 5 explores potential sources of \Ht\ excitation and in Section 6 we discuss possible links to evolution within compact groups. Section 7 presents our discussion, with conclusions summarised in Section 8. Throughout this paper we assume a cosmology with Hubble constant $H_{\rm o}=73\, \rm{km}\, \rm{s}^{-1} Mpc^{-1}$, matter density parameter $\Omega=0.3$, and dark energy density  $\Omega_\Lambda=0.7$.

\section{The Sample}\label{samp}

\begin{figure*}[!thp]
\begin{center}
\subfigure[]{\includegraphics[width=8cm]{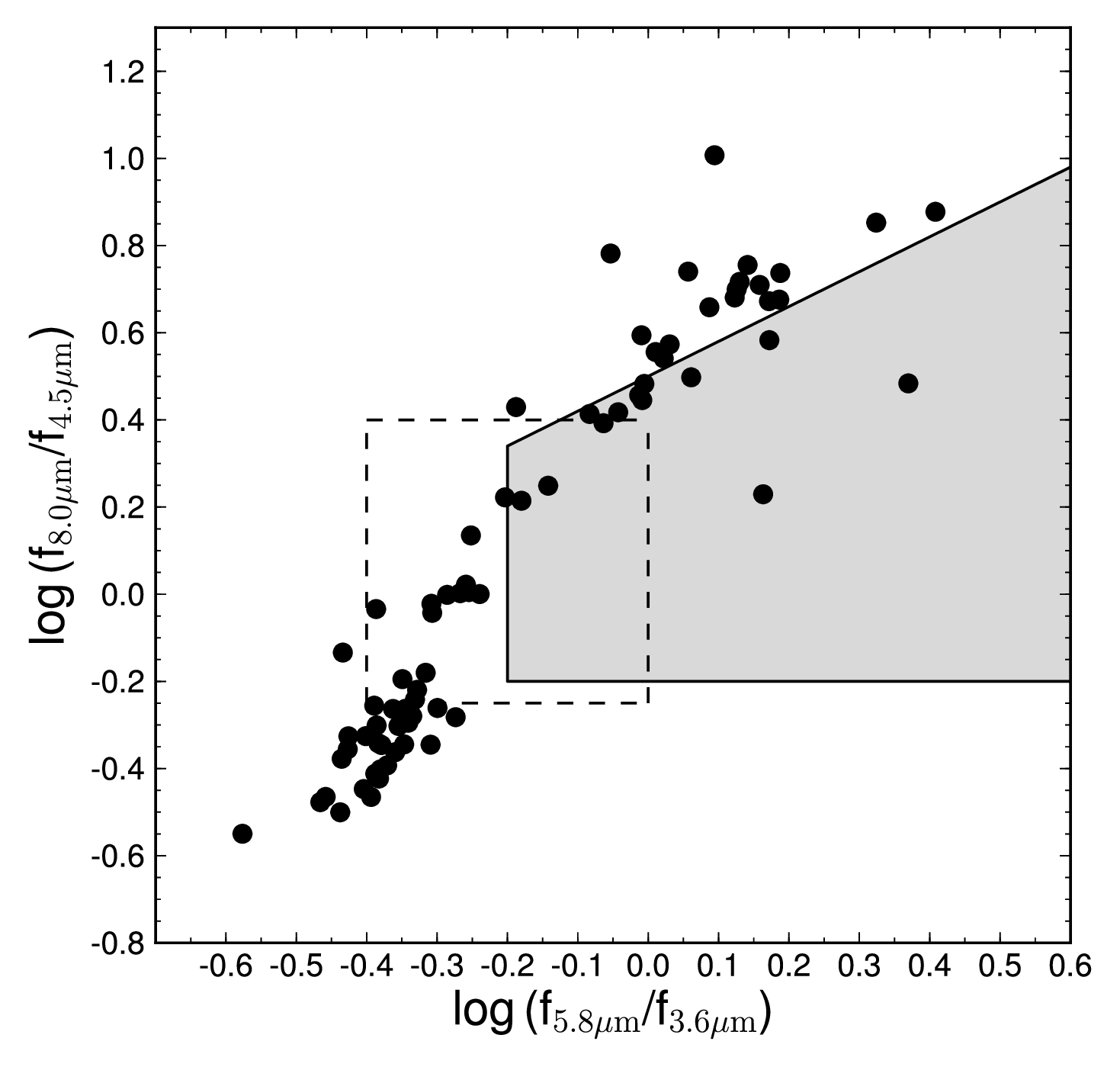}}
\hfill
\subfigure[]{\includegraphics[width=9cm]{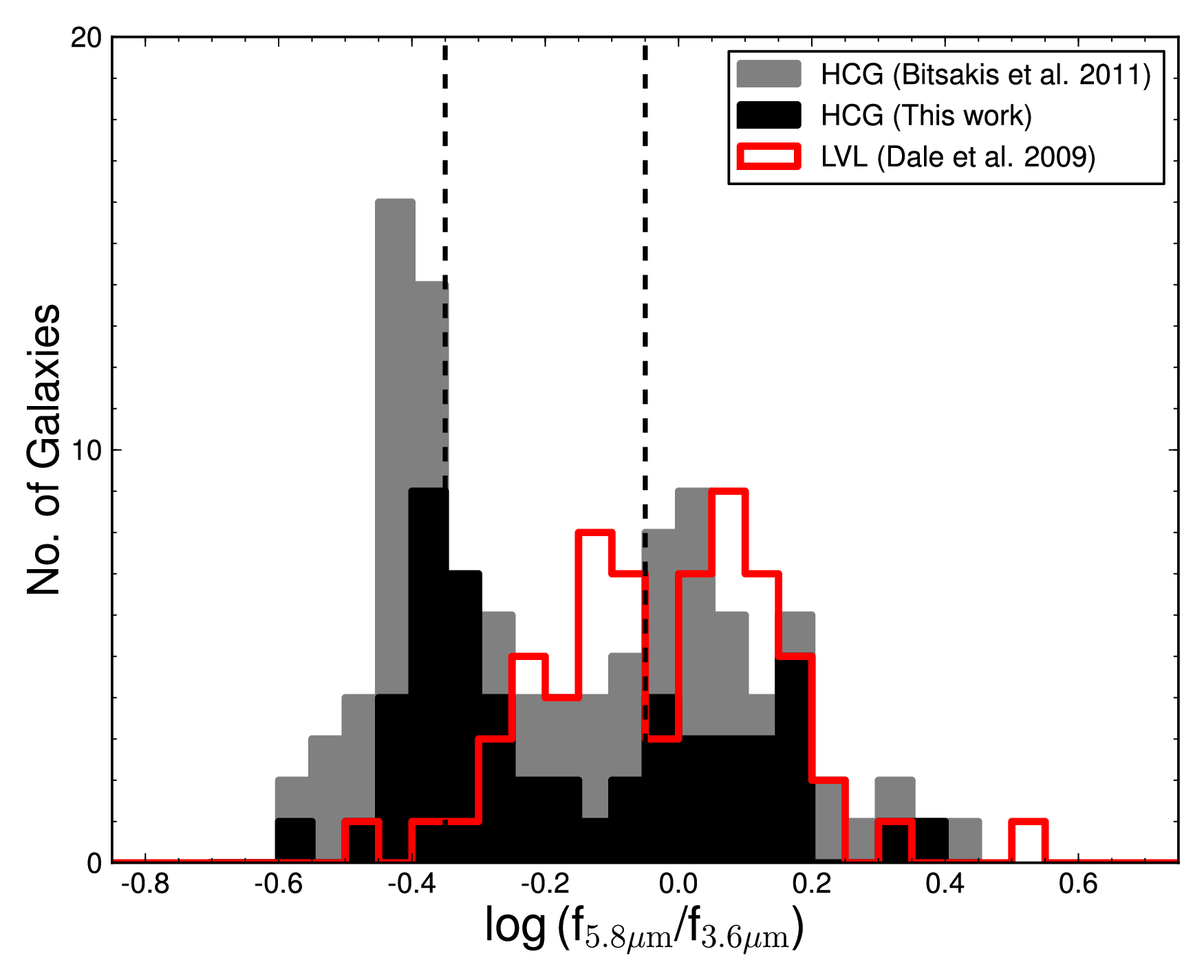}}
\caption{
a) The IRAC color-color diagram of the galaxies in our sample showing the mid-infrared colours of the chiefly intermediate \HI-deficiency groups. 
The shaded grey region indicates the AGN locus, as defined by \citet{Lacy04}, and the black dashed region shows the \citet{Joh07} ``gap" region. b) A histogram comparing the
IRAC ${\rm [f}_{5.8\micron}/{\rm f}_{3.6\micron}]$ colors within our sample (black filled; $z<0.035$ galaxies only to avoid colors affected by redshifted spectral features), versus the larger HCG sample
of \citet[][grey filled; also excluding galaxies at $z>0.035$]{Bit11}, as well as the distribution 
from the Local Volume Limited (LVL) sample of nearby galaxies of \citep[][red unfilled]{Dal09} after applying a luminosity cut (see text). The black dashed lines indicate galaxies with intermediate mid-infrared colours, as determined from the distribution of \citet{Bit11}.}
\label{fig:samp}
\end{center}
\end{figure*}

We have selected 23 compact groups from the HCG catalog of \citet{Hick82}, representing 25$\%$ of the physically associated groups, as targets for probing active transformation.
The aim of the project was to search for evidence of extended
molecular hydrogen emission by selecting compact groups with
intermediate \HI\ deficiencies (like SQ) which were
reasoned more likely to be in an active phase of gas
stripping. 

The sample of compact groups studied by \citet{VM01} 
contained 72 groups with \HI\ deficiencies ranging from -0.8 $<$ log\, [M(\HI)$_{\rm pred}$]$-$log [M(\HI)$_{\rm obs}$] $<$ 1.56. From this list, we selected galaxy groups with
an ``intermediate HI deficiency" i.e. defined as having deficiencies in the range
0 $<$ log[M(\HI)$_{\rm pred}$]$-$log[M(\HI)$_{\rm obs}$] $<$ 0.9. This resulted in
50 out of 72 groups with a median \HI\ deficiency of 0.48, which is close to the
value of 0.49 for Stephan's Quintet. We note that the median deficiency for
the full (original) sample is 0.27. However, as shown by \citet{VM01}, the spread in the distribution is quite large, and our sample
selection has resulted in effectively clipping the extreme ends of the
distribution for all groups. The term ``intermediate \HI\ depletion" should
therefore be seen in that context.  In order to form a practical sample for
observation with the IRS on {\it Spitzer}, and to further maximize the chances of
finding dynamically active systems like Stephan's Quintet, we only
considered those groups that showed visible signs of tidal interaction, specifically disturbed optical disks and tidal tails, in two or more members. This
resulted in 20 groups. We added an extra group, HCG 40, on the grounds that its X-ray morphology was quite similar to that of Stephan's Quintet, although its \HI\ deficiency is higher (log\,[M(\HI)$_{\rm pred}$]$-$log [M(\HI)$_{\rm obs} =$0.97) than most of the groups selected. In addition we added HCG 55 and 75 as groups with close associations that contain signatures of interaction.
The final sample of 23 groups is listed in Table \ref{tab:prop}; galaxies that are not group members (i.e. with discordant redshifts) are indicated and not included in this analysis.

\begin{table}[!thb]
{\scriptsize
\caption{HCG Sample \label{tab:prop}}
\begin{center}
\begin{tabular}{c c r c c }
\hline
\hline
\\[0.5pt]
Group      &  $z^{a}$    &  \HI\ Deficiency$^b$  & Galaxies & Designation$^c$ \\
  &   &    &  Sampled &\\
\\
\hline
\\
HCG 6     &  0.0379   &  0.33   &  4   &  A, B, C, D$^\diamond$\\
HCG 8     &  0.0545   &  $>0.04$  &   3   & A, C, D\\
HCG 15     &  0.0228   &  0.62 &  3   & A, C, D\\
HCG 25     &  0.0212   &   0.26   &  4   & B, C$^\dagger$, D, F\\
HCG 31     &  0.0135   &   0.18   & 2  & A+C$^\bullet$, B\\
HCG 40     &  0.0223   &   0.97   &  4   & A, B, C, D\\
HCG 44     &  0.0046   &   0.69   & 3 &  A, B, D \\
HCG 47     &  0.0317   &   0.28   &  3  & A, B, D\\
HCG 54     &  0.0049   &   0.49   &  3   & A, B, C\\
HCG 55     &  0.0526   &   --   &   5   & A, B, C, D, E$^\dagger$\\
HCG 56     &  0.0270  &   0.73   &  4   & B, C, D$^\star$, E\\
HCG 57     &  0.0304   &  0.86  &   5   & A, B, C$^\diamond$, D$^\diamond$, E\\
HCG 62     &  0.0137   &   $>0.46$   &  3   & A, B, C\\
HCG 67     &  0.0245   &   0.27  &   3  & A, B, D$^\diamond$\\
HCG 68     &  0.0080   &   0.48   &   3  & A, B, C\\
HCG 75     &  0.0416   &   --    &   4   & A, C, D$^\diamond$, E\\
HCG 79     &  0.0145   &   0.41  &  4  & A, B, C, E$^\dagger$ \\
HCG 82     &  0.0362   &   $>0.76$   & 3  & A, B, C\\
HCG 91     &  0.0238   &   0.24    &   3  & A, C, D\\
HCG 95     &  0.0396   &   $>0.21$   &  3   & A, B$^\ddag$, C\\
HCG 96     &  0.0292   &   $>0.17$   &  3  & A, B, C\\
HCG 97     &  0.0218   &   0.89   &   3   & A, C, D\\
HCG 100   &  0.0170   &   0.5  & 3  & A, B, C\\

\\
\hline
\\[0.5pt]
\multicolumn{5}{l}{$^a$ From NED} \\
\multicolumn{5}{l}{$^b$ log[M(\HI)$_{\rm pred}$]$-$log[M(\HI)$_{\rm obs}$]; Verdes-Montenegro et al. (2001)}\\
\multicolumn{5}{l}{$^c$ From Hickson (1982)}\\
\multicolumn{5}{l}{$^\dagger$ Discordant redshift (from Hickson et al. 1992)}\\
\multicolumn{5}{l}{$^\ddag$ Discordant redshift (from Iglesias-P{\'a}ramo \& V{\'{\i}}lchez 1998)}\\
\multicolumn{5}{l}{$^\bullet$ Merging Object \citep{Gal10}}\\
\multicolumn{5}{l}{$^\star$ SL coverage only}\\
\multicolumn{5}{l}{$^\diamond$ LL coverage only}\\
\\[0.5pt]
\\
\end{tabular}
\end{center}
}
\end{table}


Although our primary selection criterion is based on the groups
exhibiting intermediate \HI-depletion \citep[based on the original definition
of][]{VM01}, the sample spans a large range of galaxy properties shared by the more complete samples of
HCG groups \citep[e.g.][]{Bit11, Walk12}. As shown in the IRAC color-color diagram (log[f$_{8.0\mu {\rm m}}$/f$_{4.5\mu {\rm m}}$] vs log[f$_{5.8\mu {\rm m}}$/f$_{3.6\mu {\rm m}}$]; Figure \ref{fig:samp}a), the intermediate \HI-depletion in these groups is not biased towards mid-infrared blue or red populations, but rather spans the entire range of mid-infrared colour seen in the studies of \citet{Bit11} and \citet{Walk12}. 

This color-color space has been shown to separate late-type, star-forming
galaxies (top right) from early-type galaxies (bottom left). 
The dashed box shows the underpopulated region found in the smaller sample of
HCGs studied by \citet{Joh07} and \citet{Walk10}. \citet{Bit11} observe a similar lowering of density of galaxies with intermediate mid-infrared colors; 
they attribute this distribution to the natural result of morphological transformation as galaxies evolve from star forming to passively evolving systems. 
 We will show later in this paper that galaxies which fall within
this intermediate region of mid-infrared color preferentially show signs of shocked molecular hydrogen
emission. This may  support the idea that the ``gap galaxies''
represent a transitional population.

The mid-infrared color properties of our sample, both in the context of the larger HCG
group population and local galaxy populations, is shown in Fig. \ref{fig:samp}b. Here we compare the IRAC colors of our intermediate \HI-deficient sample, to the larger \citet{Bit11} HCG sample of 32 groups which was not selected for deficiency. Also shown are galaxies taken from the Local Volume-Limited (LVL) sample
of \citet{Dal09}, as a comparison ``field" control sample. The LVL sample consists of 256 galaxies within 11 Mpc and is dominated by spirals and irregulars; we apply a luminosity cut \citep[as motivated by][]{Walk12} of log(L$_{4.5\micron}$[erg/s/Hz])$>$ 27.5 to compare galaxies with similar characteristics, which leaves 65 galaxies. As expected from the morphology-density relation \citep[e.g.][]{Dress80}, the LVL field sample contains very few early-type systems.

By comparison to the LVL sample, HCG group galaxies appear to
have a well-defined red and blue sequence in this color space, with a deficiency of galaxies at intermediate color not seen in the
volume-limited sample of nearby galaxies, i.e. between \begin{eqnarray}
-0.35 \le {\rm log}\, {\rm f}_{5.8\micron}/{\rm f}_{3.6\micron}\le-0.05, 
\end{eqnarray}
This deficiency, although not
a complete ``gap'', is what led \citet{Walk10} to suggest HCG galaxies
rapidly evolved through this intermediate color region. 

\section{Observations and Data Reduction}\label{obs}

\subsection{{\it Spitzer} IRS Spectroscopy}\label{irs_obs}

The galaxies and groups  listed in Table 1 were targeted by the {\it Spitzer} IRS instrument \citep{Hou04} using the low-resolution Short-Low ($R\sim60-127$; $5.2-14.5$ \micron) and Long-Low
($R\sim 57-126$; $14.0-38.0$ \micron) modules. Observations were carried out as part of  GO-5 PID 50764 and taken between 2008, June 29 and 2009, January 19.

\begin{figure}[!tbh]
\begin{center}
\includegraphics[width=4.2cm]{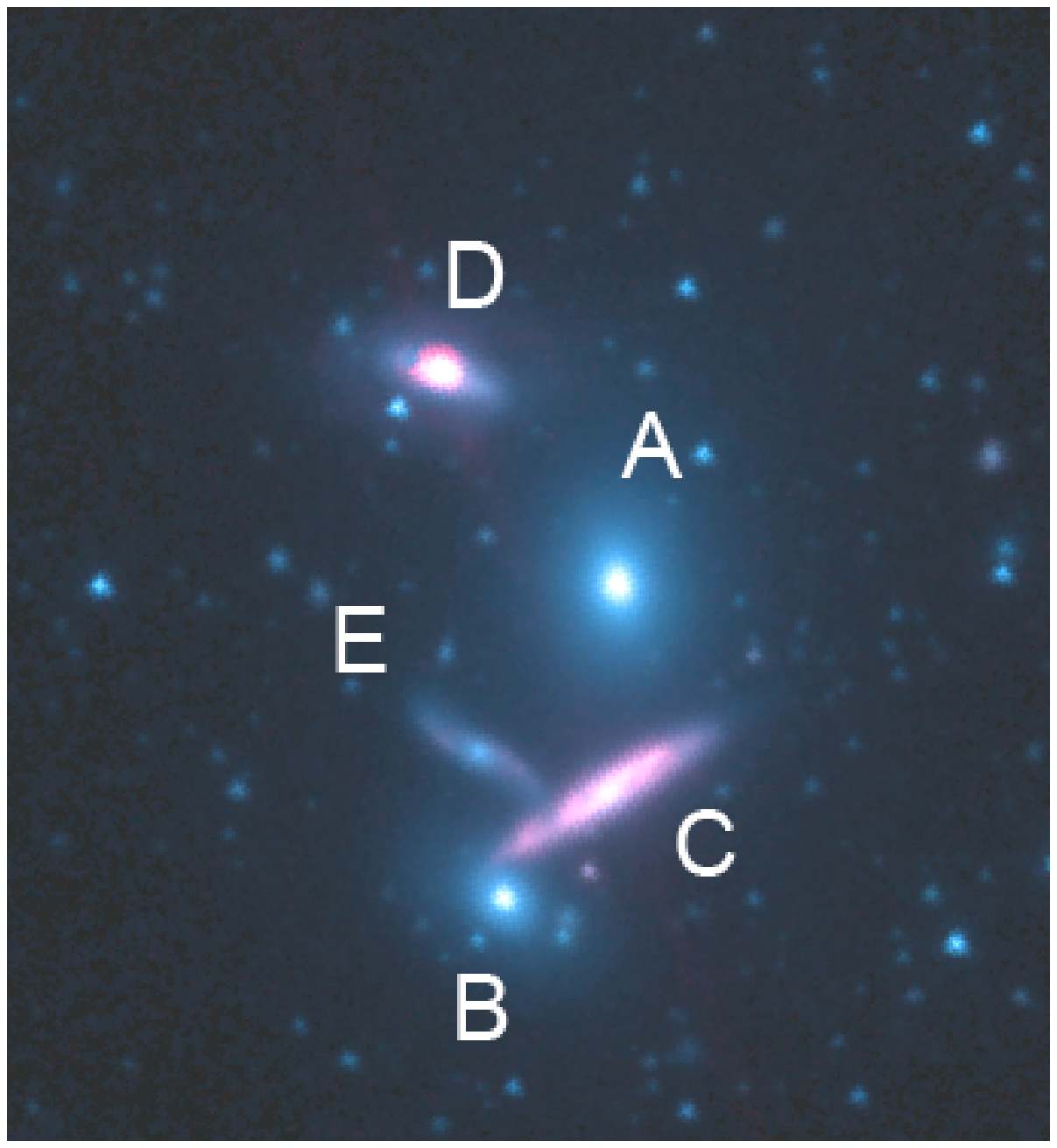}
\includegraphics[width=4.2cm]{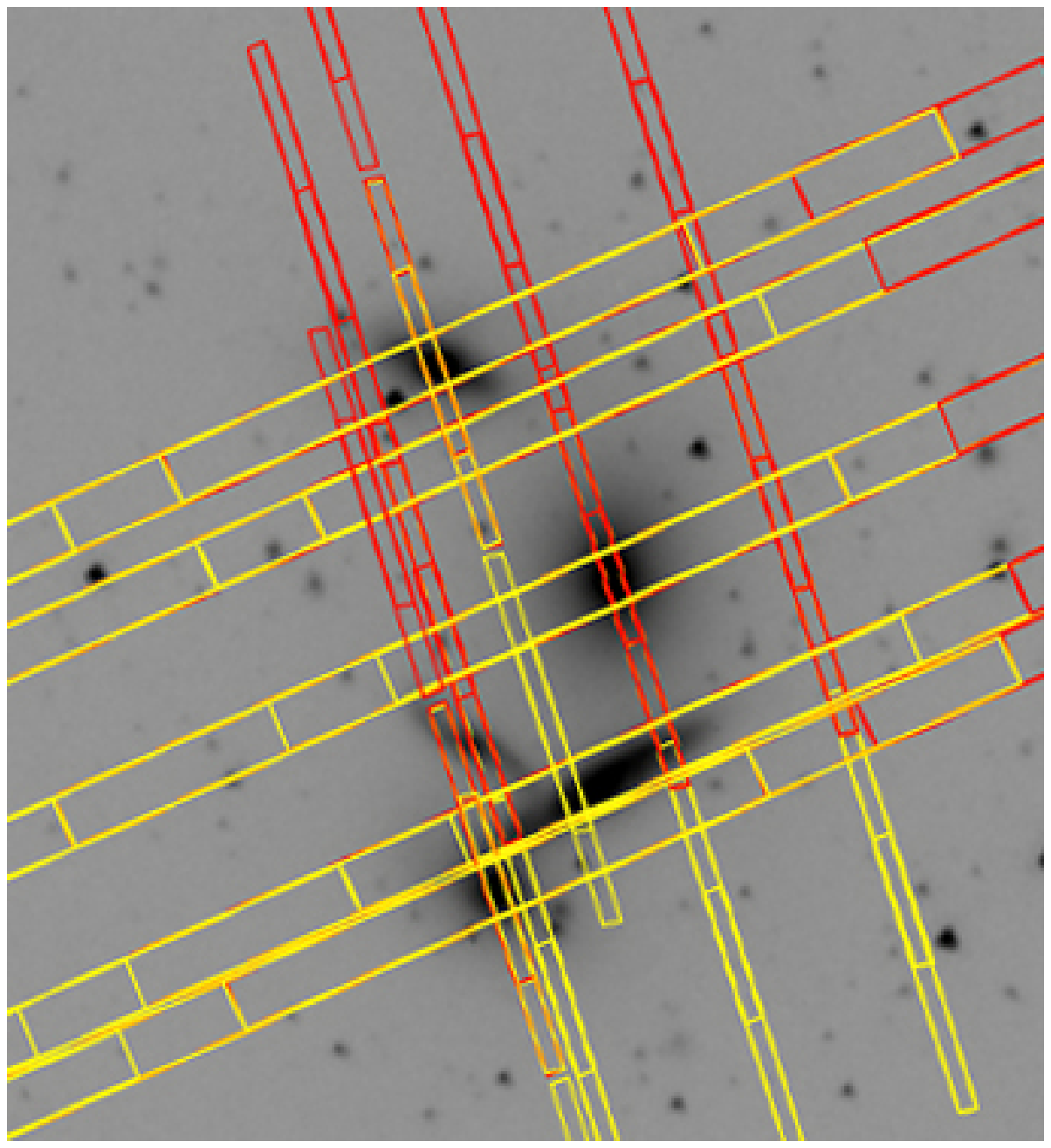}

\caption{The IRS sampling (Long-Low; $\sim$ 10.5\arcs\  and Short-Low; $\sim$ 3.6\arcs\ slits) shown on HCG 40 (IRAC four-color image of $\sim$4\arcmin $\times$ 4\arcmin). 1$^{\rm st}$ order spectral coverage is shown in yellow with 2$^{\rm nd}$ order in red. It should be noted the mapping combined with staring strategy probes the intragroup region, as well as the galaxies themselves.}
\label{fig:HCG_eg}
\end{center}
\end{figure}

Since it was not
known apriori where in the group environment shock-excited
\Ht\ emission may be located, the primary observations employed a
sparse mapping strategy -- centering a 3-leg grid\footnote{In most cases this was a 2$\times$3 sparse map
with typical step sizes of 30\arcs\ parallel (to the slit) steps and 35\arcs\ perpendicular for SL, and 70\arcs\ and 35\arcs\ parallel and perpendicular 
in the LL module.} on the most disturbed member of the
group (see Figure \ref{fig:HCG_eg}). The scale was adjusted for each group to provide
good coverage of the inner group in both modules. The typical coverage was
linear $\sim$70\,kpc for SL and  $\sim$180\,kpc for LL. In addition, a further two
member galaxies were targeted in IRS ``Staring Mode'', where the target center is placed at the $\tfrac{1}{3}$ and $\tfrac{2}{3}$ position along the length of the slit. 
Because of the way the IRS performs a staring mode observation (by sampling first the 2nd order spectrum followed by a similar observation in first order), this increases the coverage of the group, providing further information about potential extended emission in the groups, as well as often intersecting by chance (depending on the roll-angle of the focal plane at the time of the observations) additional group members. 
As a result, the SL and LL  slits typically sampled $\sim$800\,kpc$^2$ and $\sim$6000\,kpc$^2$, respectively, covering the IGM and the galaxies themselves.   

Primary data reductions were performed by the {\it Spitzer} Science Center
(SSC) pipeline, version S18.0.2-18.7.0, which performs standard spectral reductions such as wavelength and flux calibration, ramp fitting, dark current subtraction and detector droop and non-linearity linearity corrections. 
Basic Calibrated Data (BCDs) frames, output from the pipeline, were combined within the SSC tool CUBISM \citep{Smi07a}, optimised for extended sources, with each AOR forming one spectral cube. This was done to ensure proper background subtraction, using backgrounds taken close in time to the observations. This was achieved using either dedicated backgrounds or from the ``off" position BCDs with coverage outside of the group.

Pixel outlier rejection was done using the CUBISM algorithm (with a conservative 8$\sigma$ clipping) and then by visual inspection of the spectral cubes to ensure that no weak signals are lost. The cubes were inspected for any \Ht\ emission in the IGM and spectra extracted for each galaxy with coverage. Since we are particularly interested in emission from the disk, we were careful to extract spectra along the slit with an aperture large enough to capture most of the source's light while still maximising the signal to noise of the spectrum. Extraction areas for SL and LL for each galaxy are listed in Table \ref{tab:extract}. Due to the coverage obtained through the sparse mapping mode, several galaxies have ``off-nuclear" coverage where the slits have not been centred on the nucleus; these are indicated in Table \ref{tab:extract} and are treated as separate spectra to those that cover the nuclear region. 

Spectra were extracted for all but 4 galaxies in the sample (HCG 6A, 6C, 6D and 79C) where no signal above the noise was detected. Spectra were converted to flux densities (mJy) using the extraction areas, and SL scaled to LL to match continuum levels in cases that required it (see Table \ref{tab:extract}). This accounts for beam resolution differences 
between the long and short order; here we make the assumption that the emission in each slit is uniform. The scaling of spectra is further discussed in Section \ref{samp_scale} of the Appendix.

PAHFIT was used to characterize our spectra, but each visually inspected one by one to determine which lines were detected reliably and which were marginal detections ( $<$2.5$\sigma$).
PAHFIT is a spectral decomposition package \citep{Smi07b} that fits emission lines, bands and dust continua to stitched LL and SL spectra. 

The spectra RMS (root mean square) for determining upper limits were measured using ISO Spectral Analysis Package (ISAP)\footnote{The ISO Spectral Analysis Package (ISAP) is a joint development by the LWS and SWS Instrument Teams and Data Centers. Contributing institutes are CESR, IAS, IPAC, MPE, RAL and SRON.}.

\subsection{{\it Spitzer} IRAC and MIPS Photometry}\label{irac_obs}

The {\it Spitzer} IRAC \citep{Faz04} and MIPS \citep{Riek04} instruments were used to obtain imaging at 3.6, 4.5, 5.8 and 8.0\micron, and 24\micron, respectively, of our sample and observed as part of PID 50764 (P.I. Appleton), PID 40459 (P.I. Le Floc'h), PID 631 (P.I. Mazzarella) and PID 101 (P.I. Kennicutt). 

IRAC photometric measurements (to obtain colours) for 21 galaxies in our sample are taken from \citet{Bit11} as indicated in Table \ref{tab:phot}. 
Fluxes for the remainder of galaxies were carefully measured to improve the deblending systematics where contamination from nearby stars and galaxies may affect the photometry. For these systems, the data was reduced using the SSC science pipeline version S18.5.0 and 18.7.0. Galaxy photometry was performed using a matched elliptical aperture, determined by the 1$\sigma$ isophote in IRAC 3.6\micron, after foreground contaminating stars were masked from all images and replaced by the corresponding isophotal value of the source. Nearby contaminating galaxies were similarly masked. The local background was determined from the median pixel value distribution within a surrounding annulus. Aperture corrections were applied as specified by the IRAC Handbook. 
The formal photometric uncertainties are $\sim$5\% for the IRAC calibration error.

Potentially saturated sources, particularly in IRAC\,3.6\micron\ and 4.5\micron, namely HCG 56B, 91A, 96A and 100A were investigated. HCG 96A was saturated in the cBCDs at  IRAC\,3.6\micron, 4.5\micron\ and 5.8\micron. For this system the 1.2\,s HDR (High Dynamic Range) exposures were used to determine the IRAC fluxes. For HCG 100A the source counts for all bands was nominal, although at the full well capacity. For 56B and 91A the source counts for IRAC 5.8\micron\ and 8.0\micron\ were nominal and below the saturation limit at 3.6\micron\ and 4.5\micron. However, since the peak pixel flux is within the non-linear regime at 3.6\micron\ and 4.5\micron, the integrated fluxes may be slightly underestimated (as indicated in Table \ref{tab:phot}), but the mid-infrared colors are likely unaffected. 

For galaxies located at $z>0.035$ the shifting emission features, in particular the 6.2\micron\ PAH, affect the observed colours of a galaxy. For the groups in our sample affected by this (HCG 6, 8, 55, 75, 82 and 95) we have corrected the IRAC fluxes for redshift (i.e. ``k-corrected") using the empirical template library from M. Brown et al. (in prep.). Consisting of 125 galaxy templates of local, well-studied and morphologically diverse galaxies (e.g. SINGS, the {\it Spitzer} Infrared Nearby Galaxy Survey), these are generated using optical and {\it Spitzer} spectroscopy with matched aperture photometry from {\it GALEX}, {\it XMM} UV, SDSS, 2MASS, {\it Spitzer} and WISE, synthesized with MAGPHYS \citep{dacun08}. For galaxies with recessional velocities $<9000$\,km\,s$^{-1}$, k-corrections do not appreciably affect the mid-infrared colours, or analysis derived thereof, presented in this work.

MIPS 24\micron\ data was processed through SSC science pipeline versions 18.1.0, 18.12.0 and 18.13.0, achieving a spatial resolution of $\sim$6\arcs. The LL spectral extraction areas output from CUBISM were used to make matched aperture photometric measurements using the IRAF\footnote{IRAF is distributed by the National Optical Astronomy Observatory, which is operated by the Association of Universities for Research in Astronomy, Inc., under cooperative agreement with the National Science Foundation} task, POLYPHOT. For star-forming and AGN-dominated spectra showing continua with a strong power-law dependence, we applied colour corrections as recommended in the MIPS Handbook. This correction is of the order of $\sim$5\%. The MIPS calibration error is of the order of $\sim$10 -- 20\%. . 

In the cases of 31AC, 40C, 44A and 75D we have \Ht\ detections in LL coverage without matching SL coverage. In order to measure the 7.7\micron\ PAH emission, we use the prescription of \citet{Hel04} and measured matched (to the LL extraction) apertures of the IRAC 3.6\micron\ and IRAC 8\micron\ bands. By subtracting a scaled version of IRAC 3.6\micron\ from IRAC 8\micron\, we can compensate for stellar light contamination and use this as a measure of the strength of the aromatic emission within the band \citep[see, for example,][]{Rous07}. This was also done for the IGM detections discussed in Section \ref{igm}.

\section{Results}

\subsection{Emission from Warm Molecular Hydrogen}

The pure rotational transitions of molecular hydrogen covered by the IRS SL and LL spectral windows at the redshifts of the HCGs, are the 0-0 S(0), S(1), S(2), S(3), S(4) and S(5) lines at 28.22, 17.03, 12.28,
9.67, 8.03 and 6.91\micron, respectively. These can be excited in a variety of astrophysical processes including, UV pumping and collisional heating in photodissociation regions associated with star formation \citep[e.g.][]{HT}, X-ray heating in XDRs (X-ray dominated regions) particularly those associated with AGN \citep[e.g.][]{DW}, cosmic ray heating \citep[e.g.][]{Dal99} and heating by turbulence or shocks \citep[e.g.][]{SH}. We detect ($>2\sigma$) two or more lines of warm \Ht\ in 32/74 galaxies in our sample; this includes marginal detections (i.e. between 2 and 3 $\sigma$), but we do not include these in the analysis that follows. In the group IGM, we find evidence of two locations with tentative detections of excited \Ht\ -- both suffer diminished signal to noise of the S(1) line due to an artifact latent in the first set of BCDs.

Many of the galaxies located in HCGs are star forming systems, generating a UV radiation field capable of heating very small grains (VSGs) and exciting PAH molecules, thus producing distinctive dust features in the mid-infrared.  We shall focus on systems where the \Ht\ emission is enhanced relative to UV-excitation. In order to separate mechanical heating of \Ht\ from UV-heating within photodissociation regions (PDRs), we use the 7.7\micron\ PAH emission band as a discriminator \citep[described by][]{Og10}. The ratio of \Ht\ luminosity (summed over the $0-0$\,S(0)--S(3) lines) to the 7.7\micron\ PAH luminosity allows us to determine which systems have enhanced 
\Ht\ relative to SINGS star-forming galaxies \citep{Rous07} and thus exhibiting ``MOHEG-type" emission. 
This value ($\ge$ 0.04) of \Ht\,S(0)-S(3)/7.7\micron\,PAH, hereafter referred to as \Ht/7.7\micron\ PAH for brevity, separating PDR-dominated \Ht\ heating from other heating sources has been demonstrated by radiation modeling \citep{Guil12b}.


\subsubsection{\Ht\ in the Group IGM}\label{igm}

\begin{figure}[!t]
\begin{center}
\subfigure[]{\includegraphics[width=7cm]{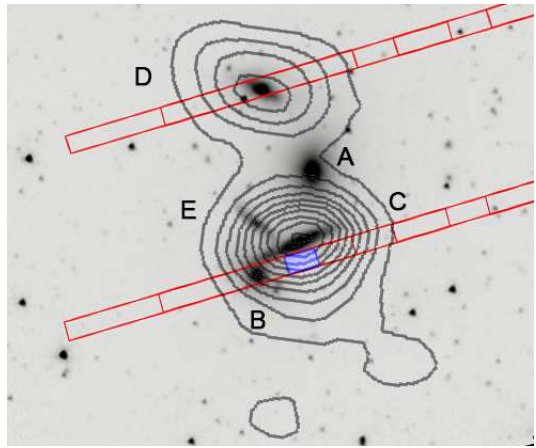}}
\hfill
\subfigure[]{\includegraphics[width=8cm]{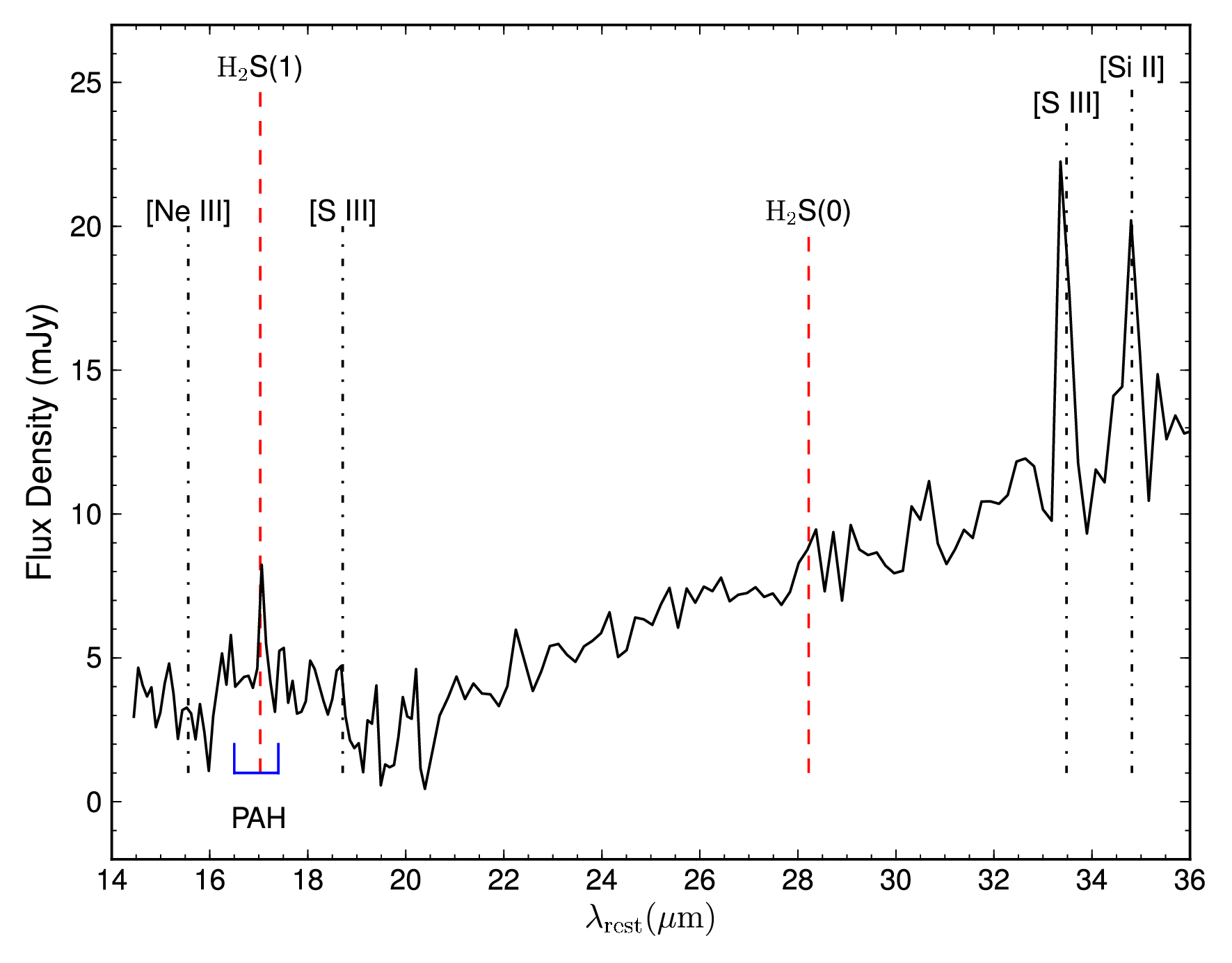}}
\caption
{a) HCG 40 IRAC 3.6\micron\ image ($\sim$4\arcmin $\times$ 4\arcmin) with LL slit coverage overlaid; the blue box shows the position of the extraction. The VLA integrated \HI\ distribution (Jy/beam/s) from Verdes-Montenegro et al. (private communication) is shown as overlaid contours; image reproduced with kind permission. North is up and East is left. b) LL Extraction centered on 09$^{\rm h}$38$^{\rm m}$53.45$^{\rm s}$, -04$\arcdeg$51$\arcmin$47.2\arcs. Due to a latent-induced flux bias in the first set of BCDs, the signal-to-noise of the S(1) line is diminished.}
\label{fig:spectra_out}
\end{center}
\end{figure}

\begin{figure}[!th]
\begin{center}
\subfigure[]{\includegraphics[width=8cm]{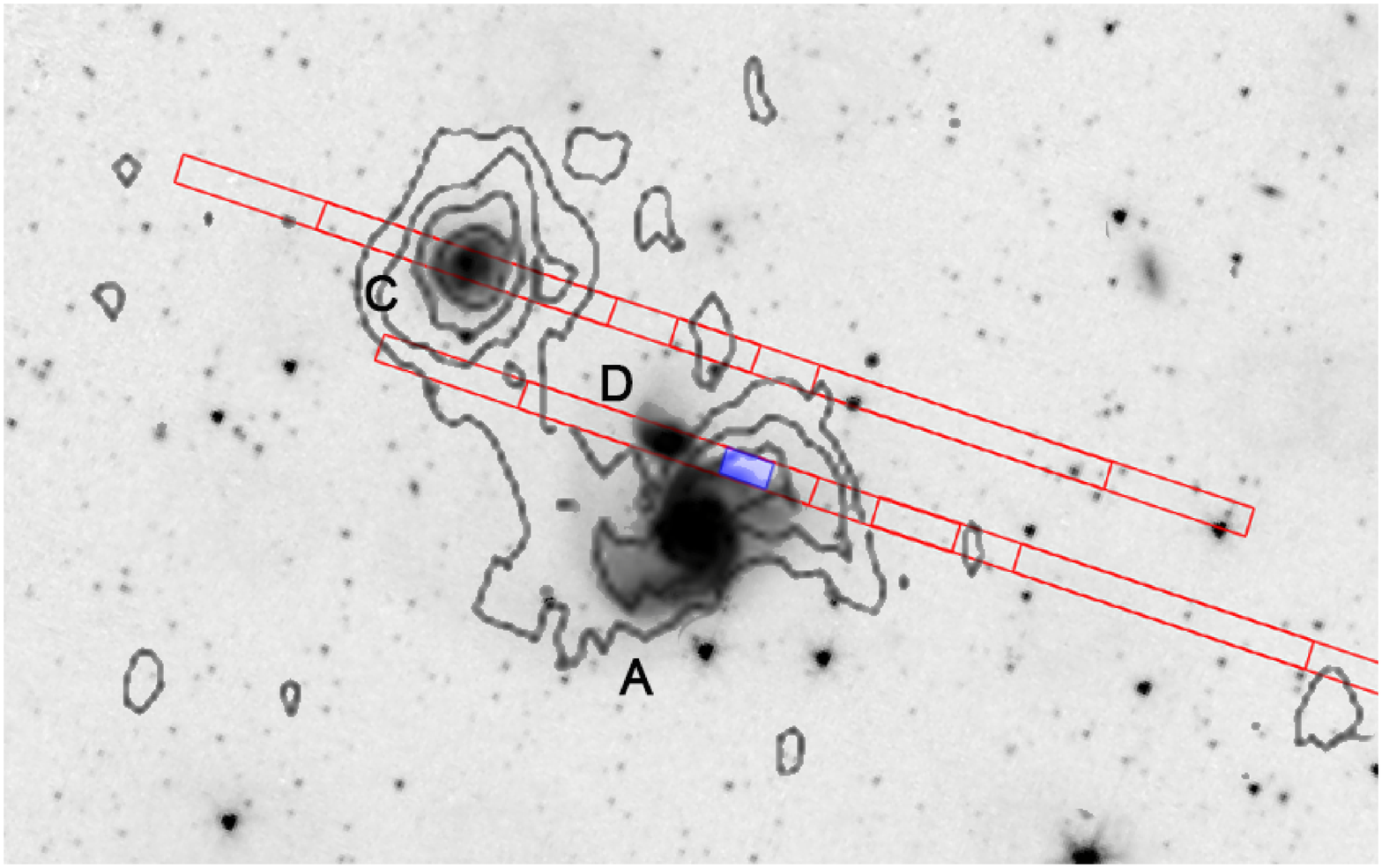}}
\subfigure[]{\includegraphics[width=8cm]{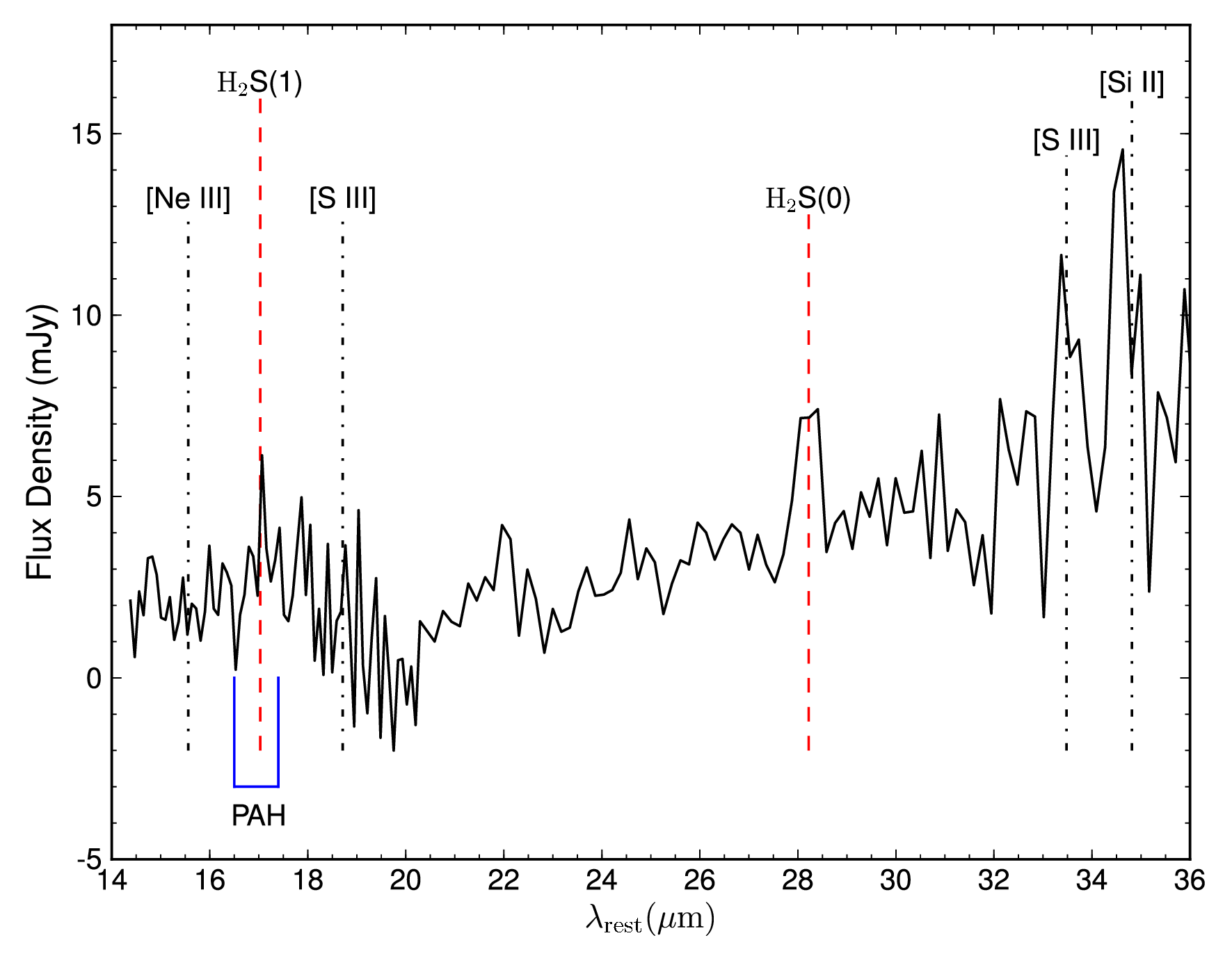}}
\caption
{a) HCG 91 IRAC 3.6\micron\ image ($\sim$6\arcmin $\times$ 10\arcmin) with LL slit coverage overlaid; the blue box shows the position of the extraction. The ATCA \HI\ distribution from \citet{Bar01} are shown as contours. Image reproduced with kind permission from the authors. North is up and East is left. b) LL extraction centered on 22$^{\rm h}$09$^{\rm m}$06.20$^{\rm s}$, -27$\arcdeg$48$\arcmin$09.1\arcs. Due to a latent effect in the first set of BCDs, the signal-to-noise of the S(1) line is diminished.}
\label{fig:spectra_out2}
\end{center}
\end{figure}

A key aim of this project was to determine the prevalence of warm \Ht\ emission in the IGM of HCGs most likely to be in an active stage of transformation.
From the sample of 23 groups, we have discovered two locations showing warm \Ht\ detections in the IGM. These are both detected in the LL spectra and appear associated with the edges of galaxy disks. Figure \ref{fig:spectra_out} and \ref{fig:spectra_out2} show the detections in HCG 40 and HCG 91, respectively. We consider these preliminary, demanding follow-up observations for confirmation.

In HCG 40, we detect the S(1) line (Figure \ref{fig:spectra_out}) outside the Sb galaxy HCG 40C (the positions of the IRS slits are shown in Fig. \ref{fig:HCG_eg}).
Using the IRAC 3.6\micron\ and 8.0\micron\ coverage to provide an estimate of the 7.7\micron\ PAH emission  (4.77$\times 10^{-16}$\,W.m$^{-2}$) and the combined S(0) and S(1) flux of 1.02$\times 10^{-17}$\,W.m$^{-2}$, we find a \Ht/7.7\micron\ PAH ratio of $>0.021$. This can regarded as a lower limit due to missing SL coverage, as well as the poor signal to noise of the S(1) line due to the latent-induced flux bias in the first set of BCDs.

The VLA \HI\ distribution (from Verdes-Montenegro et al., private communication) is shown as contours in Figure \ref{fig:spectra_out}a and indicates the presence of \HI\ around 40B and 40C, with a tail towards 40D. 
Interactions with a tidal tail could account for this emission, and would also explain the (albeit weak) warm \Ht\ signal from 40B presented in the next section. Follow-up IRAM CO observations (Lisenfeld et al., in prep.) find indications for extended molecular CO emission at the location of the IGM detection within the group.

In HCG 91, we find a detection at what appears to be the edge of the disk of 91A, a star-forming galaxy (Figure \ref{fig:spectra_out2}). The LL2 spectrum suffers from a similar latent effect as above (due to observing a bright target prior to this observation) and the first BCD is contaminated, resulting in a lower signal to noise for the S(1) detection, yet there is a strong S(0) detection. 
Moreover, the estimated \Ht/7.7\micron\ PAH ratio (calculated as above, with an \Ht\ flux of 1.43$\times 10^{-17}$\,W.m$^{-2}$ and 7.7\micron\ dust estimate of 3.51$\times 10^{-16}$\,W.m$^{-2}$) is $>$0.041 and therefore would be classified as a MOHEG even without SL coverage.

\citet{Bar01} find two \HI\ knots centered around HCG 91A and 91C with a connection between the two through a gas bridge (shown as contours in Fig. \ref{fig:spectra_out2}a), since there is a common velocity between the southern part of HCG 91C and the northern part of HCG 91A. \citet{Am03} find that the H$\alpha$ distribution shows a tidal arm pointing from HCG 91A towards HCG 91C. They also find a double gaseous component for HCG 91C strongly suggestive of a past interaction. They propose a scenario where HCG 91C is passing through the group forming the tail of HCG 91A.

The location of these two detections suggest a possible connection with disks interacting with the group IGM.

\subsubsection{\Ht\ in Individual Group Galaxies}

The mapping strategy employed in this study resulted in 74 compact group galaxies with an IRS spectrum (either full or partial). In Table \ref{tab:phot} of the Appendix, we indicate whether a warm \Ht\ detection was made in an individual galaxy.
The fluxes determined for the \Ht\ emission lines are presented in Table \ref{tab:Ht};
we note that several systems have their SL and LL lines presented separately due to the regions sampled by the IRS not overlapping (and therefore not joined together). In addition, spectra that are not centred on the nucleus (and therefore dominated by emission from the disk) are indicated. For completeness, extraction areas for \Ht-detected galaxies are listed in Table \ref{tab:extract}.

The strengths of the PAH complexes and atomic emission lines for the \Ht-detected systems are presented in Tables \ref{tab:pah} and \ref{tab:FS}, respectively. Upper limits for the \Ht\ S(0)--S(3) emission for galaxies without \Ht\ detections, and measurements of their PAH features, are included in the Appendix.

\begin{table}[!t]
{\scriptsize
\caption{\Ht-detected Galaxies: \Ht\ and 24\micron\ Data \label{H2_lum}}
\begin{center}
\begin{tabular}{l c c c  r c}
\hline
\\
\hline
\\[0.5pt]
Source   & \Ht\,S(0)--S(3)   &  log\,$L({\rm H_2})$ & \Ht/7.7\micron$^a$  &   $F_{24\mu\,{\rm m}}$ & log\,$L_{\rm 24}\,^b$\\ 
   &   (W\,m$^{-2}$)    & $(L_\sun)$ &    &   mJy  & $L_{\sun}$  \\
\\
\hline
\\

6B     &   1.633e-17   &    7.093  & 0.198  & 2.14   & 8.31 \\  
 15A     &   8.012e-18      &    6.321 & 0.044  &  3.46  &  8.05 \\  
 15D     &   2.280e-17    &    6.776 & 0.112   & 3.25    &  8.03  \\  
 25B$^\star$     &   3.299e-18   &    5.883 & 0.098  & -- \\  
 25B$^{\diamond\dagger}$     &   1.159e-17    &    6.429  & --  &  3.08  & 7.95 \\  
 31A+C$^\diamond$    &   2.004e-17    &    6.289 & 0.002 &   286.87 & 9.54 \\  
 40B     &  5.619e-18    &    6.227 & 0.083  &   4.72 &  8.25 \\  
 40C$^\diamond$     &   9.166e-17   &    7.439  & 0.010   &  46.52 &  9.24 \\  
 40D     &   4.707e-17  &    7.150 & 0.013 &   67.71 & 9.40 \\  
 44A$^{\diamond \ast}$     &   1.338e-16  &    6.355 & 0.027  &  98.74 & 8.32  \\  
 44D     &   6.202e-17    &    6.021 & 0.021 &  35.22 & 7.87  \\  
 47A     &   2.556e-17    &    7.181  & 0.035  &   31.96 & 9.38 \\  
 55C     &   8.395e-18    &    7.126  & 0.018  &  5.31 & 9.02 \\  
 56B     &   1.829e-17   &    6.885 & 0.053 &  138.52 & 9.86 \\  
 56C     &   1.205e-17   &    6.704 & 0.037  &  $^c$3.59 &  8.27 \\  
 56D$^\star$     &   1.338e-17   &    6.749  & 0.018 &  6.03 & 8.50  \\  
 56E     &   3.169e-18   &    6.124 & 0.012  &  4.88 & 8.41 \\  
 57A$^\dagger$     &   2.501e-17     &    7.140 & --  &  -- & -- \\  
 57A$^\diamond$     &   8.385e-17    &    7.665 & 0.174 &   8.52 & 8.77 \\  
 67B     &   2.330e-17     &    6.920 & 0.006  &  20.84 &   8.97 \\  
 68A     &   4.275e-17     &    6.228  & 0.741  &  13.63   &  7.83   \\  
 68B     &   8.100e-18    &    5.505 & 0.073  &    6.77 &     7.52 \\  
 68C     &   1.311e-16    &    6.714  & 0.044  &  137.13 & 8.83  \\  
 75D$^\diamond$     &   9.676e-18   &    6.985 &  0.010  &  8.59 & 9.03 \\  
 79A     &   2.693e-17   &    6.503 & 0.011  &   15.13 &  8.35 \\  
 79B     &   1.045e-17   &    6.092 & 0.010  &    18.11 & 8.43 \\  
 82B     &   8.947e-18  &    6.821 & 0.045   &  3.26 & 8.48 \\  
 82C     &   5.567e-17    &    7.614 & 0.014   &  62.51 & 9.76 \\  
 91A     &   5.931e-17   &    7.227  & 0.030  &   159.93 & 9.75 \\  
 91C     &   1.770e-17    &    6.702 &  0.016  &  13.35 & 8.68 \\  
 95C     &   3.354e-17    &    7.438 & 0.071   &  13.35 & 9.14 \\  
 96A     &   8.378e-17    &    7.553  & 0.007   &   914.78 & 10.69  \\  
 96C     &   3.928e-17    &    7.224  & 0.018  &  30.45  & 9.21  \\  
 100A     &   5.459e-17   &    6.910 &  0.013  & 102.11 &  9.28 \\

\\
\hline
\\[0.5pt]
\multicolumn{5}{l}{$^a$ MOHEG is assigned based on \Ht/7.7\micron\ $\ge 0.04$ }\\
\multicolumn{5}{l}{$^b$ $L_{\rm 24} \equiv \nu L_{\nu}$(24\micron)}\\
\multicolumn{5}{l}{$^\star$ $S$(2), $S$(3) lines (SL coverage)}\\
\multicolumn{5}{l}{$^\diamond$ $S$(0), $S$(1) lines (LL coverage)}\\
\multicolumn{5}{l}{$^\dagger$ Off nuclear extraction}\\ 
\multicolumn{5}{l}{$^\ast$ Known MOHEG from \citet{Rous07}}\\ 
\multicolumn{5}{l}{$^c$ Contamination from 56B and 56D}\\

\\
\end{tabular}
\end{center}
}
\end{table}

\begin{figure*}[!tbh]
\begin{center}
\subfigure[HCG 57A - including the nuclear region]{\includegraphics[width=8cm]{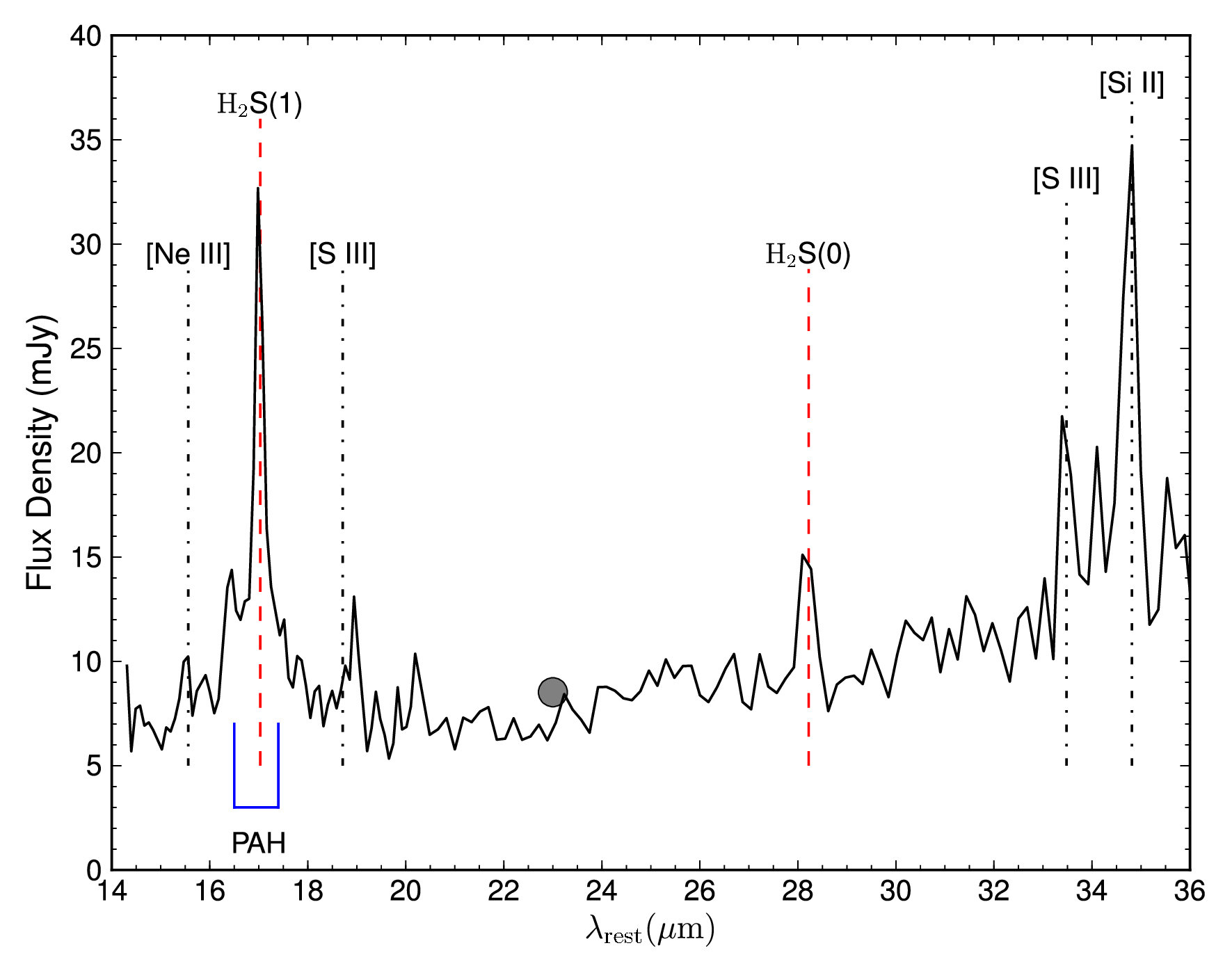}}
\hfill
\subfigure[HCG 57A - off-nuclear extraction]{\includegraphics[width=8cm]{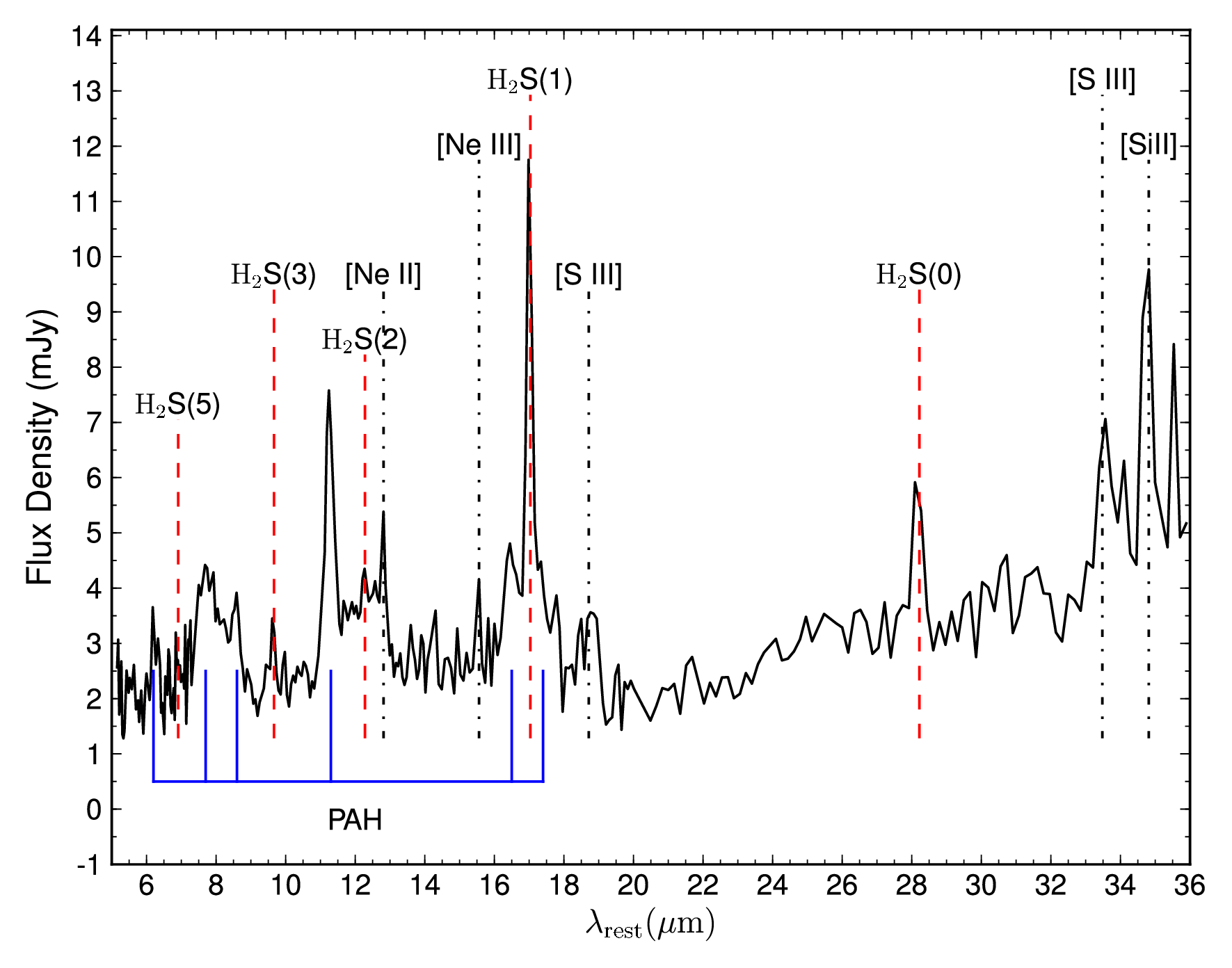}}
\caption{Spectra of HCG 57A. The matched MIPS 24\micron\ photometry is shown as a filled grey circle.}
\label{H2_57}
\end{center}
\end{figure*}

In Table \ref{H2_lum} we list the summed \Ht\ fluxes, \Ht/7.7\micron\ PAH ratio and MIPS 24\micron\ fluxes (measured within a matched aperture). Upper limits and marginal detections are not included in the summed \Ht\ fluxes. Our sample contains 13 systems with enhanced \Ht\ emission, satisfying the MOHEG criteria (\Ht/7.7\micron\, PAH $\ge 0.04$); here we include HCG 56C with its \Ht/7.7\micron\, PAH ratio of 0.037. The remaining systems (19 in total), have nominal \Ht\ emission, consistent with UV photoionization. There is one exception to list: HCG 44A has limited IRS coverage, rendering the \Ht/7.7\micron\ PAH ratio as an upper limit. However, from the SINGS study, we know it has a \Ht/7.7\micron\ PAH value of 0.042 \citep{Rous07} and is therefore a weak MOHEG.
We therefore include it as a MOHEG in our sample and use this value and its value of L(\Ht\,S(0)-S(3))/L$_{24} = 0.022$ \citep{Rous07} in future analyses.


\begin{table}[!tbph]
{\scriptsize
\caption{Morphologies and Optical Nuclear Classifications \label{tab:morph}}
\begin{center}
\begin{tabular}{l l   l  r}
\hline
\hline
\\[0.5pt]
Galaxy    &  Morphology &  Nuclear class. & T- \\ 
      &   & (optical) & Type$^\dagger$ \\

\hline
\\
\multicolumn{4}{c}{ MOHEGs with $-0.35 \le {\rm log[f}_{5.8\mu \rm m}/{\rm f}_{3.6\mu \rm m}] \le -0.05$}\\
\hline
\\
6B  & S?$^\diamond$, Sab$^\star$ &  TO$^a$  & --  \\
15A  &  S0$^\diamond$ &   dLINER$^b$  &  -2\\
15D & S0$^\diamond$ & LINER$^{b}$   &  -3\\
25B  &  SBa$^\star$  & \HII$^c$ &  -2\\
40B   & SA0(r) pec$^\diamond$  &  No emission$^b$
  &  -3\\
44A  & SA(s)a pec$^\diamond$    & AGN$^a$   &  1\\
56C  & S0/a pec$^\diamond$ & No emission$^b$
&  0\\
57A  & Sab? pec$^\diamond$, Sb$^\star$  &   AGN$^a$   &  2\\
68A  & S0?$^\diamond$, S0$^\star$    &  No emission$^b$
 &  -2\\
68B  & S0$^\diamond$   &   TO$^a$, dLINER$^b$   &  -2\\
82B & SB0?$^\diamond$, Sa$^\star$  &  TO$^a$   & -2\\
95C &  Sm$^\star$  &   AGN$^a$  &  --\\
\hline
\\
\multicolumn{4}{c}{ MOHEGs with ${\rm log[f}_{5.8}/{\rm f}_{3.6}]>-0.05$}\\
\hline
\\
68C  & SB(r)b$^\diamond$  &  TO$^a$, LINER$^b$
&  3\\
56B$^\ddagger$  &SB0$^\star$    &  AGN$^a$, Sy 2$^b$
&  --\\

\hline
\\
\multicolumn{4}{c}{non-MOHEG \Ht-galaxies}\\
\hline
\\
31A+C   & Sm/Im$^\star$   &  \HII$^a$  & --\\
40C  &  SB(rs)b pec$^\diamond$   &  TO$^a$, \HII$^b$ &  3\\
40D  & SB(s)0/a pec$^\diamond$  & TO$^a$, \HII$^b$ &  0.3\\
44D  & SB(s)c pec$^\diamond$   &  TO$^a$   & 5\\
47A  & SA(r)$^\diamond$   &  AGN$^a$ & --\\
55C  & SBa pec?$^\diamond$, E$^\star$  &  --  &  1\\
56D  & SA(s)0/a pec?$^\diamond$, S0$^\star$, $\bullet$  & TO$^a$, \HII$^b$
& 0\\
56E  & SB0 pec?$^\diamond$, S0$^\star$, $\bullet$   &  \HII$^a$, \HII?$^b$   & -2.1\\
67B  & Sb$^\diamond$   &  \HII$^a$
&  3\\
75D   & Sd$^\star$    &  \HII$^c$ 
&  7\\
79A  & Sa pec$^\diamond$    & TO$^a$, LINER$^b$ &  1\\
79B  & S0 pec$^\diamond$   &   \HII?$^b$ &  -2\\
82C  & S?$^\diamond$, Im$^\star$  & TO/AGN$^a$  & --\\
91A  & SB(s)bc pec?$^\diamond$, SBc$^\star$    & AGN$^a$
&  4.3\\
91C  & S?$^\diamond$, Sc$^\star$    &  -- &  --\\
96A$^\ddagger$   & SA(r)bc pec$^\diamond$  & AGN$^a$
  & 4\\
96C  & S?$^\diamond$, Sa$^\star$  & TO$^a$ &  --\\
100A & S0/a$^\diamond$   & TO$^a$  & 0\\
\\
\hline
\\
\multicolumn{4}{l}{TO indicates line ratio's intermediate between AGN and \HII}\\
\multicolumn{4}{l}{dLINER designation indicates a LL (low luminosity) AGN}\\
\multicolumn{4}{l}{$^\dagger$ From the RC3 Catalogue of \citet{deV92}}\\
\multicolumn{4}{l}{$^\star$ From the Optical Classification of \citet{Hick89}}\\
\multicolumn{4}{l}{$^\diamond$ From the Classification of \citet{deV91}}\\
\multicolumn{4}{l}{$\bullet$ Misclassification \citet{Bit11}}\\
\multicolumn{4}{l}{$^\ddagger$ AGN-dominated Mid-Infrared Spectrum}\\
\multicolumn{4}{l}{$^a$ Optical Classification from \citet{Mar10}}\\
\multicolumn{4}{l}{$^b$ Optical Classification from \citet{Coz04}}\\
\multicolumn{4}{l}{$^c$ Optical Classification from \citet{Brin04}}\\
\\
\end{tabular}

\end{center}
}
\end{table}

As shown in Section \ref{samp}, the distribution of \citet{Bit11} indicates a lowering in the density of galaxies, compared to the LVL galaxy colors, in the region given by Equation (1). 
If we separate the MOHEGs according to mid-infrared color, we find 12 located between $-0.35 \le {\rm log[f}_{5.8\mu \rm m}/{\rm f}_{3.6\mu \rm m}] \le -0.05$ (i.e. intermediate mid-infrared colors); these systems have enhanced \Ht\ relative to their star formation, and their mid-infrared colors reflect that globally they are not dominated by star formation (or AGN emission); this is a clear indication that the warm \Ht\ is not UV-excited. This is discussed further in Section \ref{midcol}. 

The morphological types of the \Ht-detected systems are given in Table \ref{tab:morph}, where \citet{deV91} classifications are given for most of the sample, or \citet{Hick89} for those that were absent or assigned unknown classification. We include the nuclear classifications from \citet{Coz04}, \citet{Brin04}, \citet{Mar10}, as well as the T-Type from the RC3 catalogue of \citet{deV92}.

We include spectra for the non-MOHEG \Ht\ galaxies in Section \ref{non_spec} of the Appendix, as well as any discussion pertaining to individual systems. Spectra for the MOHEG galaxies are presented in the following section.

\subsubsection{Spectra of MOHEG Sources}\label{moheg_sources}

The power in the \Ht\ emission lines compared to the mid-infrared continuum is particularly noticeable in several MOHEGs.
For example, in Figure \ref{H2_57} we show two spectra of HCG 57A, one centered on the nuclear region (LL only) and a matched, off-nuclear extraction in the disk of the galaxy, made possible by the sparse mapping strategy employed (see Section \ref{irs_obs}). The nuclear extraction shows powerful \Ht\ S(1) and S(0) emission and, although we have no coverage of the PAH bands, we do not see a steeply rising mid-infrared continuum typical of star-forming and AGN-dominated systems. 

The off-nuclear extraction shows powerful \Ht\ emission with relatively weak 6.2 and 7.7\micron\ PAH emission. This spectrum suggests a non-star-forming mechanism able to excite the \Ht\ in the disk of galaxy, offset from the nucleus.

\begin{figure*}[!p]
\begin{center}
\vspace{-1cm}
\subfigure[HCG 6B]{\includegraphics[width=8cm]{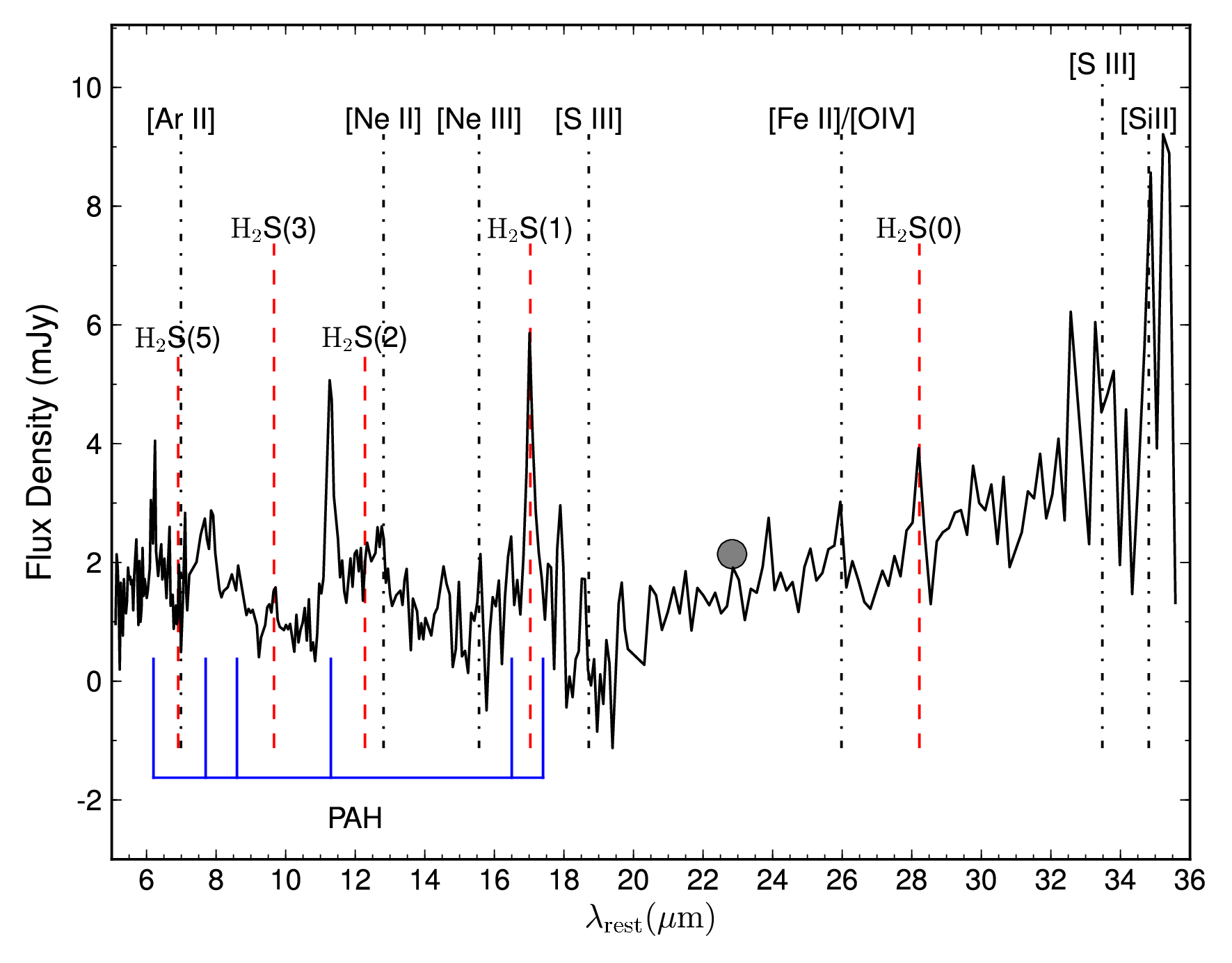}}
\hfill
\subfigure[HCG15A]{\includegraphics[width=8cm]{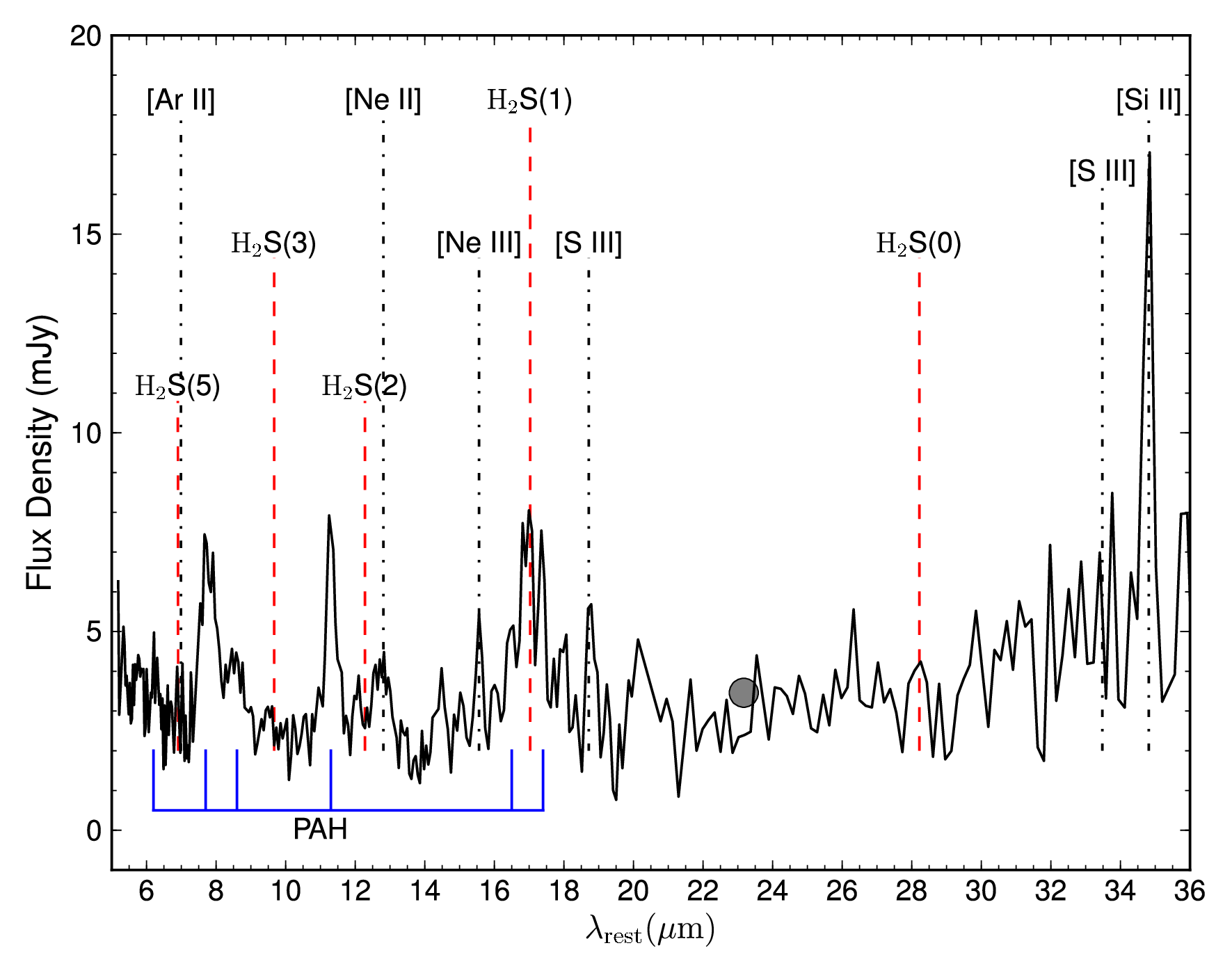}}
\vfill
\subfigure[HCG15D]{\includegraphics[width=8cm]{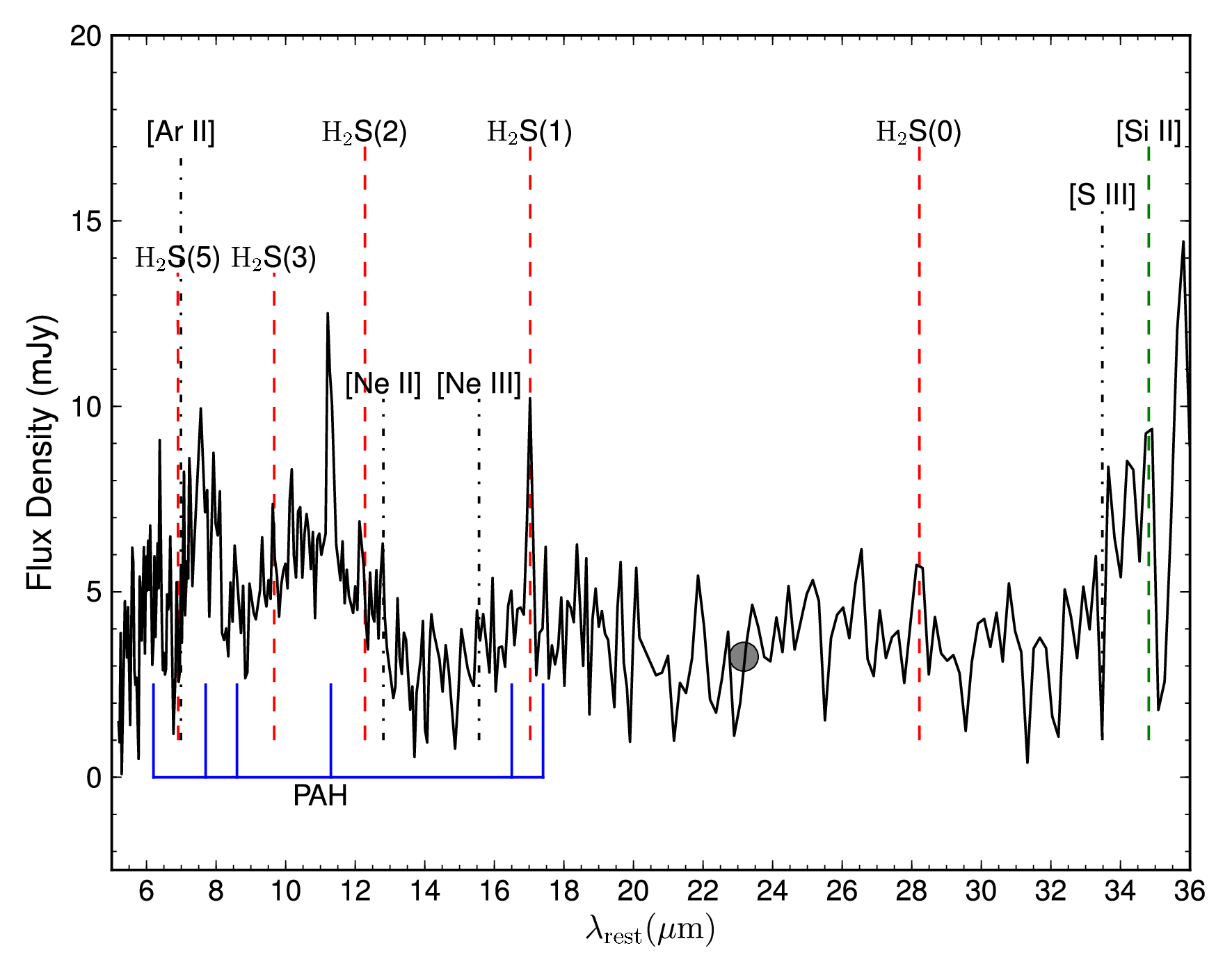}}
\hfill
\subfigure[HCG 40B]{\includegraphics[width=8cm]{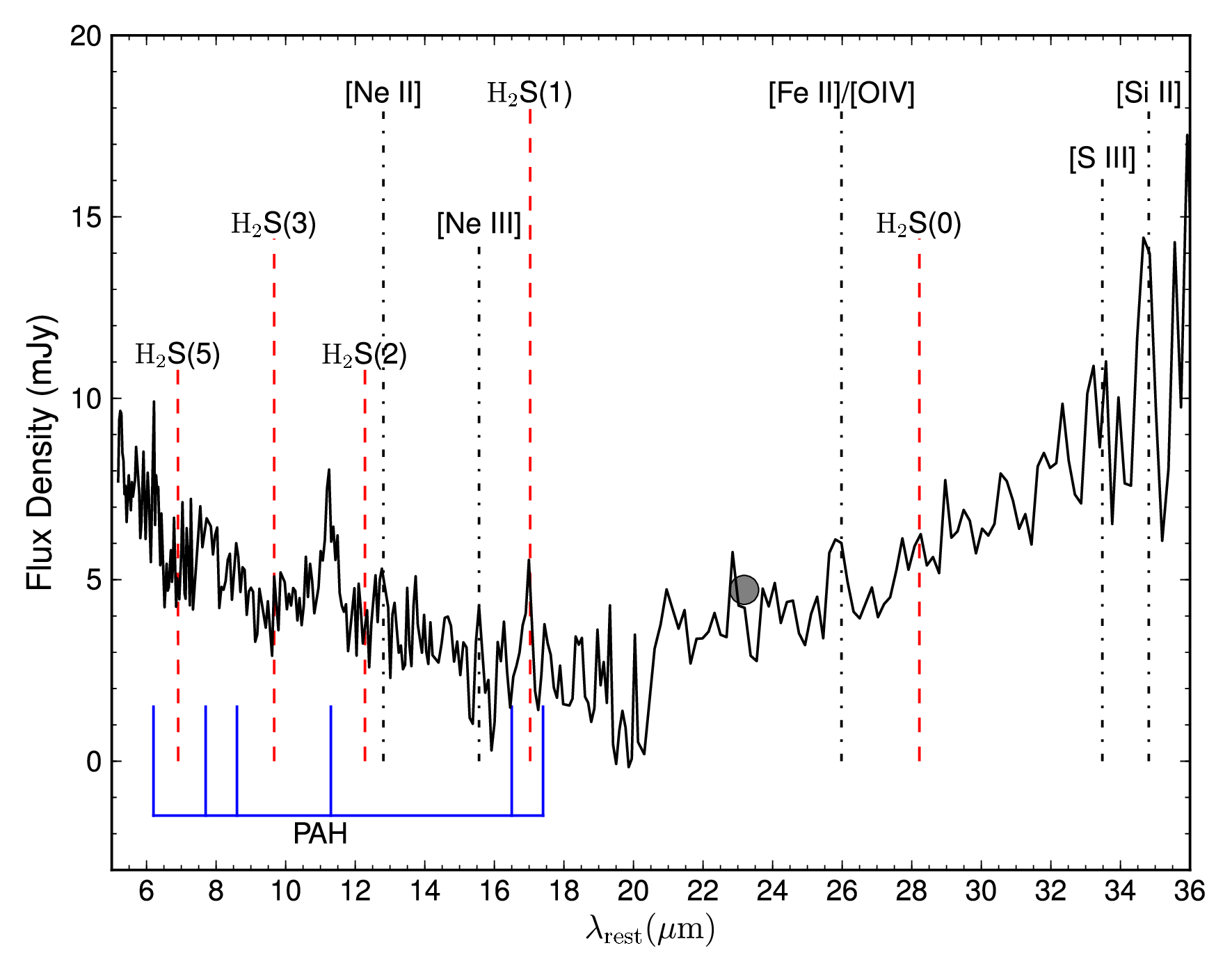}}
\vfill
\subfigure[HCG25B - including the nuclear region]{\includegraphics[width=8cm]{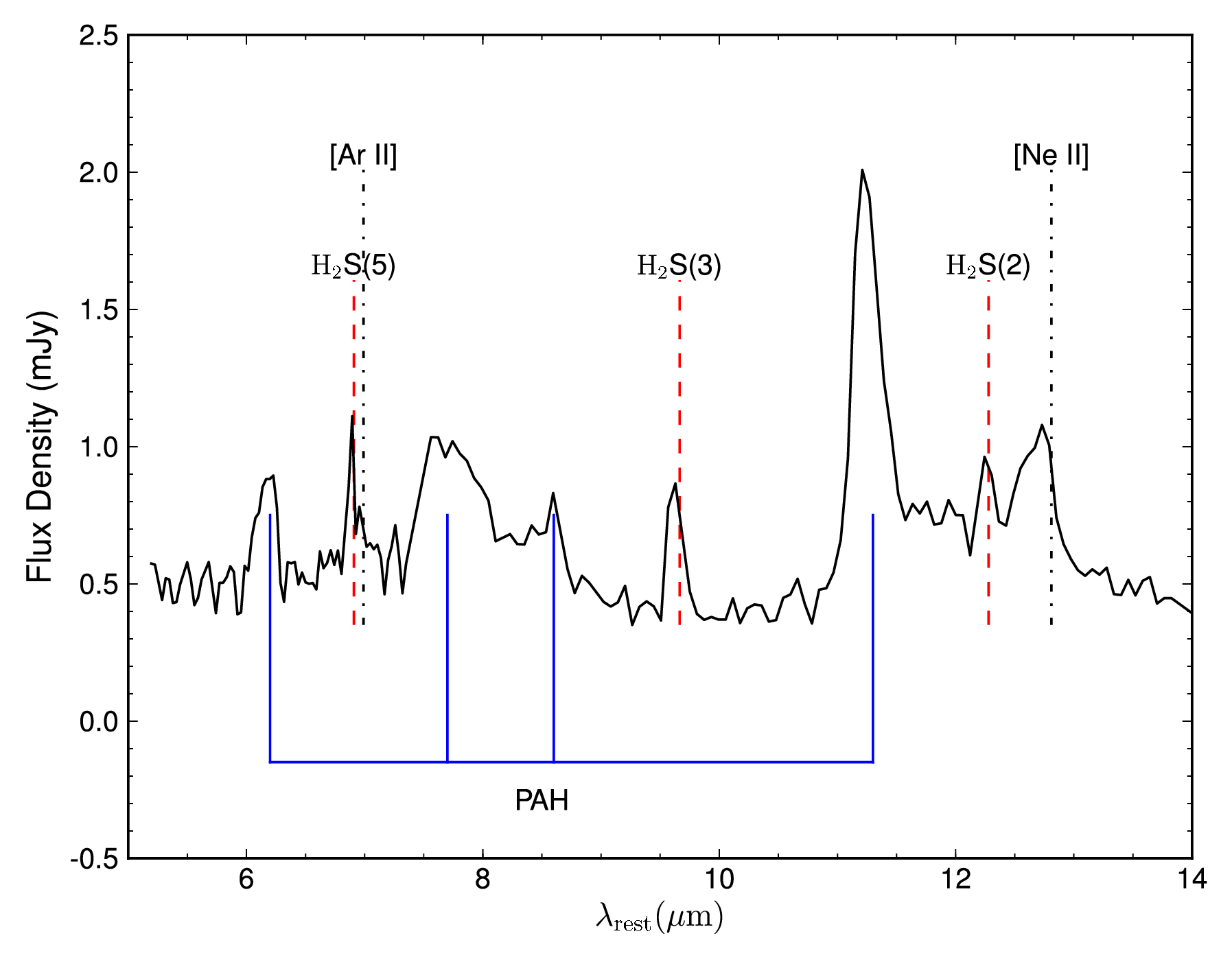}}
\hfill
\subfigure[HCG25B - off-nuclear extraction]{\includegraphics[width=8cm]{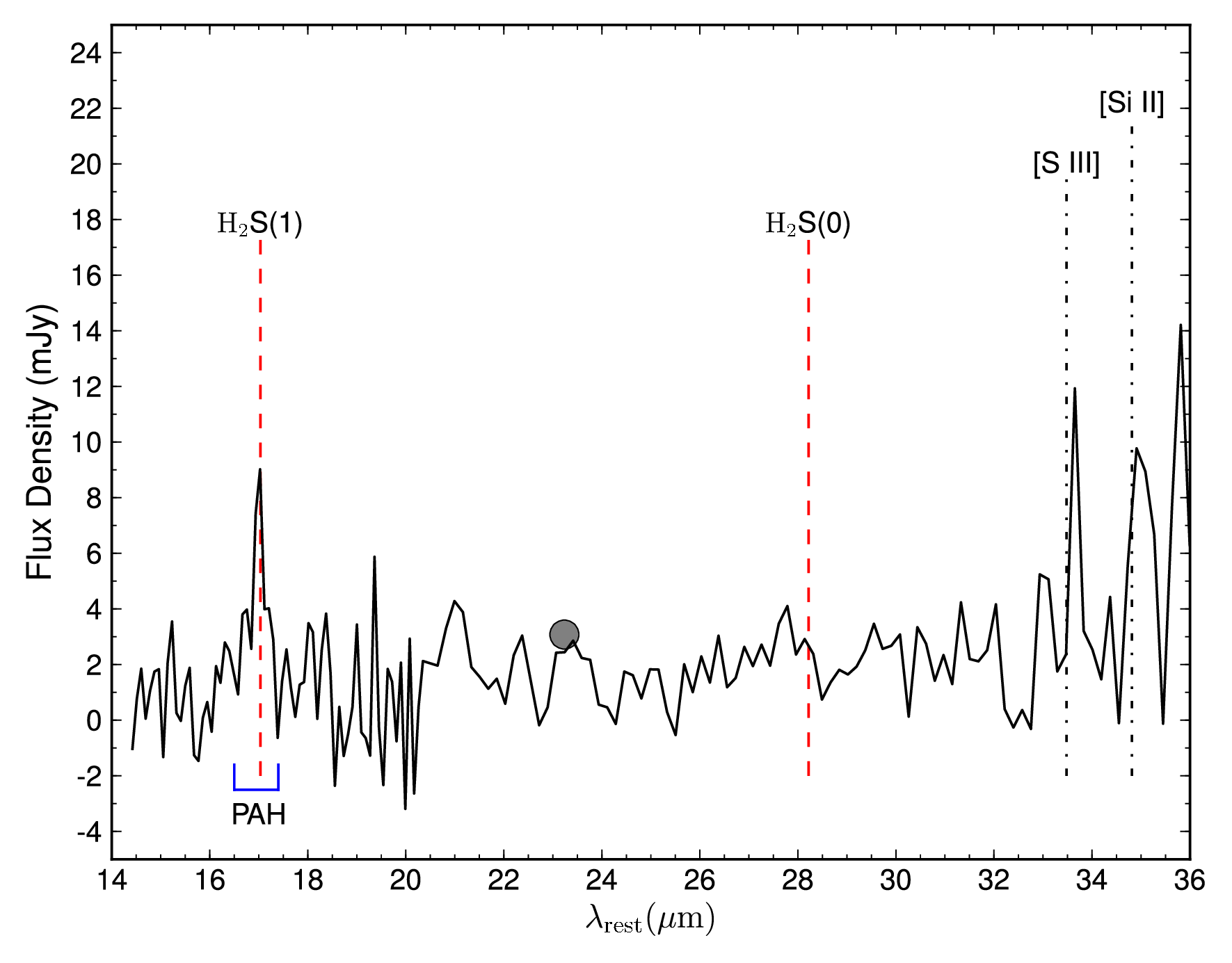}}

\caption[]
{MOHEGs with Intermediate Mid-infrared Colours. The matched MIPS 24\micron\ photometry is shown as a filled grey circle.}
\label{fig:spectra}
\end{center}
\end{figure*}

\begin{figure*}[!p]
\begin{center}
\vspace{-1cm}
\subfigure[HCG 44A - SL (Off-Nuclear)]{\includegraphics[width=8cm]{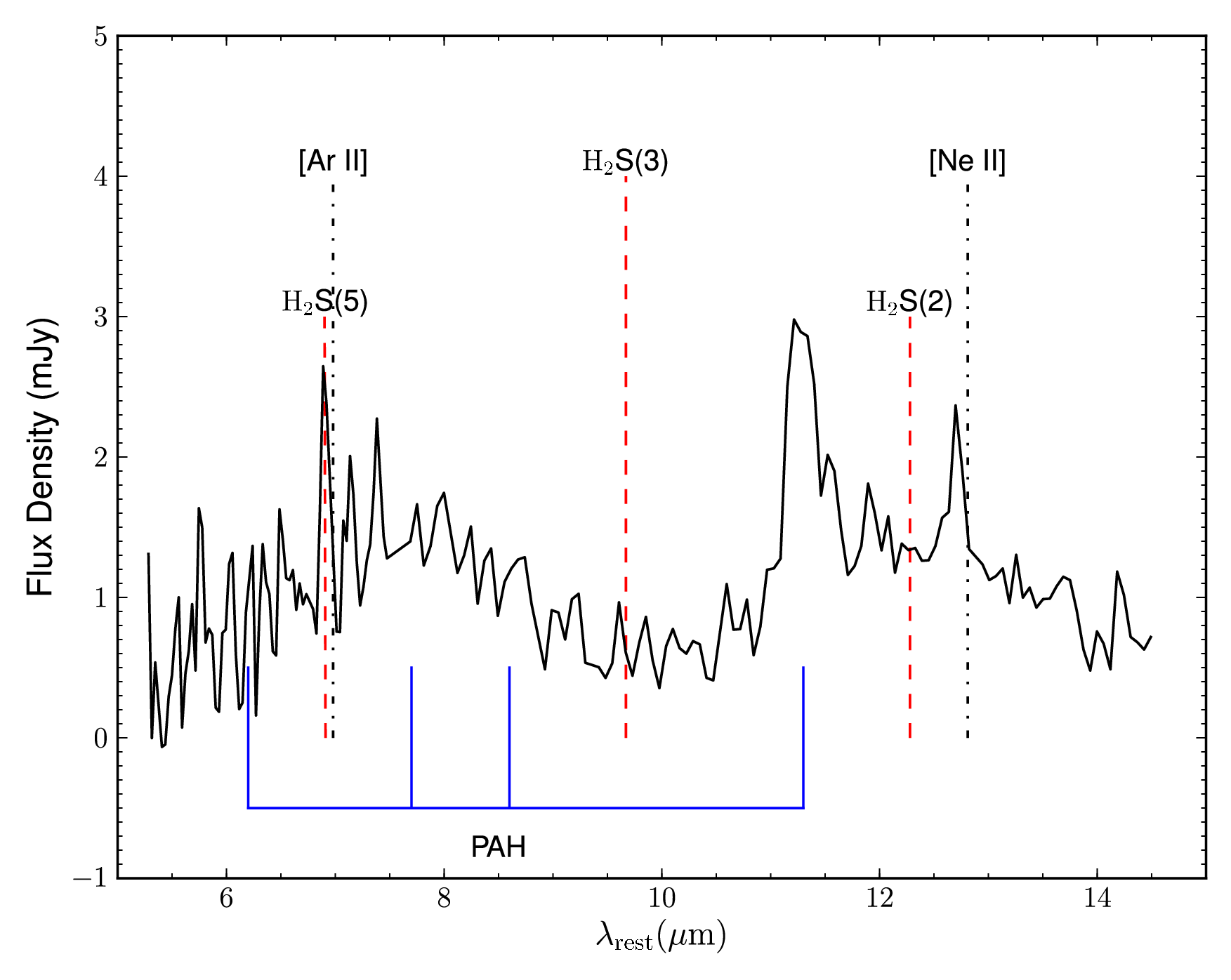}}
\hfill
\subfigure[HCG 44A - LL (Nuclear)]{\includegraphics[width=8cm]{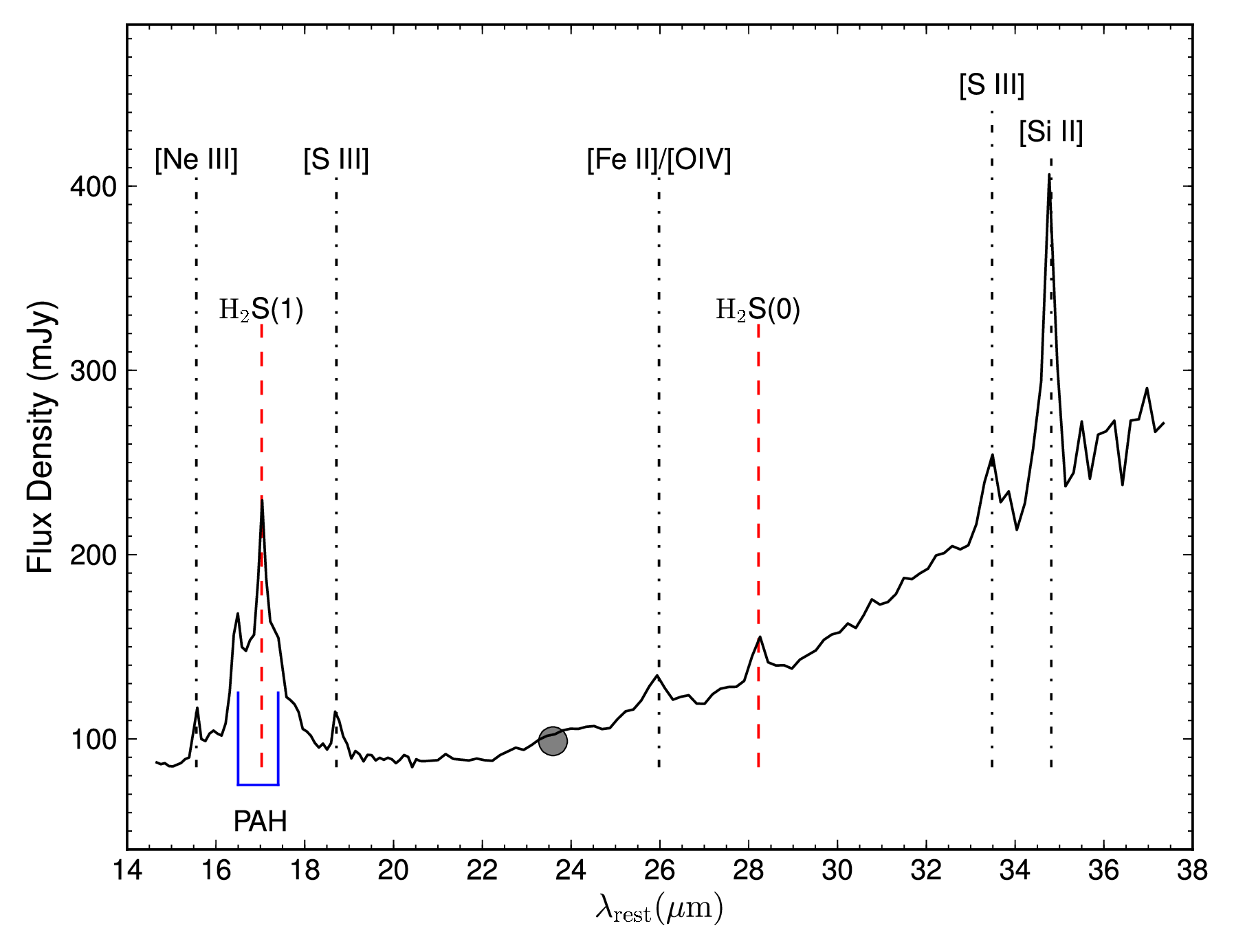}}
\vfill
\subfigure[HCG 56C]{\includegraphics[width=8cm]{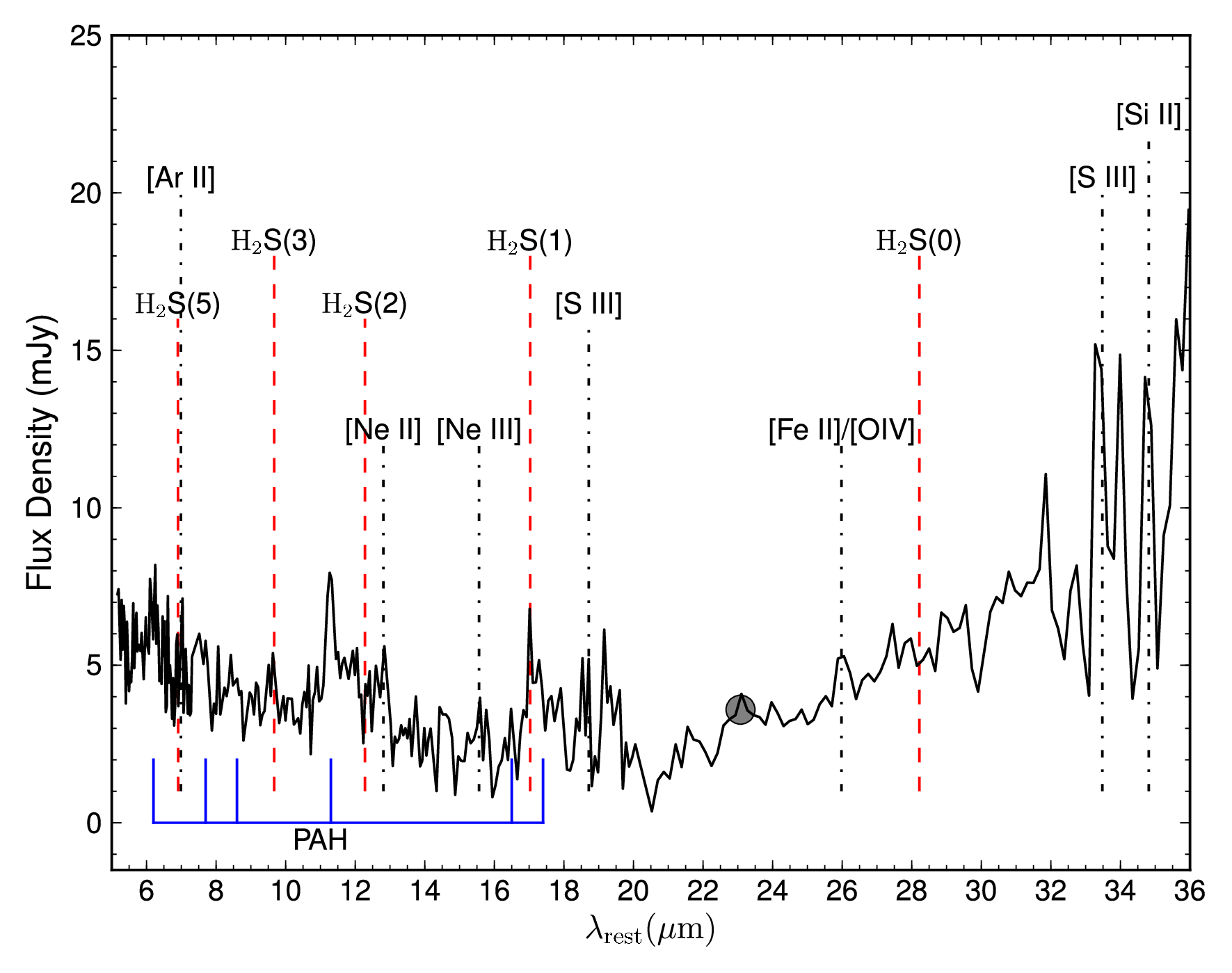}}
\hfill
\subfigure[HCG 68A]{\includegraphics[width=8cm]{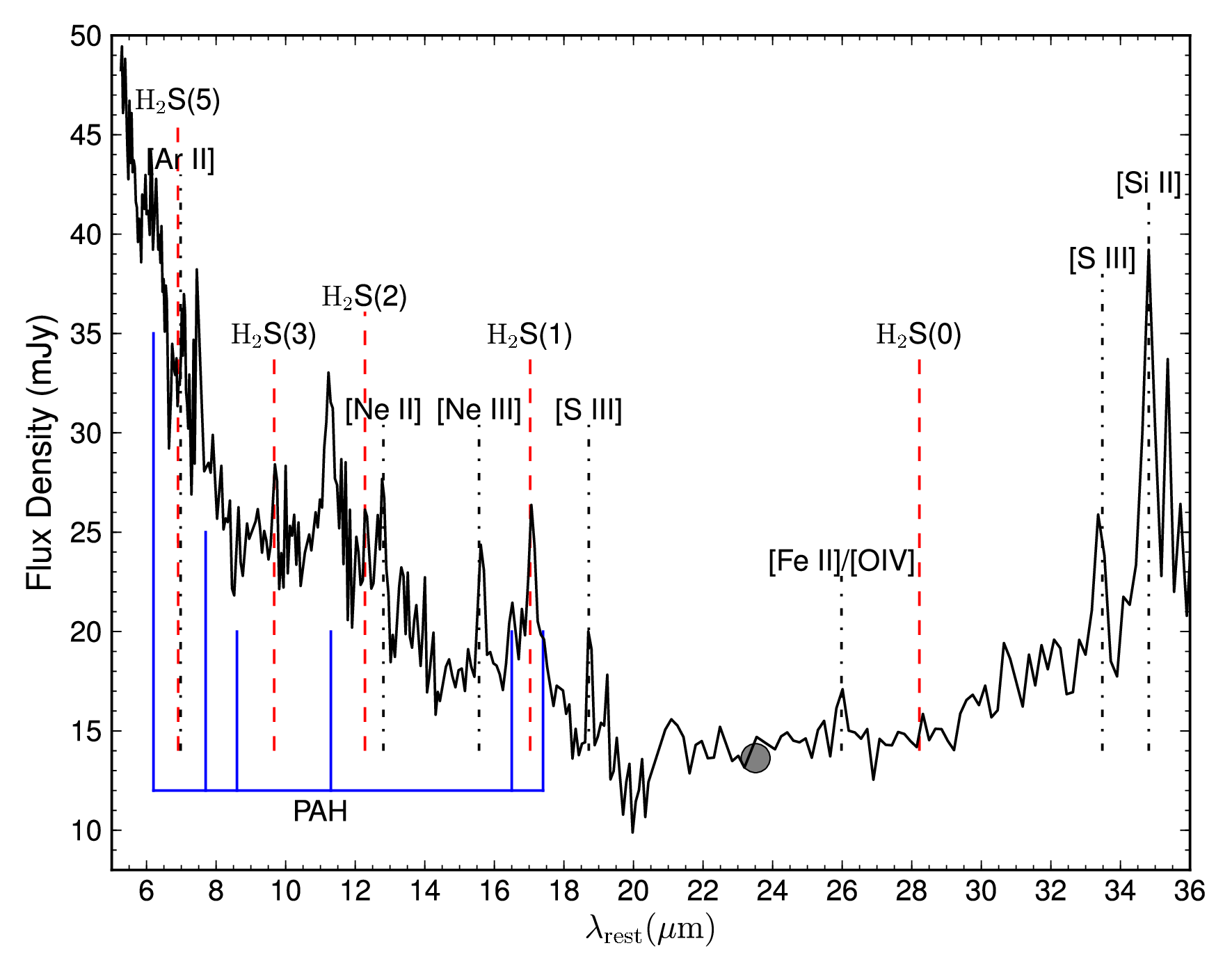}}
\vfill
\subfigure[HCG 68B]{\includegraphics[width=8cm]{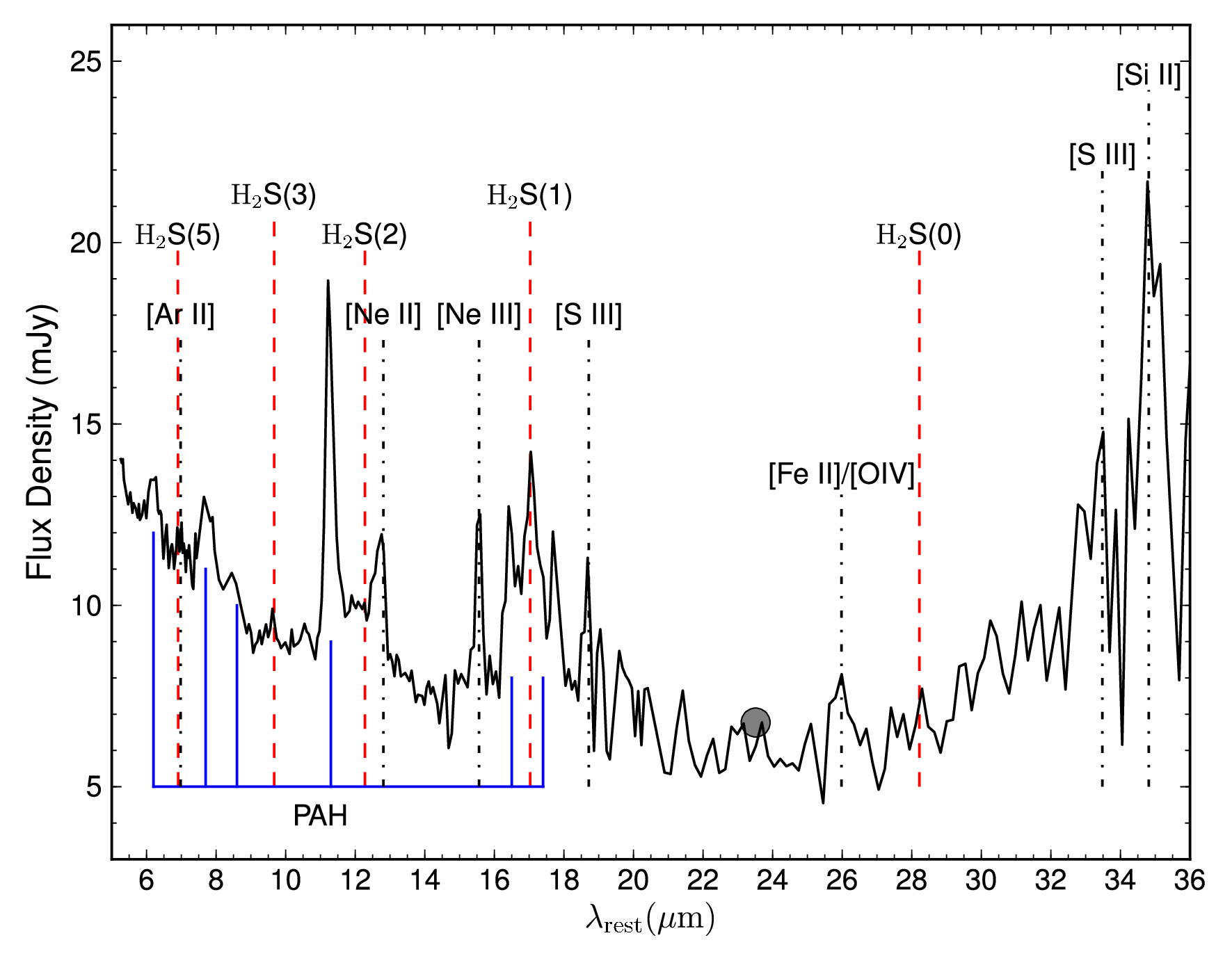}}
\hfill
\subfigure[HCG 82B]{\includegraphics[width=8cm]{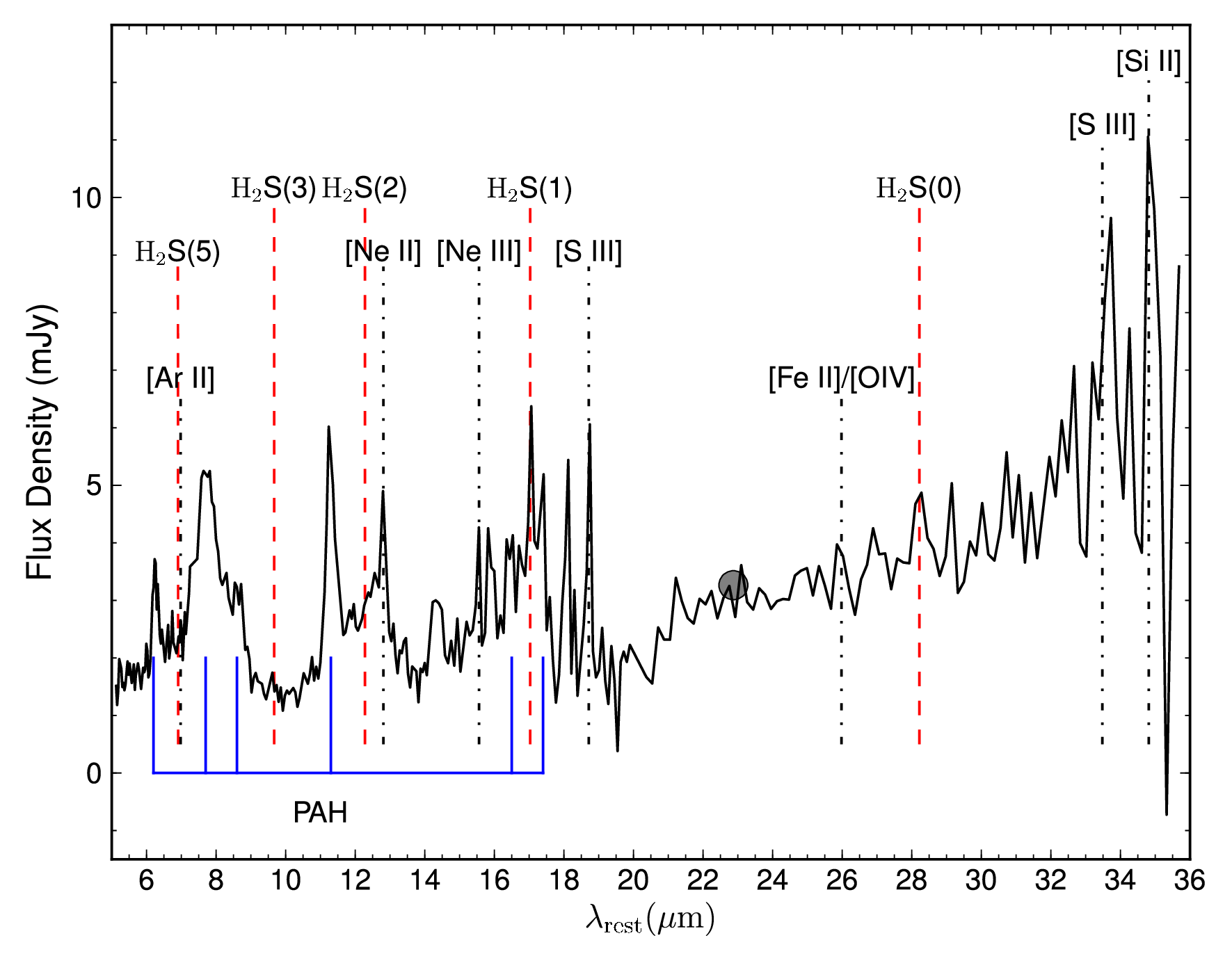}}
\caption[]
{(a)$-$(f) MOHEGs with Intermediate Mid-Infrared Colours; matched MIPS 24\micron\ photometry is shown as a filled grey circle.}
\label{fig:spectra2}
\end{center}
\end{figure*}

\begin{figure*}[!htb]
\begin{center}
\vspace{-1cm}

\subfigure[HCG 95C]{\includegraphics[width=8cm]{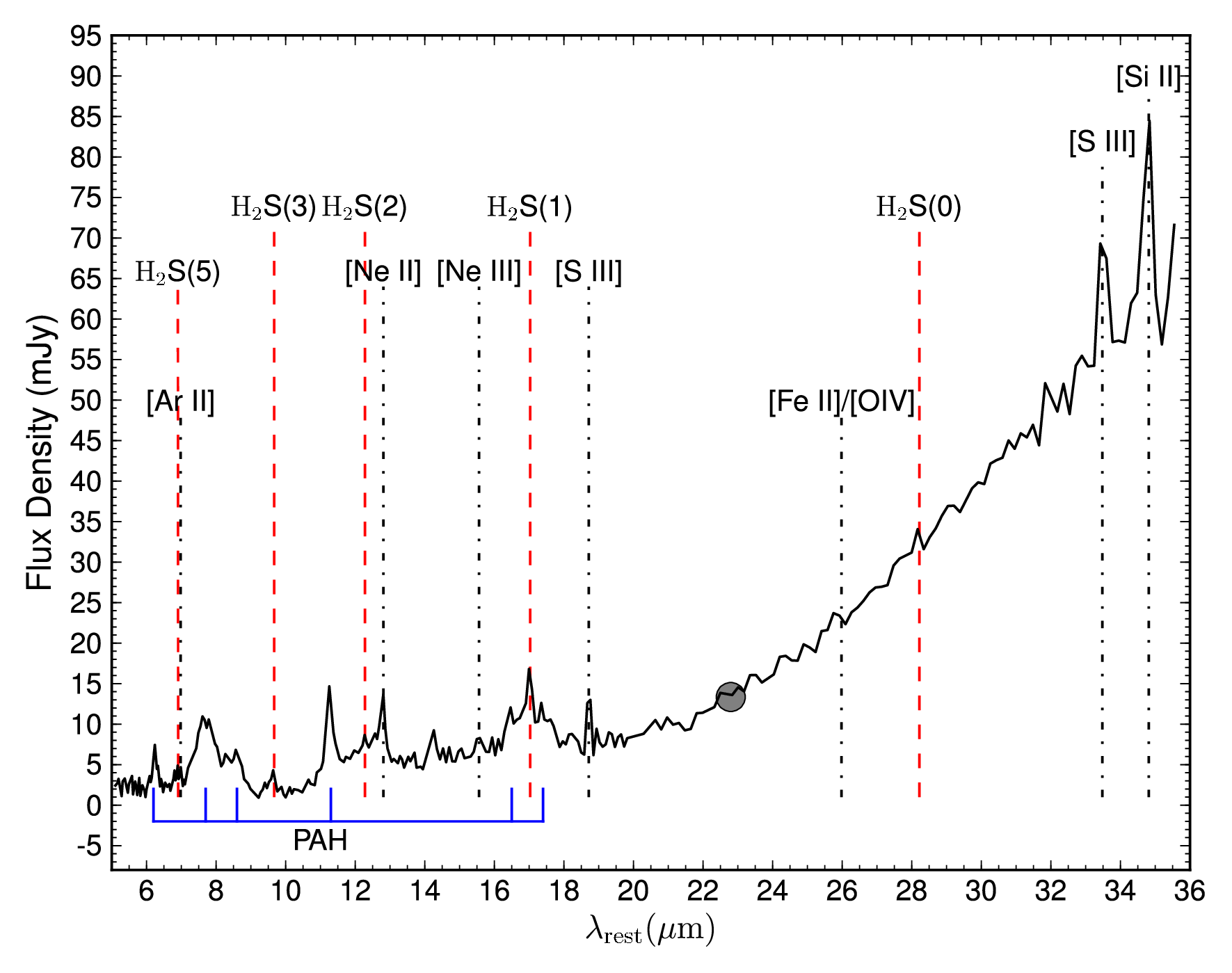}}
\hfill
\subfigure[HCG 95C ]{\includegraphics[width=7cm]{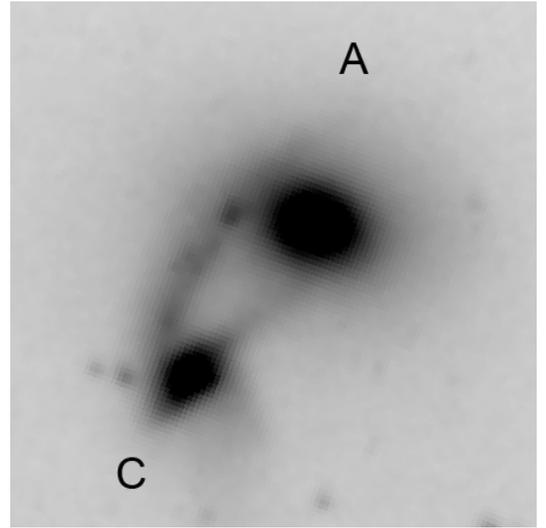}}
\caption[]
{(a) Spectrum of MOHEG HCG 95C with matched MIPS 24\micron\ photometry shown as a filled grey circle. b) IRAC 3.6\micron\ image of HCG 95C (field of view of $\sim$1\arcmin$\times$1\arcmin, North is up and East is left).}
\label{fig:spectra3}
\end{center}
\end{figure*}

In Figures \ref{fig:spectra}, \ref{fig:spectra2} and \ref{fig:spectra3} we plot the spectra of the other HCG galaxies classified as MOHEGs that lie at intermediate mid-infrared colors (Equation 1) as used in Table \ref{tab:morph}). In particular, HCG 6B (Fig. \ref{fig:spectra}a) and HCG 15D (Fig. \ref{fig:spectra}c)
show the distinctive \Ht\ S(1) emission line dominating the spectrum, with little 6.2 and 7.7\micron\ PAH emission, reflected in Figure \ref{H2_PAH}.

\begin{figure*}[!htb]
\begin{center}
\subfigure[HCG 68C -- Star Formation-dominated spectrum]{\includegraphics[width=8cm]{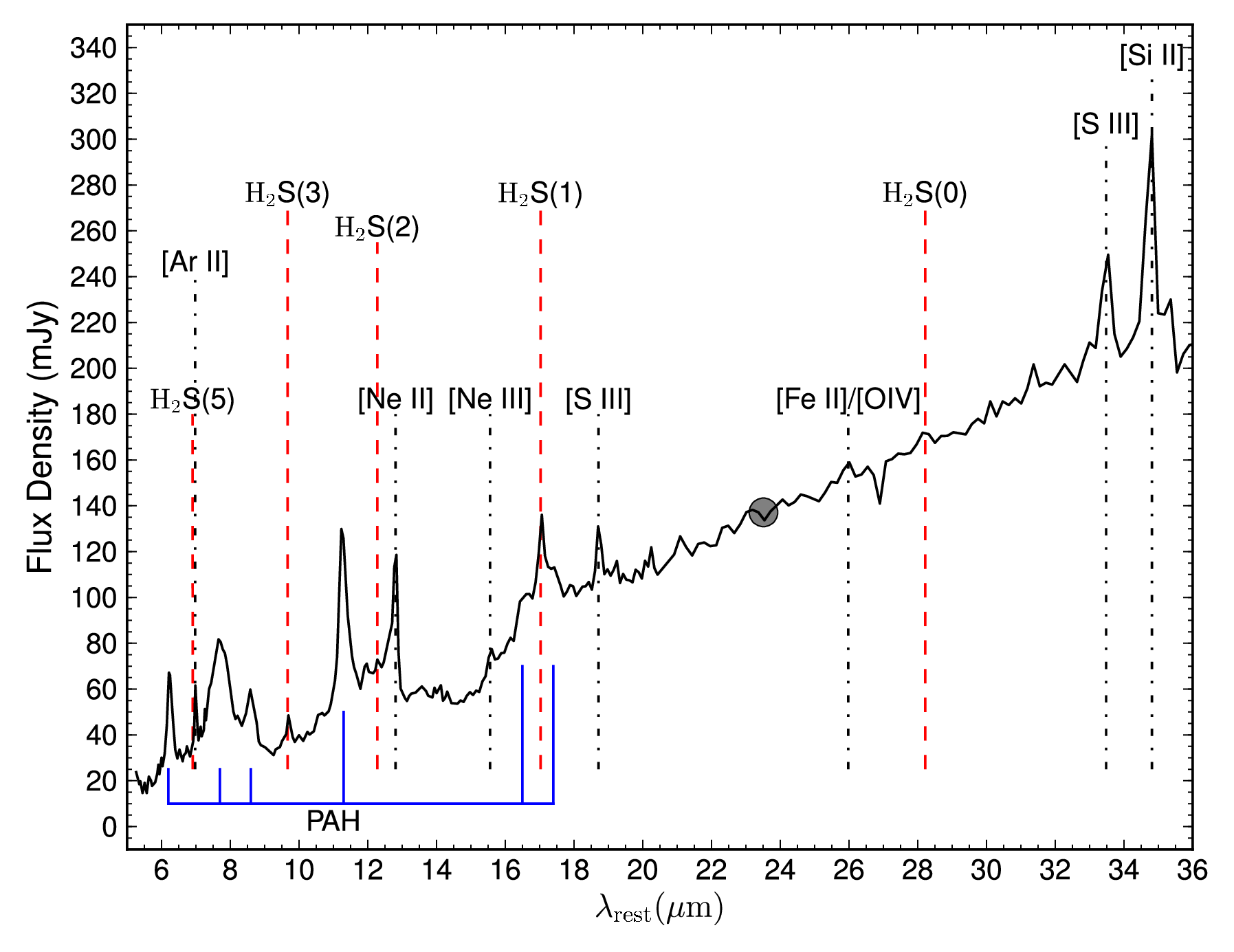}}
\hfill
\subfigure[HCG 56B -- AGN-dominated spectrum]{\includegraphics[width=8cm]{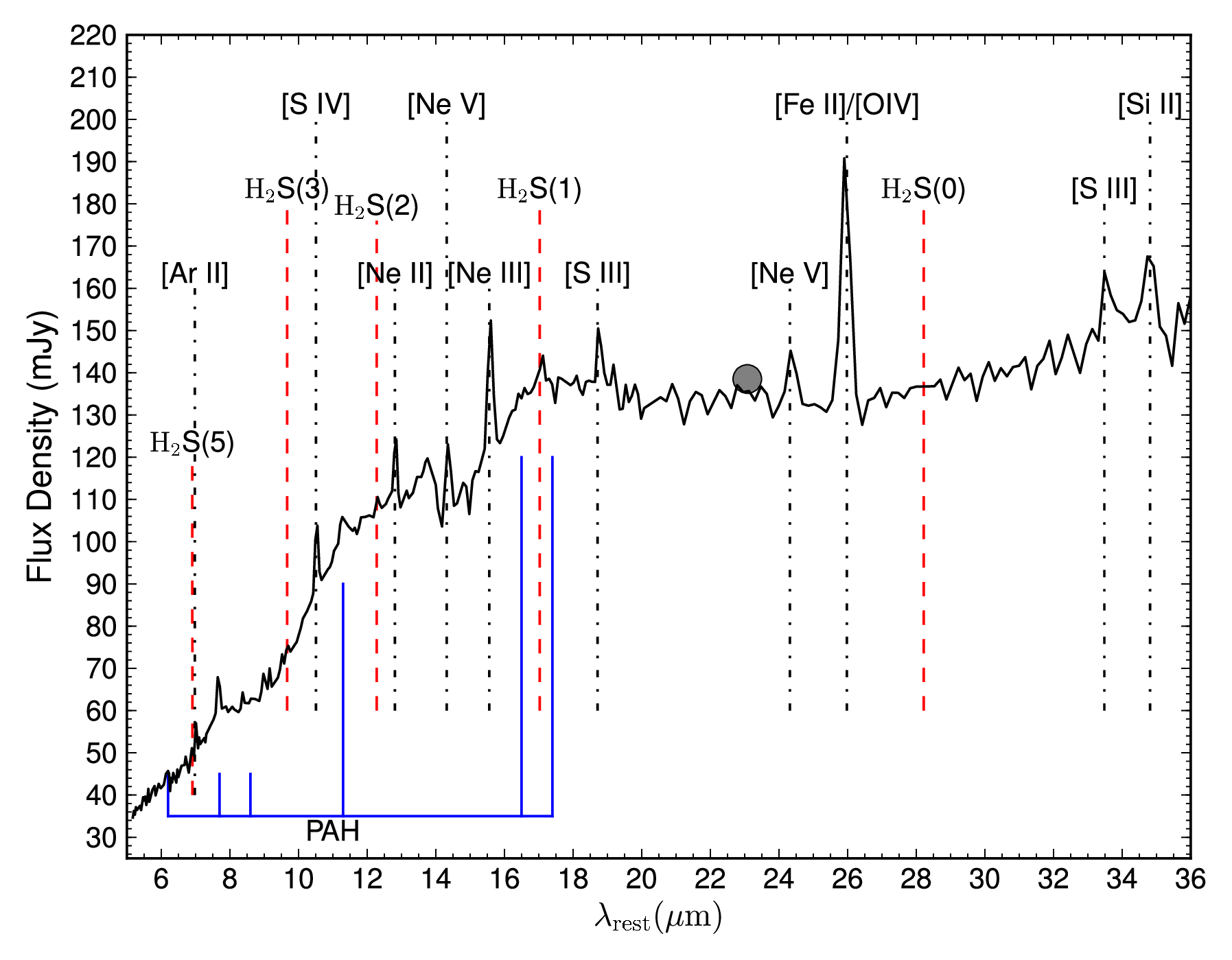}}

\caption[]
{MOHEGs with Star Formation-/AGN-Dominated Colours. The matched MIPS 24\micron\ photometry is shown as a filled grey circle. }
\label{fig:spectra4}
\end{center}
\end{figure*}

HCG 25B has SL coverage of the nuclear region (Fig. \ref{fig:spectra}e) and we see high signal to noise \Ht\ S(2), S(3) and S(5) emission (the weaker S(4) transition is not seen above the PAH emission at 8\micron). Similar to HCG 57A, 25B has coverage of the disk region -- it covers a larger region compared to the SL extraction, and is noisier, yet we see the S(1) clearly above the weak mid-infrared continuum, again suggestive of a mechanism influencing regions away from the nucleus.

For HCG 68A and 68B (Fig. \ref{fig:spectra2}d and e), we see the strong contribution from the stellar continuum ($\lambda<18$\micron), indicative of their early type morphology (S0), yet the \Ht\ S(1) and S(3) ortho transitions feature strongly. The 7.7\micron\ PAH is particularly weak in HCG 68A and only an upper limit exists. In HCG 68B the PAH is clearly defined, but the power in the \Ht\ lines is striking.

By contrast, the galaxy HCG 40B (Fig. \ref{fig:spectra}d) has the S(1) line weakly detected  with similarly low 7.7\micron\ PAH emission. 
The fact that the 11.3\micron\ PAH often features prominently in these spectra is not surprising as the large, neutral PAHs can be excited by soft radiation from evolved stars \citep{Kan08} and this would be consistent with their early-type morphologies.

In contrast to the typical MOHEG spectra presented in this paper, the MOHEG HCG 95C exhibits an exceptional, likely tidally-induced, star formation spectrum (Fig. \ref{fig:spectra3}). The strong PAH emission and steeply rising 24\micron\ continuum is indicative of a system dominated by star formation. Figure \ref{fig:spectra3}b shows that the galaxy is highly disrupted due to an interaction with 95A, with its current state classified as an Sm morphology. \citet{Ig98} find evidence for two nuclei and suggest it is either part of an ongoing merger between two galaxies, or is seen in projection with, each system having interacted with 95A individually to produce the bridges and tails.
Interestingly, the mid-infrared color of 95C -- after applying a k-correction -- is not strongly star-forming and indeed, its log[f$_{5.8\mu \rm m}$/f$_{3.6\mu \rm m}$] ratio of -0.08 places it at intermediate mid-infrared color, but closest to the mid-infrared ``red" population. 

Finally, there are two galaxies in our sample that show MOHEG-like emission with global mid-infrared colours that suggest they are dominated by star formation, or AGN activity (${\rm log[f}_{5.8\mu \rm m}/{\rm f}_{3.6\mu \rm m}]>-0.05$). The spectrum of HCG 68C (Figure \ref{fig:spectra4}a) appears typical of a star-forming galaxy, with a steeply rising mid-infrared continuum and clear 6.2 and 7.7\micron\ PAH emission. Whereas HCG 56B shows a distinctive AGN-dominated mid-infrared spectrum (Fig. \ref{fig:spectra3}b) with a substantial warm dust continuum and the high ionisation [Ne{\sc v}]\,24.32\micron\ and [O{\sc iv}]\,25.89\micron\ (blended with [Fe{\sc ii}]\,25.98\micron\ in LL) emission lines. This galaxy is classified as a Seyfert 2 \citep{Kh74}. Given the powerful continuum emission, the warm \Ht\ lines do not appear very strong, but the weak 7.7\micron\, PAH band gives rise to a high \Ht/7.7\micron\, PAH ratio. This system may be exhibiting excited \Ht\ produced within an XDR associated with the AGN \citep{Mal96} or, alternatively, due to jet interactions with the ISM \citep{Og10}.

\subsubsection{Comparison to Mid-Infrared Color}\label{midcol}

\begin{figure}[!tb]
\begin{center}
\includegraphics[width=8.5cm]{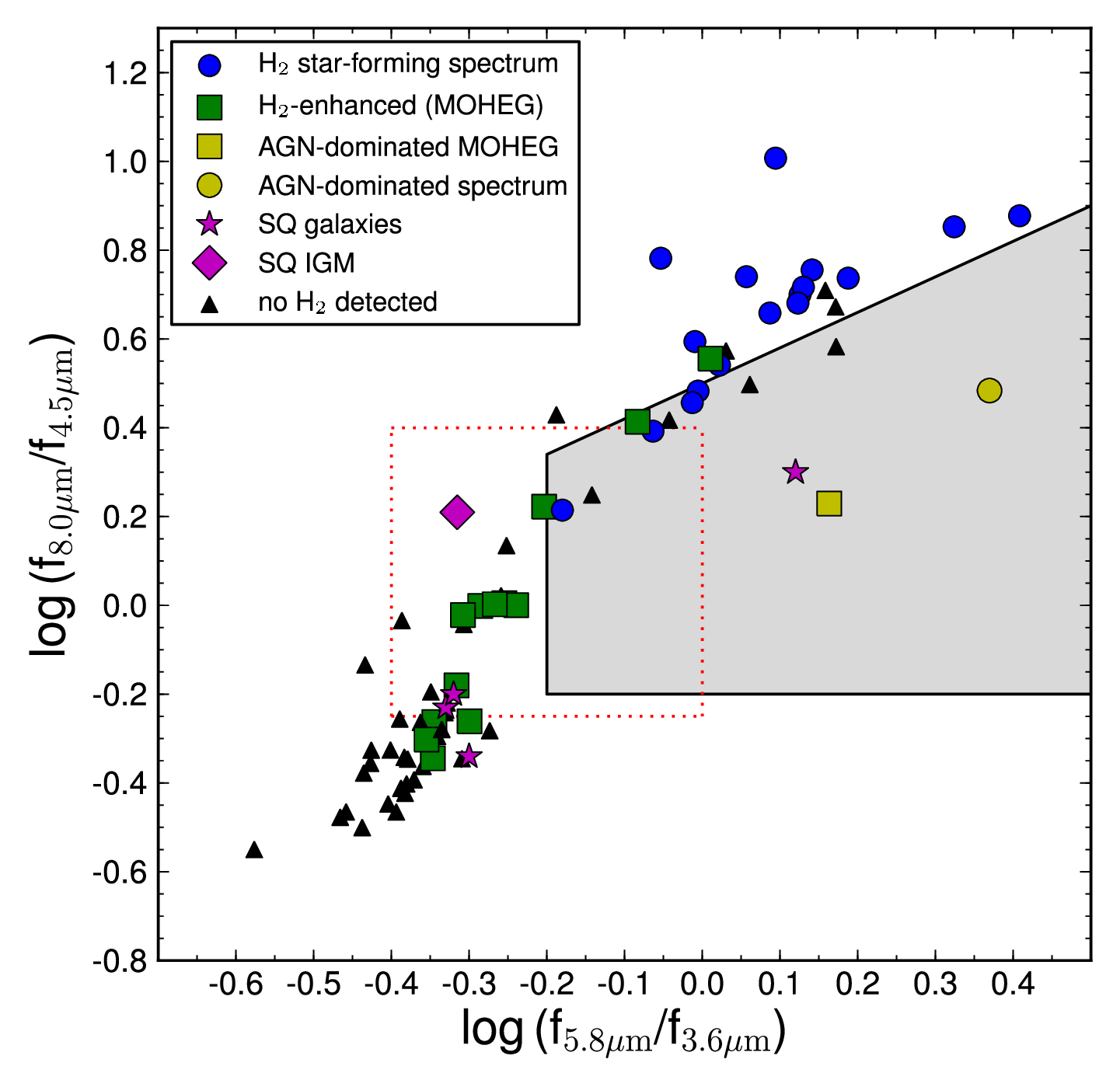}
\caption{IRAC colour-colour diagram for the sample; H$_2$-enhanced galaxies (MOHEGs) are shown as green squares, H$_2$-detected galaxies are shown as blue circles. The red dashed box shows the underpopulated region in color-color space from \citet{Walk10}. The shaded area corresponds to the AGN colour selection criteria of \citet{Lacy04}; AGN-dominated systems (as deterrmined by their mid-infrared spectra) are in yellow. The purple diamond is the location of the shocked IGM in SQ \citep[from][]{Clu10}, with SQ member galaxies shown as magenta stars.}
\label{HCG_col}
\end{center}
\end{figure}

In Figure \ref{HCG_col} we plot the IRAC color-color diagram (log[f$_{8.0\mu {\rm m}}$/f$_{4.5\mu {\rm m}}$] vs log[f$_{5.8\mu {\rm m}}$/f$_{3.6\mu {\rm m}}$]) for our entire sample of 74 galaxies, color-coded according to their mid-infrared spectral features as follows: no warm \Ht\ detected (black triangles), \Ht\ detected in line with UV-heated PDR emission (blue circles), \Ht\ detected in a galaxy with an AGN-dominated mid-infrared spectrum (yellow circle), \Ht-enhanced galaxies i.e. MOHEGs (green squares), and MOHEGs with AGN-dominated mid-infrared spectra (yellow squares). For reference, we include the region defined by \citet{Walk10} as the gap in color-color space.
We note that our mid-infrared spectra sample the nuclear and disk regions and reflect the galaxies' global properties -- optical nuclear classifications from the literature are provided in Table \ref{tab:morph}. This is likely why several galaxies with star-forming, mid-infrared colours are not detected in \Ht; a discussion of this is included in Section \ref{non-detect} of the Appendix.


\begin{figure}[!th]
\begin{center}
\subfigure[]{\includegraphics[width=8cm]{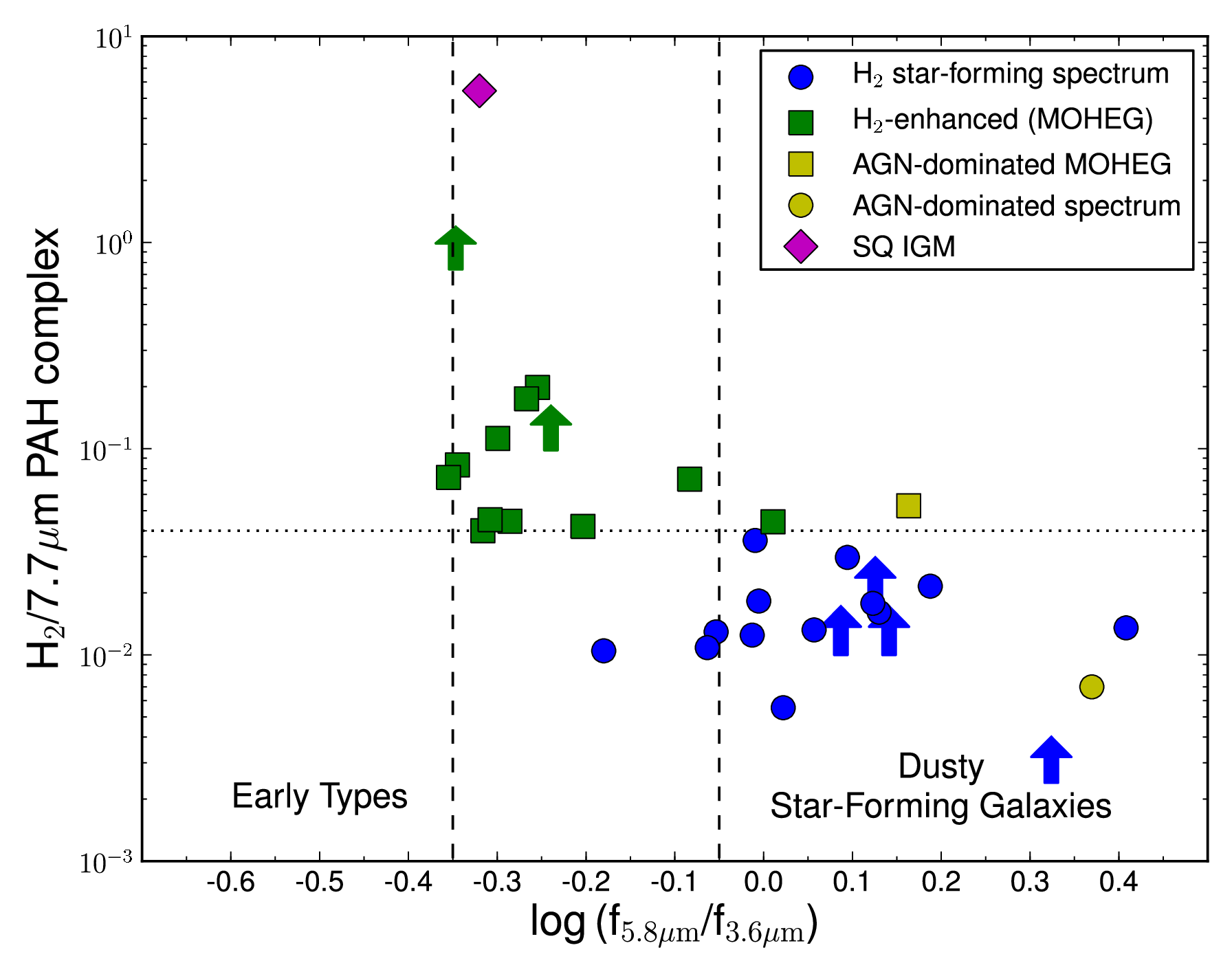}}
\hfill
\subfigure[]{\includegraphics[width=8.2cm]{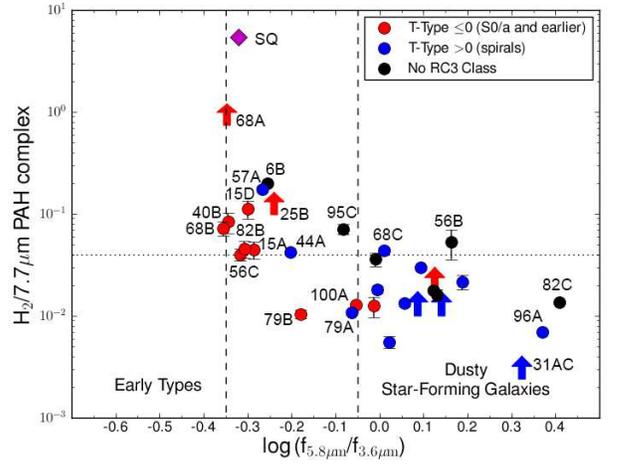}}
\caption{a) \Ht/7.7\micron\ PAH ratio as a function of IRAC color (${\rm f}_{5.8\micron}/{\rm f}_{3.6\micron}$) with the same symbols as in Figure \ref{HCG_col}. Lower limits are shown as arrows and the location of the IGM shock in SQ is plotted as a magenta diamond. The dashed lines indicate intermediate mid-infrared colours (determined in Section \ref{samp}) and the dotted horizontal line indicates MOHEG \Ht-enhancement \citep{Og10}. b)  \Ht/7.7\micron\ PAH ratio as a function of ${\rm [f}_{5.8\micron}/{\rm f}_{3.6\micron}]$ color color-coded according to T-Type. Error bars from the combined line measurements are indicated, although for several systems they are comparable to the size of the marker.}
\label{H2_PAH}
\end{center}
\end{figure}

For comparison, we also plot in Figure \ref{HCG_col} the IRAC colors of the shocked region in SQ \citep[shock sub-region; see ][]{Clu10} and the SQ member galaxies.

The \Ht-enhanced systems (green squares) are located preferentially in and around the lower left part of the red box, which suggests a possible connection to transformation from optically blue star-forming to red sequence, passively evolving systems. 
To explore this further, we plot in Figure \ref{H2_PAH}, the \Ht/7.7\micron\ PAH ratio as a function of mid-infrared color ({\it Spitzer} IRAC 5.8$\mu$m/3.6$\mu$m bands); here \Ht\ represents the sum of S(0)-S(3) lines, as per the definition of MOHEG, as listed in Table \ref{H2_lum}. We note that HCG 68A has an upper limit for its 7.7\micron\ PAH detection, the other lower limits are due to not having a complete spectrum to determine the total warm \Ht\ emission.
The dashed lines indicate the intermediate mid-infrared color space given in Equation (1) and which we 
shall refer to as the mid-infrared ``green valley". 

Apart from the AGN-dominated systems (in yellow), the \Ht/7.7\micron\ PAH ratios appear to increase towards blue IRAC color, with MOHEGs predominately found at intermediate color in what we now term the mid-infared green valley. A discussion of the link between the optical and mid-infrared green valley for HCG MOHEGs is included in Section \ref{green}.

\subsubsection{\Ht\ and the Mid-Infrared Continuum}

VSGs (very small grains) reprocess UV radiation to give rise to the mid-infrared continuum at 24\micron. Since a UV radiation field can be produced by star formation, AGN emission and radiative shocks \citep{Dop96}, a paucity of 24\micron\ emission relative to warm \Ht\ is a strong indicator of non-radiative (e.g. shock) heating of \Ht. In star-forming galaxies, L(\Ht)/L$_{24}$ appears to decrease with increasing L$_{24}$ \citep{Rous07}. This effect is similar to the observed decrease in L(PAH)/L$_{24}$ ratio, as L$_{24}$ increases, seen in starbursts and ULIRGS possibly due to increasing AGN heating in these systems \citep[e.g.][]{Des07}. Depletion of VSGs due to shocks may also be a factor.

\begin{figure*}[!ht]
\begin{center}
\subfigure[]{\includegraphics[width=8.1cm]{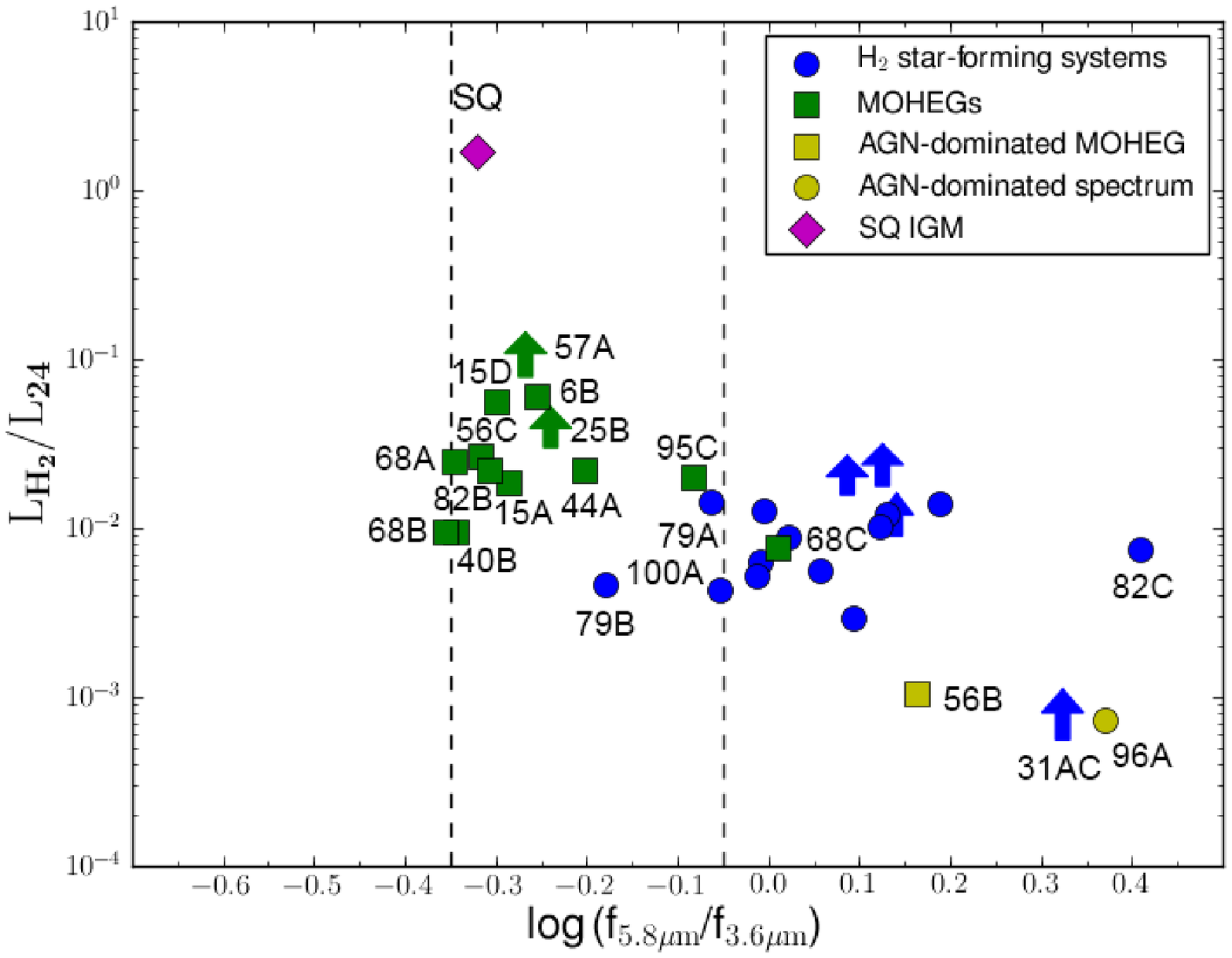}}
\hfill
\subfigure[]{\includegraphics[width=8.1cm]{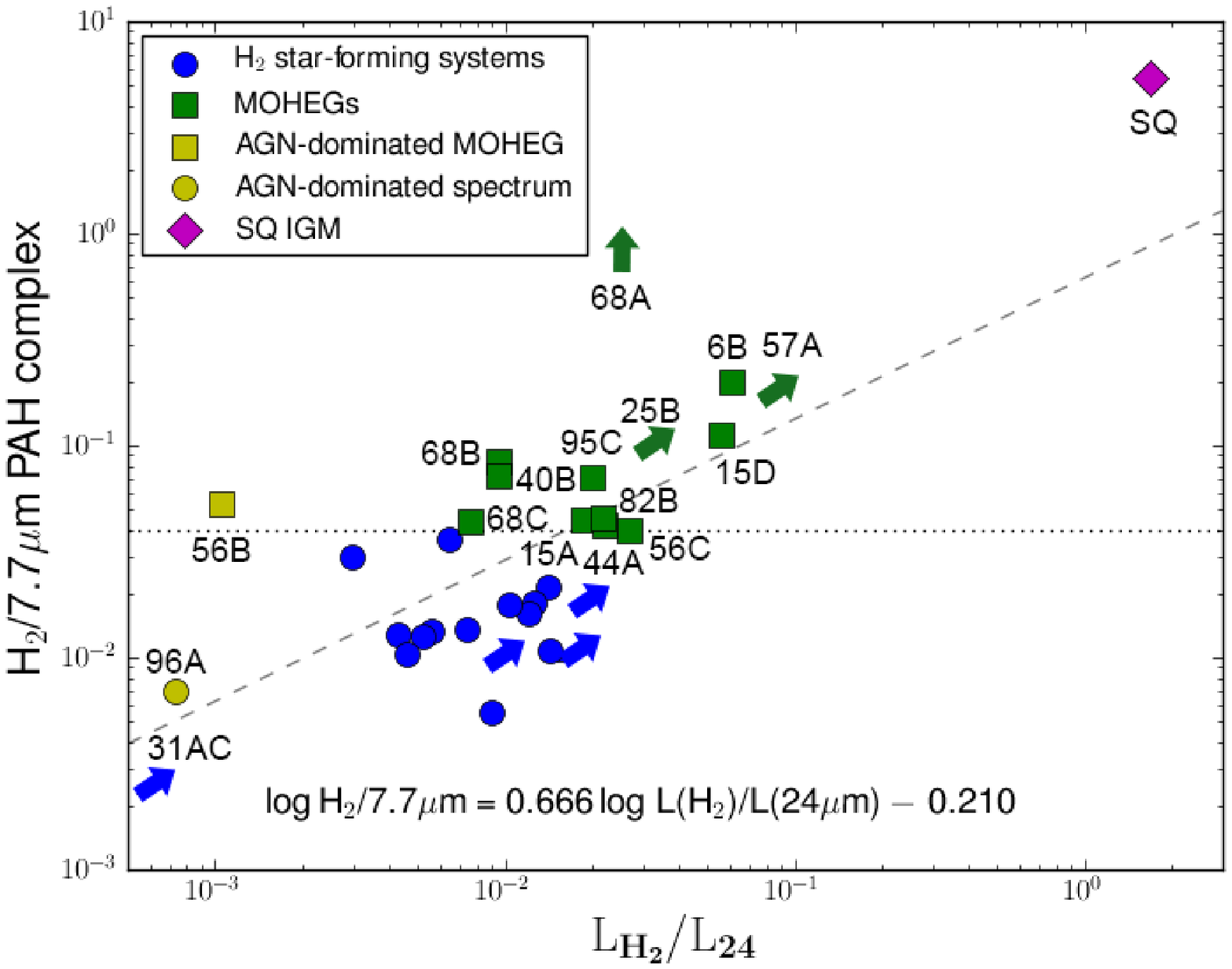}}

\caption{a) L$_{{\rm H}_2\, \rm S(0)-S(3)}$/L$_{24}$ luminosity ratio for \Ht-detected systems as a function of ${\rm log[f}_{5.8}/{\rm f}_{3.6}]$ color with the same symbols as in Figures \ref{HCG_col} and \ref{H2_PAH}a. Arrows represent systems where the \Ht\ luminosity is a lower limit and the vertical dashed lines are from Equation 1. The diamond shows the location of the shocked IGM of SQ \citep[from][]{Clu10}. b)  \Ht/7.7\micron\ PAH ratio as a function of  L$_{{\rm H}_2\, \rm S(0)-S(3)}$/L$_{24}$ with the same color-coding as on the left. The dotted line indicates MOHEG \Ht-enhancement \citep{Og10}. Arrows indicate systems with lower limits for their \Ht\ fluxes (due to not having the full SL$+$LL coverage) and are not included in the least-squares fit (grey dashed line); the best-fit relation is given at the bottom of the figure.}
\label{fig:H2_IR}
\end{center}
\end{figure*}

We plot in Fig \ref{fig:H2_IR}a the ratio of \Ht\ to 24\micron\ luminosity as a function of ${\rm log[f}_{5.8\mu \rm m}/{\rm f}_{3.6\mu \rm m}]$ color and we see a similar distribution to that of the  \Ht/7.7\micron\ PAH ratio. The increased L(\Ht)/L$_{24}$ values (on average) of the MOHEG systems compared to non-MOHEG star-forming systems indicates the importance of non-radiative heating, most likely due to shocks. 
We find that systems with AGN-dominated mid-infrared colors (HCG 56B and HCG 96A) have low L(\Ht)/L$_{24}$ values, but the highest  L$_{24}$ luminosities in the sample, lying within the locus found for radio galaxies \citep{Og10}. This is in agreement with their mid-infrared spectra. 

The \Ht/7.7\micron\ PAH versus L(\Ht)/L$_{24}$ ratios (Fig. \ref{fig:H2_IR}b) indicate that the 7.7\micron\ PAH emission decreases in line with a decrease in L$_{24}$. Omitting lower limits and mid-infrared AGN-dominated systems, we show a least-squares fit to the data as the grey dashed line in Fig \ref{fig:H2_IR}b. The clear outlier is HCG 56B with a large L$_{24}$, yet high \Ht/7.7\micron\ PAH emission indicative of MOHEG activity. The deficit of 24\micron\ continuum relative to \Ht\ emission in the HCG MOHEGs (except for HCG 56B) strongly suggests shock heating.

\subsection{Atomic Emission Lines}\label{FS}

\begin{figure*}[!htb]
\begin{center}
\subfigure[]{\includegraphics[width=8.1cm]{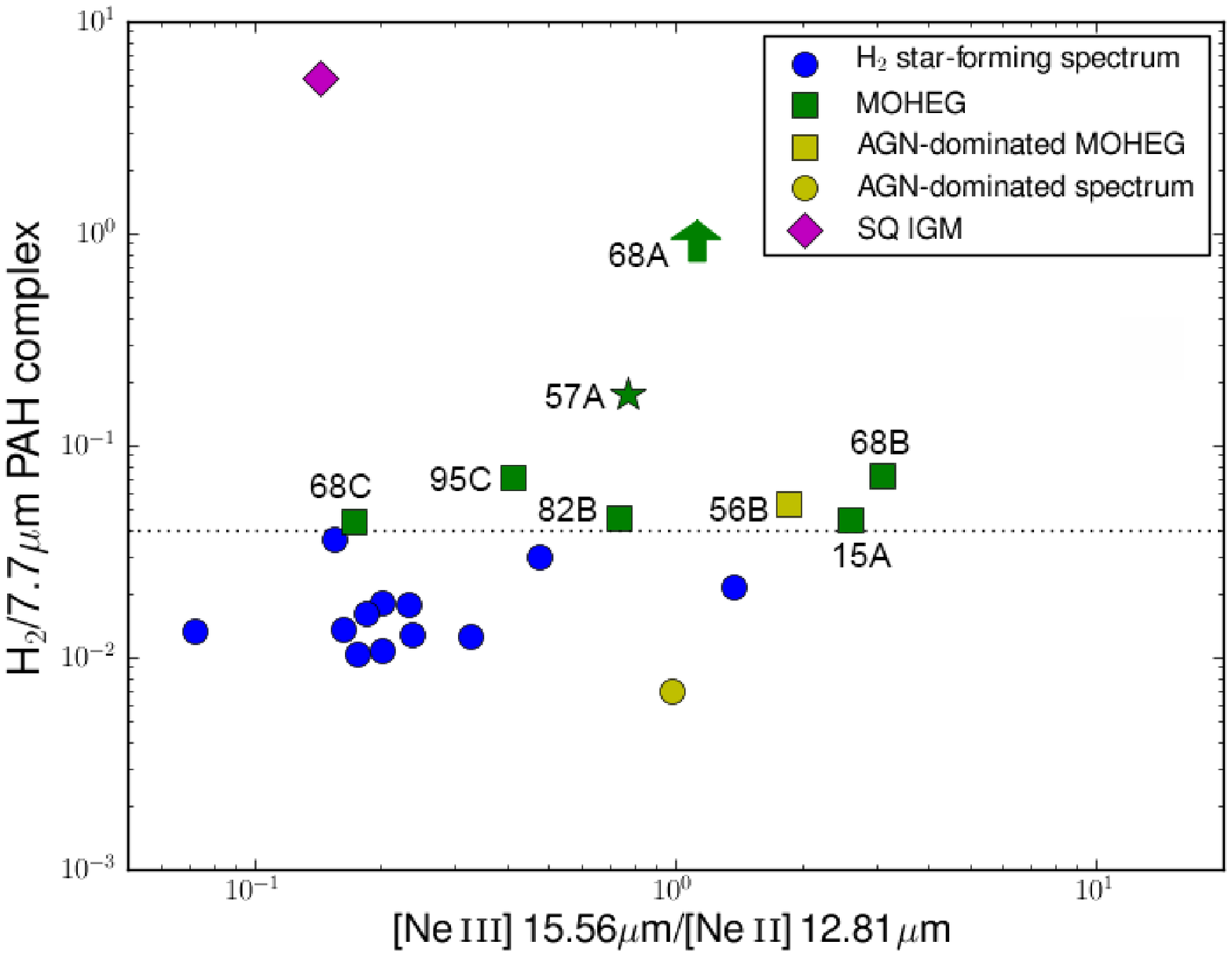}}
\hfill
\subfigure[]{\includegraphics[width=8.1cm]{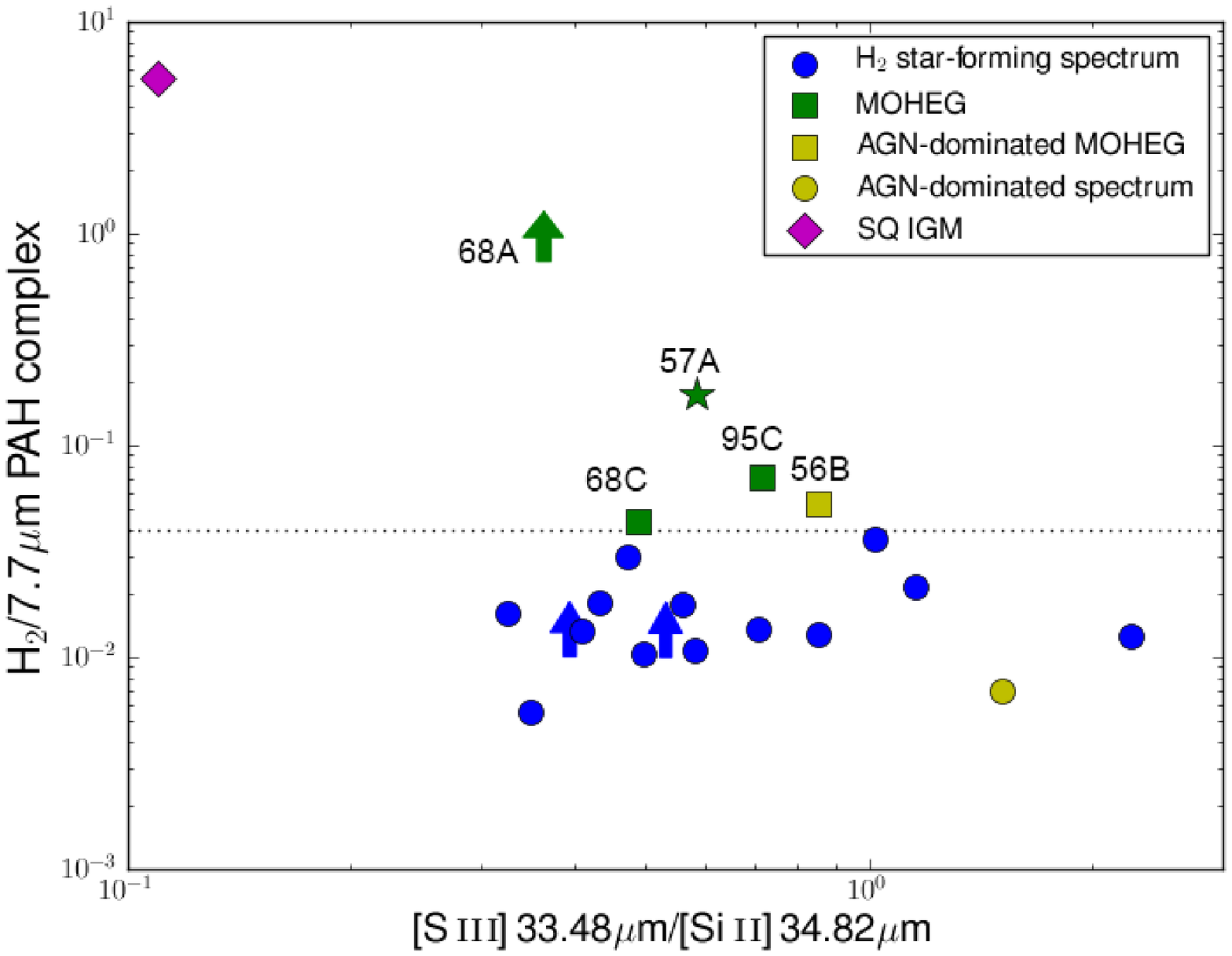}}
\caption{a) \Ht/7.7\micron\ PAH plotted as a function of [Ne{\sc iii}]\,15.56\micron/[Ne{\sc ii}]\,12.81\micron\ with the same color-coding as Figures \ref{HCG_col} and \ref{H2_PAH}a. The off-nuclear extraction of HCG 57A is shown as a star. b) \Ht/7.7\micron\ PAH plotted as a function of [S{\sc iii}]\,33.48\micron/[Si{\sc ii}]\,34.82\micron\ with the same color designation as on the left.} 
\label{fig:FS}
\end{center}
\end{figure*}

In this section we discuss the ionic emission of several key systems, as well as the sample overall. 
In Figure \ref{fig:FS}a and b we plot the \Ht/7.7\micron\ PAH ratios as functions of $[$Ne{\sc iii}$]$\,15.56\micron/$[$Ne{\sc ii}$]$\,12.81\micron\ and $[$S{\sc iii}$]$\,33.48\micron/$[$Si{\sc ii}$]$\,34.82\micron, respectively. We note that for the majority of HCG MOHEGs, the $[$Ne{\sc ii}$]$\,12.81\micron, $[$Ne{\sc iii}$]$\,15.56\micron, $[$S{\sc iii}$]$\,33.48\micron\ and $[$Si{\sc ii}$]$\,34.82\micron\ are upper limits or marginal detections and are not shown.

The [Ne{\sc iii}]\,15.56\micron/[Ne{\sc ii}]\,12.81\micron\ ratios shown in Fig. \ref{fig:FS}a occupy the same range as what is found in the SINGS sample \citep{Dal09} and starburst systems \citep{Ber09}. 
HCG 68B and 15A have the highest ratios of  [Ne{\sc iii}]\,15.56\micron/[Ne{\sc ii}]\,12.81\micron\ (3.1 and 2.6, respectively) and are classified as low-luminosity LINERS from their optical spectra \citep{Coz04}. We note that 68B and 15A are NVSS catalogued radio sources (see Table \ref{tab:xray_nuc}). 

HCG 68A also has a relatively high ratio, but \citet{Coz04} assign it as having no clear optical emission lines, however, it is an NVSS source (see Table \ref{tab:xray_nuc}). Since [Ne{\sc iii}]\,15.56\micron\ and [Ne{\sc ii}]\,12.81\micron\ can also be shock-excited \citep{All08} and is observed associated with the high-velocity shock in SQ \citep{Clu10} and in supernovae remnants \citep{Neu07}, in the absence of an AGN [Ne{\sc iii}]\,15.56\micron\ and [Ne{\sc ii}]\,12.81\micron\ emission may be arising due to shocks.
HCG 56B has a similarly high [Ne{\sc iii}]\,15.56\micron/[Ne{\sc ii}]\,12.81\micron\ ratio within this sample ($\sim 1.86$) signifying a hard radiation field. As discussed in Section 4.1.3. it is a known Seyfert 2 galaxy.

For HCG 57A's off-nuclear extraction, the [Ne{\sc iii}]\,15.56\micron/[Ne{\sc ii}]\,12.81\micron\ ratio is $\sim 0.77$ indicating a radiation field typical of star-forming and Seyfert sources \citep{Dal06}. 

The nebular line [S{\sc iii}]\,33.48\micron\ can be used to determine the relative enhancement of [Si{\sc ii}]\,34.82\micron, usually occurring in active galaxies with hard radiation fields, associated with XDR emission and, finally, due to shocks returning Si to the gas phase \citep[as seen in SQ;][]{Clu10}. [S{\sc iii}]\,33.48\micron/[Si{\sc ii}]\,34.82\micron\ ratios of between 0.5 and and 2 (Fig. \ref{fig:FS}b) are expected for star-forming regions \citep{Dal09}, with an average of $\sim$0.8. [S{\sc iii}]\,33.48\micron/[Si{\sc ii}]\,34.82\micron\ values of $<0.4$ generally suggest enhanced [Si{\sc ii}]\,34.82\micron\ emission, indicative of the presence of an AGN or shocks. Very few HCG MOHEGS can be plotted on this figure, mostly due to the [S{\sc iii}]\,33.48\micron\ line having only an upper limit (see Table \ref{tab:FS}). However, HCG 68A has a [S{\sc iii}]\,33.48\micron/[Si{\sc ii}]\,34.82\micron\ value of 0.36, significantly lower than the SINGS ratios \citep{Dal06}. This may be an indication of silicon atoms being liberated from dust grains driven by shocks. The galaxy is an NVSS source, but is not classified as an AGN optically \citep{Coz04}.
HCG 57A's off-nuclear spectrum has a  [Si{\sc ii}]\,34.82\micron/[S{\sc iii}]\,33.48\micron\ of $\sim 1.71$, which is marginally higher than the average found for typical, star-forming galaxies \citep[$\sim 1.2$,][]{Dal06} despite not incorporating the nucleus. 


\section{Sources of \Ht\ Excitation}

\subsection{AGN and X-ray Excitation of \Ht}

Studies of AGN activity with HCG galaxies have found most to be low luminosity AGNs \citep[LLAGN;][]{Coz04, Mar10}, with a  ratio of broad-to narrow-line AGNs of only 3\%. As shown by Coziol et al. (2004), the LLAGN are either LINERS or Seyfert 2 systems based on their emission line diagnostics. From the optical classifications of \citet{Coz04} and \citet{Mar10} of the \Ht-detected galaxies (shown in Table \ref{tab:morph}), 50\% have \HII\ classifications, 50\% have a Transitional Object classification and 25\% have AGN designations. Only one \Ht-detected galaxy has a LINER classification (HCG 79A). By comparison, the MOHEG systems have one \HII\ classification (HCG 25B), 29\% classified as LINERs, 29\% classified as Transitional Objects and 29\% with AGN designations based on optical spectra of their nuclear regions. 21\% of the MOHEGs have no detectable optical emission. This indicates fewer \HII\ systems, but more LINERs objects amongst the MOHEGs compared to the \Ht-detected systems, whereas optical AGN are roughly equally represented in both. 

Powerful AGN in our sample, as measured by the
shape of the IRS spectrum and high-excitation emission lines, are rare. Only two galaxies, HCG 56B and 96A \citep[classified as Seyfert 2 systems optically,][]{Coz04}, have clear power-law, mid-infrared spectra. 
Tidal interactions are expected to trigger nuclear activity. However, since galaxies are being stripped, their neutral gas reservoir rapidly becomes depleted causing relatively low accretion rates onto the black hole \citep{Mar10}.

Ogle et al (2007; 2010) explained the powerful excited \Ht\ emission in their sample of 3CR radio galaxies to be the result of shocks associated with radio jets from these generally low-luminosity AGN interacting with the host galaxy \citep[see also][]{Guil12b}. However, in our sample, this mechanism is
unlikely (except in the case of HCG 56B) since very few are known radio sources, and those that are detected are low luminosity (see Table \ref{tab:xray_nuc}).

Molecular hydrogen can be excited directly by X-ray heating \citep{Lepp83}, either in the form of XDRs associated with AGN \citep{Mal96}, or due to the presence of hot, diffuse intragroup gas. We know from SQ that the collision of a group galaxy with previously stripped tidal material can produce hot plasma and copious X-ray emission \citep{Trin05, Guil09}. However, whereas the most \HI-deficient systems, and therefore most dynamically evolved, would be expected to be the most X-ray luminous, half are undetected \citep{Ras08}. We present archival X-ray data of the MOHEG groups in Section \ref{diffuse} and find 5 out of 8 have detections, but all with relatively low luminosities. The highest X-ray luminosity system is HCG 82 with a luminosity of  $1.9\times10^{42}$\,erg\,s$^{-1}$, whereas the typical luminosity is closer to $\sim10^{41}$\,erg\,s$^{-1}$).

To investigate X-ray heating within the galaxies themselves, we use archival Chandra and $XMM$ data available for 9 out of 14 MOHEGs (HCG 15A, 15D, 40B, 56B, 56C, 57A, 68A, 68B and 68C; Appendix \ref{xray_gal}). The X-ray emission is insufficient to excite the amount of warm \Ht\ we observe and we conclude that it is unlikely to be a significant heating mechanism.

The available radio data (see section \ref{xray_gal}) suggests that \Ht\ excitation due to cosmic rays in these systems, by analogy with SQ \citep{Guil09} and the Taffy galaxies \citep{Pet12}, is unlikely. This leaves collisional excitation induced by shocks the most
important mechanism to focus our attention and analysis on. 


Before we consider shocks, we should emphasise that it is possible that AGN activity within our sample could produce an artificial enhancement of the \Ht/7.7\micron\ PAH ratio since weak AGNs are capable of influencing the chemistry and composition of nuclear regions within galaxies. The spectra of the inner few square kiloparsecs of 59 nearby galaxies \citep{Smi07b} suggest that the radiation from AGN may modify the grain distribution, exciting unusual PAH emission. 
However, both AGN and dynamical shocks \citep{Mic10} can have impact
close to the emitting source, destroying PAHs and rendering the \Ht\ to
PAH ratio unreliable.

Since we have extracted large areas (see Table \ref{tab:extract}) that nominally include the nuclear region, but are not limited to it, the effect of AGN radiation that may be present will be lessened compared to the study of \citet{Smi07b} which focused specifically on the nuclear region. Of the HCG MOHEGs with detected optical nuclear activity, we have one Seyfert galaxy (HCG 56B, Sy2), four optically classified as AGN (HCG 44A, 57A and 95C; Table \ref{tab:morph}) and four with LINER classifications. 
For HCG 57A, however, an extraction in the disk, well away from the nucleus, clearly shows \Ht-enhancement (Fig. \ref{H2_57}b). We note that the sources of LINER emission in galaxies is still under debate \citep[see e.g.][]{Yan12}, but recent work by \citet{Rich11} find widespread LINER-like emission due to shocks in late-stage mergers.

If the 7.7\micron\ PAH is tracing star formation (and not being depleted) in the HCG MOHEGS, we would expect it to scale approximately with the 24\micron\ continuum emission (unless there is a strong contribution from AGN). Indeed, we show in Fig. 11b that elevated \Ht/7.7\micron\ PAH ratio's map to elevated \Ht/24\micron\ emission, except in the case of HCG 56B where the MOHEG emission may be due to excitation within an XDR or kinetic energy from a jet in this Seyfert galaxy. Based on its  
L$_{\rm H_2}$/L$_{\rm X}$ ratio (see section \ref{xray_gal} of the Appendix) we can rule out XDR-heating and, in combination with its radio luminosity (Table \ref{tab:xray_nuc}), dissipation of kinetic energy provided by the AGN jet appears a more likely excitation mechanism.

The evidence presented here is most consistent with shock-excitation through dynamical interaction; nevertheless our current data does not allow us to definitively distinguish between shock excitation of \Ht\ (via \HI\ cloud collisions or gas infall) or PAH depletion due to AGN, although the latter appears unlikely for the majority of cases based on their optical activity (Table \ref{tab:morph}).



\subsection{Collisional Excitation by Interaction of Galaxies with the Group IGM}

In the previous section we ruled out X-ray heating from within MOHEG galaxies as a mechanism for producing the observed enhanced \Ht\ emission. However, could interactions with the group IGM produce shock-heating of molecular hydrogen? Possible processes are ram pressure stripping due to a hot IGM, collisions with previously stripped tidal material (i.e. cold IGM), and finally material being accreted onto galaxies from the IGM.

\citet{Ras08} investigated the X-ray properties of \HI-deficient groups and found the effect of ram pressure stripping to be small, only capable of removing small amounts of cold gas from group members. Unlike clusters, the relatively shallow gravitational potential well does not produce similarly high temperatures and luminosities of X-ray emission (see section \ref{xray}). Given the relative inefficiency of stripping by hot plasma, \citet{Ras08} find  turbulent viscous stripping due to hydrodynamical interactions \citep[e.g.][]{Nul82} to be a more viable mechanism for removing gas in the HCG galaxies. The efficiency of viscous stripping may even be underestimated given that such processes may not be well treated by SPH (Smoothed Particle Hydrodynamics) schemes \citep{Kaw08}. Therefore, of the \HI-stripping mechanisms in compact groups, tidal interactions appear to dominate, but with viscous stripping potentially contributing significantly and ram-pressure stripping being at most a small effect\citep{Ras08}. Since tidal stripping could help to produce the material required for viscous stripping, we need to look for evidence of this material within the IGM of MOHEG groups.

Recent \HI\ observations of HCGs have shown that even though the galaxies themselves are \HI-deficient \citep{Will87, VM01}, a large diffuse-\HI\ component exists in their IGM \citep{Bor10}.  
We have also shown that for the MOHEG groups HCG 40 and HCG 91 (Figures \ref{fig:spectra_out} and \ref{fig:spectra_out2}, respectively), interferometric \HI\ data hints at structure in between and around the galaxies. More recent VLA data (Verdes-Montenegro et al. in prep.) reveals more: the iso-velocity counters for HCG 40A appear to range from 6225\,km\,s$^{-1}$ to 6600\,km\,s$^{-1}$, but are disturbed towards the location of HCG 40B (a MOHEG) located at $\sim$6800\,km\,s$^{-1}$. \HI\ in HCG 91 reveals streams connecting 91A and C, and further away B, as tentatively suggested by Figure \ref{fig:spectra_out2}.

Of the remaining MOHEG groups with recent VLA data (HCG 15, 25, 68 and 95) the majority, i.e. all except HCG 68, show evidence of interaction-driven disturbed \HI\ (Verdes-Montenegro et al.; private communication). However, due to limitations of the interferometric data (sensitivity limits and resolving out emission), we cannot claim to have a complete picture of the \HI\ distribution within the groups. Considering the groups individually, HCG 15A has no \HI\ detected, however 15D (MOHEG) lies close to 15F spatially and at a similar recessional velocity \citep[seen in the GBT data of][]{Bor10}. The VLA data shows that HCG 15F (with no mid-infrared spectral coverage) has misaligned \HI\ gas compared to its velocity field, indicative of a tidal stream. 
\HI\ is detected around HCG 25B (MOHEG) and F, and forms a bridge-like structure to A. The MOHEG HCG 95C has associated \HI, as well as adjacent gas corresponding to the optical velocity of 95A. 

Given both the limitations of detecting more diffuse \HI\ structures that (likely) exist in the groups, and that not all galaxies in each group were targeted by IRS (i.e. some MOHEG emission is likely missed due to the orientation of the slits), the presence of \HI\ structures within MOHEG groups with available data could be indicative of a causal connection.
As galaxies in compact groups travel through the clumpy IGM formed by numerous gravitational interactions, we could expect stochastic collisions between the inhomogeneous tidal \HI\ and the disks of the galaxies where 
interactions of the gaseous components could lead to the formation of bow shocks \citep{Nul82}. High relative velocities and collisions with more dense \HI\ concentrations in the IGM would produce stronger excited \Ht\ signatures. However, gas disks disrupted and heated due to the turbulent energy created in the collision would be more prone to viscous stripping \citep{Nul82}, due to hydrodynamic drag or hydro-instabilities, and would therefore lose gas in a similar way to traditional ram-pressure stripping by a hot medium. This may also play a role in transforming
late-type disk galaxies into early-type disk systems \citep{Quil00}.

If this is the case, the physical process producing the HCG \Ht-enhanced galaxies is similar to what is observed in Stephan's Quintet, but with much lower relative velocities and therefore lower densities and lower ambient ISM pressure. This would limit the \HI\ to \Ht\ conversion in the region experiencing the collision. Inspection of the groups included in this study show that groups with MOHEGs have velocity dispersions ranging from $\sim$100\,km\,s$^{-1}$ to 500\,km\,s$^{-1}$ \citep{Hick92}, whilst the broad \HI\ profiles of the MOHEG HCGs \citep{Bor10} indicate that high-velocity dispersion ($\Delta V\sim$\,400$-$900\,km\,s$^{-1}$) cold gas is available in the IGM (with column densities $>10^{19}$\,cm$^{-2}$) in addition to tidal streams and clouds.

Lower kinetic energy in the interaction (from lower shock speeds) would not produce as luminous \Ht\ as seen in SQ. Nevertheless, the surface area of the leading disk edge and the distribution of broad \HI\ clouds could produce significant warm \Ht\ mass emission and be detected by {\it Spitzer}. The limitations of slit coverage prevent us from making a definitive statement about the spatial distribution of shock-excited \Ht, but the extraction areas (particularly that of LL which is sensitive to the strongest \Ht\ transitions) cover large parts of the disk, so even though we may have missed stronger emission towards the edges of galaxies, we are able to maximise obtaining a detection of excited \Ht. Follow-up IRAM CO mapping (Lisenfeld et al, in prep.) in combination with optical AGN and shock diagnostics using VIRUS-P (Freeland et al, in prep.) and {\it Herschel} far-infrared spectral mapping will shed light on the mechanisms contributing to the observed evolution.

An alternative to a viscous stripping mechanism is accretion, where IGM tidal debris falls back onto group galaxies in streams that are low-to-medium velocity, thus fueling new star formation activity. If the gas is being decelerated abruptly, it would produce shock excitation, as well as additional fuel for star formation, shifting the galaxy from the red sequence to intermediate mid-infrared colours. Given the velocity dispersions in compact groups \citep{Hick92} and complex gravitational interactions, it is not clear how common such events would be and how much gas could be added in this way. 


The only HCG MOHEGs in this study that show significant star formation (based on their mid-infrared spectra) are 95C and 68C. HCG 95C has the intriguing combination of intermediate mid-infrared colours and a star forming spectrum that may indicate a recent increase in star formation relative to the galaxy's stellar mass. This may be due to tidal interactions or possibly a significant gas accretion event enhancing star formation (see Section 4.1.3). If a threshold of gas mass exists that must be accreted to generate excited \Ht\ , and an associated amount of star formation expected to be seen in these systems, detailed modelling is required to investigate this threshold.
In the majority of HCG systems, however, no enhancement in star formation is found \citep{Ig98, Mar10} and instead evidence of truncation in early-type HCG galaxies \citep{del07} and relatively low star formation rates \citep{Bit11}.

\section{The Mid-Infrared ``Green Valley": An Evolutionary Connection?}\label{green}

\begin{figure}[!t]
\begin{center}
\includegraphics[width=8.8cm]{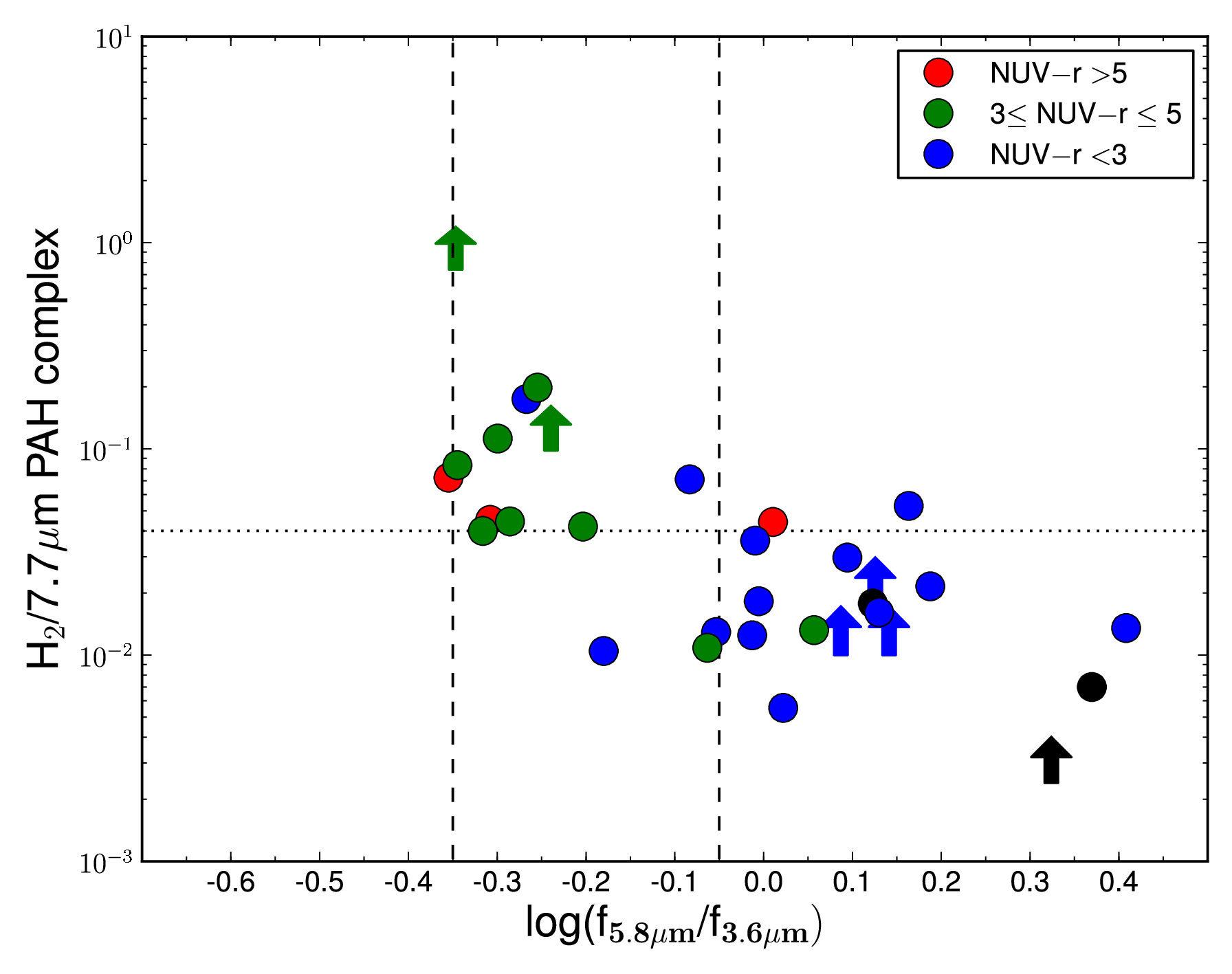}
\caption{\Ht/7.7\micron\ PAH ratio color-coded according to NUV-r color. The galaxies are divided into red sequence systems (red), blue cloud galaxies (blue) and green valley members (green) with black points representing galaxies with absent data. The vertical dashed lines indicate the mid-infrared green valley (from Equation 1).}
\label{optcol}
\end{center}
\end{figure}

The observed underdensity of HCG galaxies at intermediate mid-infrared colors hints at active evolution from dusty star-forming to early type \citep[as originally suggested by ][]{Joh07}. 
The location of MOHEG galaxies (as shown in Figs \ref{HCG_col} and \ref{H2_PAH}), lying in the mid-infrared ``green valley", provides a tantalising addition to the overall picture of transformation within HCGs.

Large area surveys have revealed an intermediate population in UV-optical color space: the ``green valley" galaxies  \citep{Blan03}. These are speculated to be undergoing active transformation from the``blue cloud" to the ``red sequence". Given that \citet{Bit11} find over 50\% of early-type and more than 60\% of late-type galaxies in ``dynamically old" groups are located in the optical green valley, there may exist a possible connection between HCG MOHEGS and the optical green valley.

Using the NUV and r-band photometry from \citet{Bit11}, we determine extinction-corrected NUV$-$r colors for the HCG MOHEGs with available data. 
In Figure \ref{optcol} we plot the \Ht/7.7\micron\ PAH values color-coded by NUV$-$r value, with $3 \le$NUV$-$r$\le 5$ in the optical green valley, NUV$-$r$> 5$ belonging to the red sequence and NUV$-$r$<3$ designating the blue cloud. A large fraction of MOHEG galaxies lie either in the green valley or on the red sequence, with only three having blue NUV$-$r colors. The predominance
of ``green'' NUV$-$r colors in the MOHEG sample is suggestive of a connection between \Ht\ enhancement and the star formation properties of the compact group systems.

\begin{figure*}[!tb]
\begin{center}

\subfigure[]{\includegraphics[width=8cm]{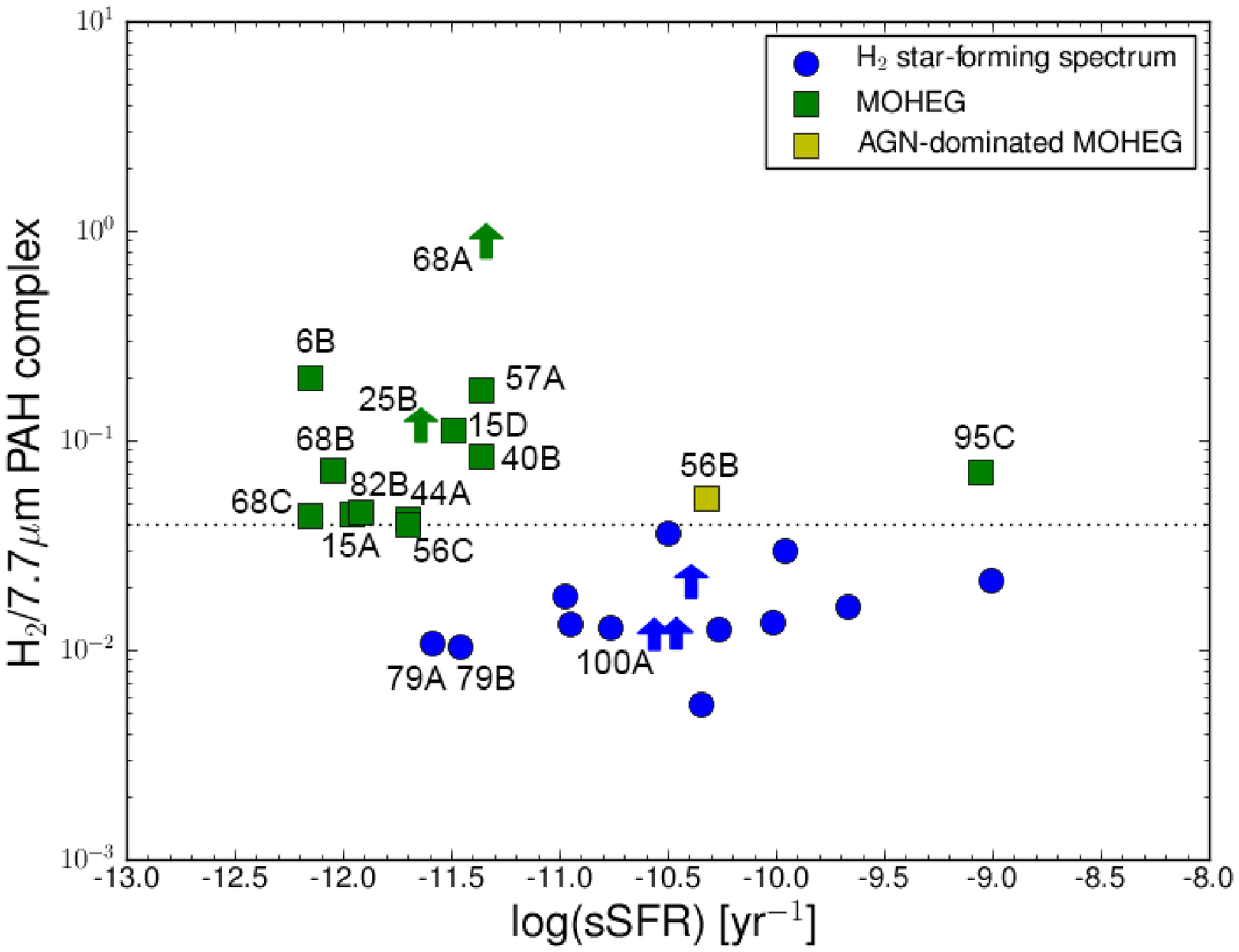}}
\subfigure[]{\includegraphics[width=7.8cm]{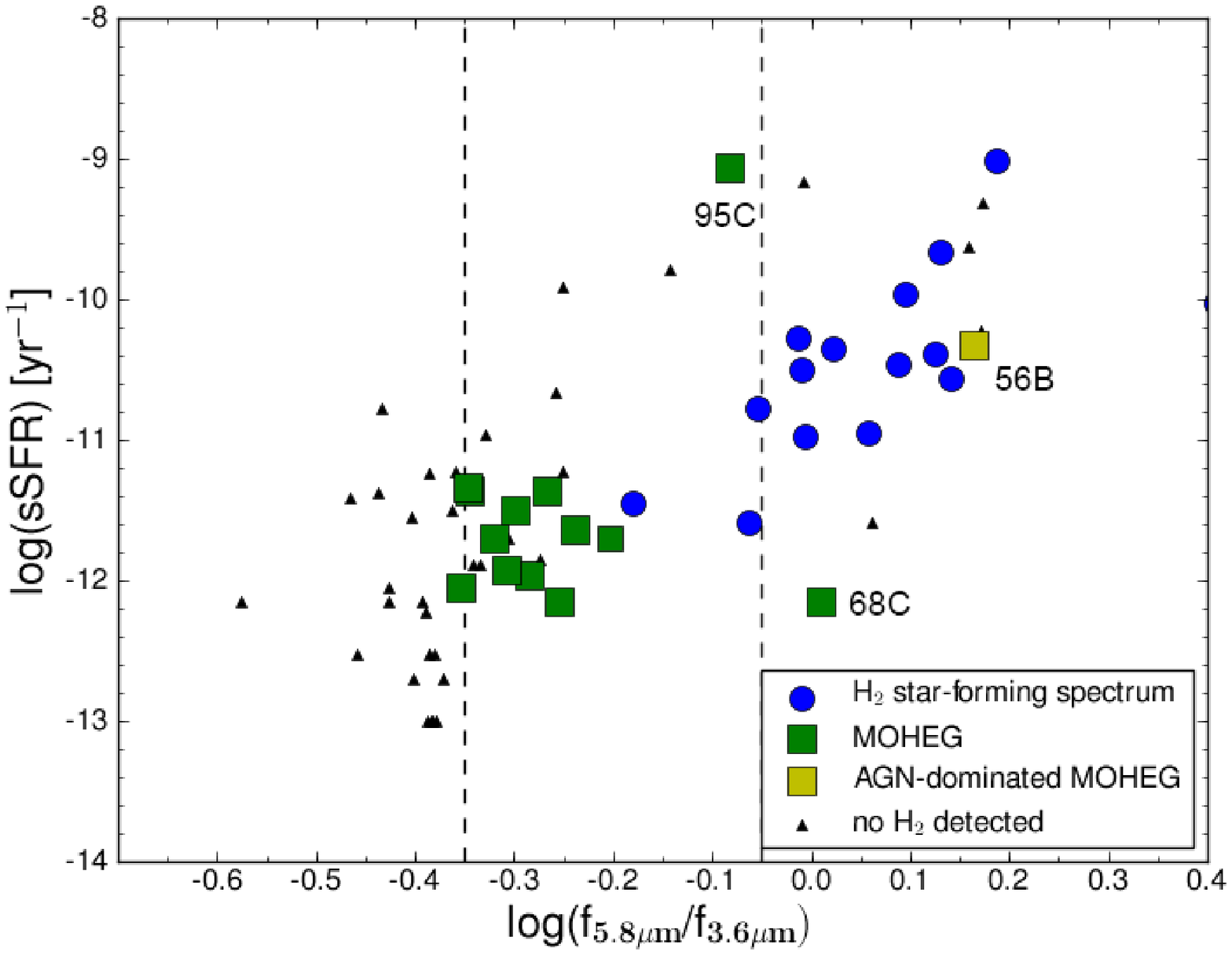}}
\caption{a) \Ht/7.7\micron\ PAH as a function of log sSFR \citep[sSFR data from][]{Bit11} with \Ht\ star-forming systems in blue, MOHEGs in green and AGN-dominated MOHEGs in yellow. b) Plotting log\,sSFR against ${\rm log[f}_{5.8\mu \rm m}/{\rm f}_{3.6\mu \rm m}]$ shows that the majority of MOHEGs (in green) may form a transitional population between star-forming and early-type systems. AGN-dominated MOHEGs are shown in yellow, \Ht-detected star-forming galaxies in blue and non-\Ht\ detections as black triangles.}
\label{ssfr}
\end{center}
\end{figure*}

Employing intermediate mid-infrared colors to define a mid-infrared ``green valley" within compact groups has the advantage of mitigating dust obscuration, which affects the NUV$-$r color heavily. In dense environments, where interactions cause gas and dust to be stripped from galaxies, this may be a key tracer of transformation \citep[e.g. the Coma Infall region; ][]{Walk12}. Furthermore, utilising mid-infrared colors from {\it Spitzer} and {\it WISE}, particularly to select dusty, early-type galaxies, may help uncover the ubiquity of active transformation in a variety of environments.

To investigate this further, we look at the specific star formation rates (sSFR) of galaxies in our sample in common with that of \citet{Bit11}. Their sSFR (SFR per unit old stellar mass) is based on SED (spectral energy distribution) fitting and 2\micron\ near-infrared luminosity. 

The majority of HCG MOHEGs occupy a narrow range in sSFR ($-12.3 < $log(sSFR)[Gyr$^{-1}$]$ < -11.3$), except for the somewhat unusual HCG 95C (see section \ref{moheg_sources}) and the AGN-dominated system HCG 56B (Fig. \ref{ssfr}a). This suggests that the mechanism producing enhanced excited \Ht\ in these systems is somewhat different to the other MOHEGs. In the case of HCG 95C, a merger or gas accretion may explain the increased star formation in relation to stellar mass, producing shock-excited \Ht\ as a by-product. In HCG 56B, a jet interacting with the ISM is a possible source of excess excited \Ht, which elevates it above the \Ht\ star-forming systems with similar sSFR.

Figure \ref{ssfr}b plots our entire galaxy group sample with sSFR taken from \citet{Bit11}. 
The ${\rm log[f}_{5.8\mu \rm m}/{\rm f}_{3.6\mu \rm m}]$ color -- sensitive to star formation through the continuum and 6.2\micron\ PAH, divided by the 3.6\micron\ emission tracing stellar mass-- appears to act as a general proxy for sSFR. This is not surprising since the ${\rm log[f}_{5.8\mu \rm m}/{\rm f}_{3.6\mu \rm m}]$ color tracks the ${\rm log[f}_{8.0\mu \rm m}/{\rm f}_{3.6\mu \rm m}]$ color closely: the 8.0\micron\ band
is dominated by the 7.7\micron\ PAH tracing star formation, normalized by the 3.6\micron\ light that tracing stellar mass. The majority of HCG MOHEGs appear to lie at low-intermediate sSFR i.e. at the red end of the early-type population. This suggests that the HCG MOHEGs may be a transitional population between the two main populations, and that the warm \Ht\ emission is connected to the mechanism driving the bimodality in mid-infrared color, actively moving systems onto the optical red sequence.

A notable exception to this picture, HCG 68C, lies well below the actively star-forming cloud in disagreement with its mid-infrared color. A possible explanation is that its star formation rate has been underestimated \citep[0.05\,M$_\sun$yr$^{-1}$;][]{Bit11} since \citet{Mart12} find a SFR of 1.46\,M$_\sun$yr$^{-1}$ and log(sSFR) of -10.70\,yr$^{-1}$, which would move it to the blue cloud of actively star-forming galaxies.

\begin{figure*}[!bt]
\begin{center}
\subfigure[]{\includegraphics[width=8cm]{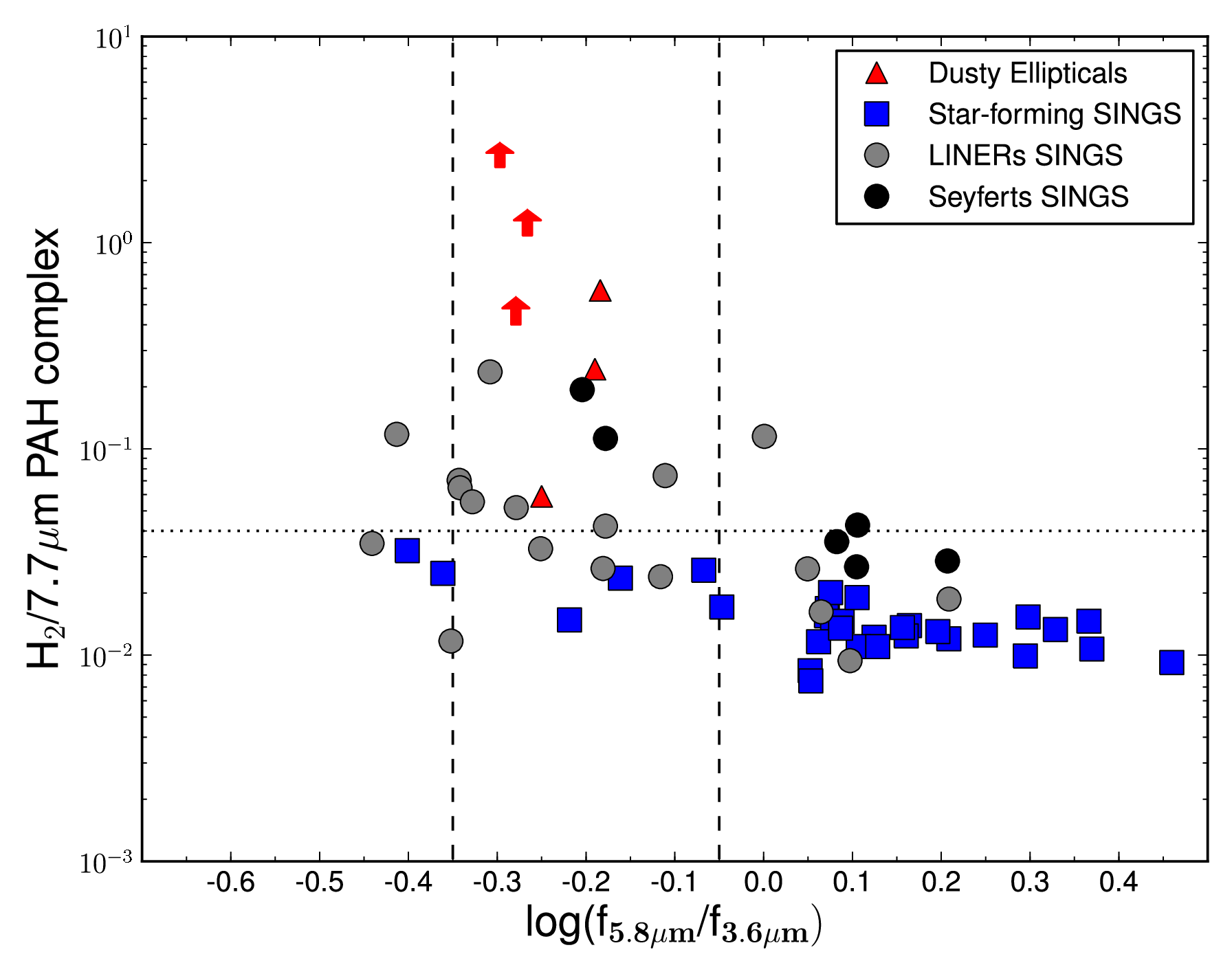}}
\subfigure[]{\includegraphics[width=8cm]{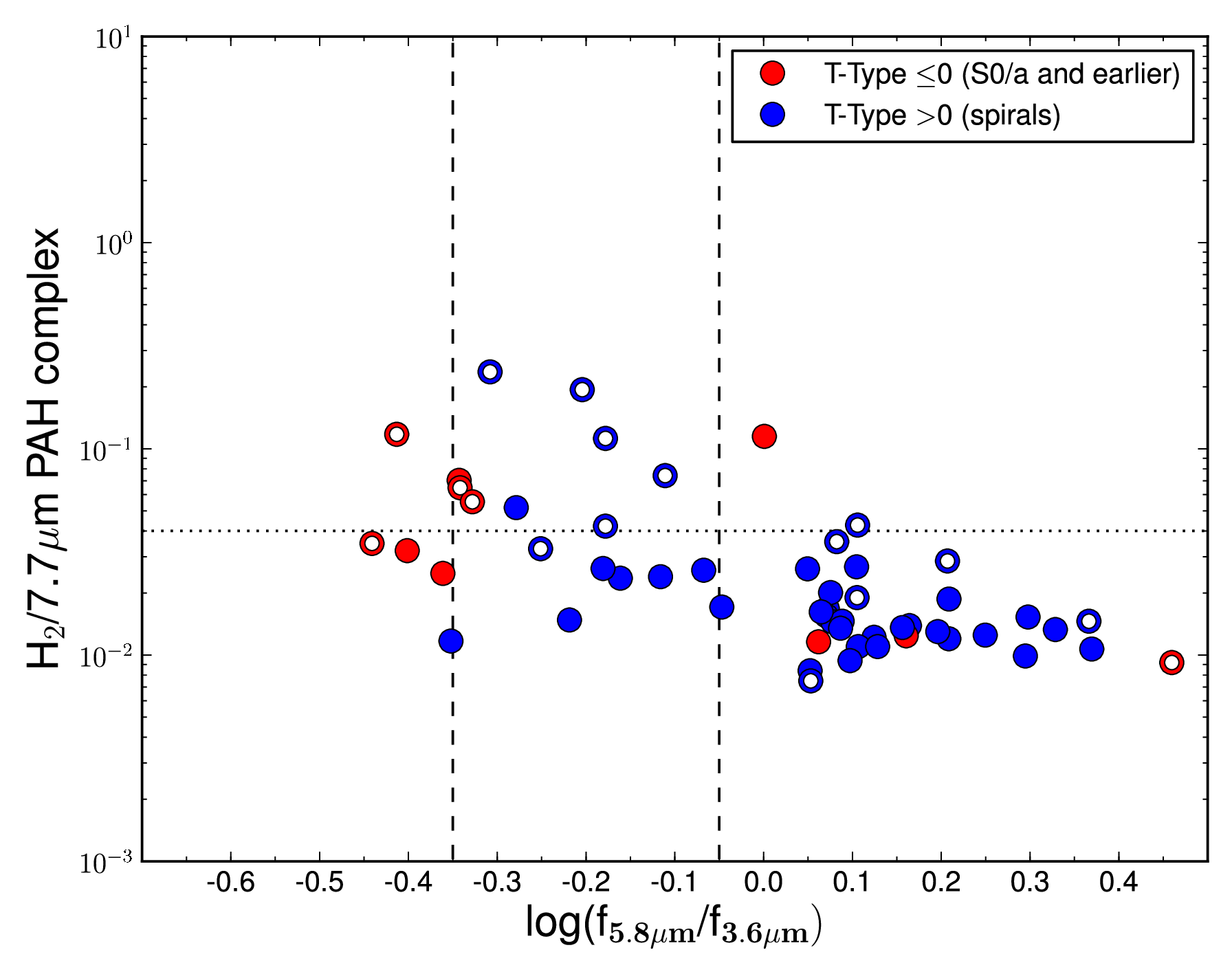}}
\caption{a) The \citet{Kan08} dusty ellipticals (red) and SINGS \Ht-detected galaxies are plotted as a function of ${\rm log[f}_{5.8\mu \rm m}/{\rm f}_{3.6\mu \rm m}]$ color, where the vertical dashed lines indicate the mid-infrared green valley (from Equation 1).  SINGS star-forming systems are shown in blue (squares), LINER systems are grey (circles) and Seyferts are in black (circles). b) The SINGS sample is plotted in terms of T-type (late-type in blue, early type in red), with open circles indicating galaxies in dense/interacting environments.}
\label{MOHEG+all}
\end{center}
\end{figure*}

\subsection{Comparison with other samples}

\citet{Og10} have shown that jet-ISM interactions can efficiently produce excited \Ht\ in radio-loud galaxies. Since we have ruled out this mechanism in the majority of HCG MOHEGs, we now consider whether the phenomenon of enhancement in warm \Ht\ emission could be linked to the environment of the host. For example, several nearby dusty early type galaxies reveal strong \Ht\ emission associated with almost negligible star formation \citep{Kan08,Vega10}. Recent accretion events have been suggested as a mechanism, particularly since some systems show star/gas counter-rotating disks. The shocks believed to give rise to the observed warm \Ht\ emission could arise from a mechanism such as gas infall or a minor merger delivering additional gas, that then fuels a rejuvenation event \citep{Vega10}. 
The ellipticals in \citet{Kan08} were chosen to be X-ray-bright, dusty systems, with some having known jets (making jet-ISM interactions a possible mechanism) and other containing LLAGN.

A connection between \Ht-enhanced systems and LINER-like optical emission is not clear. All four of the early types in the sample of \citet{Vega10} are classified as LINERs, and, within the SINGS sample, galaxies classified as LINERs as well as Seyfert galaxies show enhanced excited \Ht\ emission \citep{Rous07,Og10}.

In Figure \ref{MOHEG+all}a we plot six of the \citet{Kan08} systems (with archival IRAC photometry) and the SINGS sample \citep{Rous07,Dal05}, as a function of ${\rm log[f}_{5.8\mu \rm m}/{\rm f}_{3.6\mu \rm m}]$ color. We see that the dusty ellipticals from \citet{Kan08} and several SINGS LINERs occupy a similar phase space compared to the HCG MOHEGs. However, the SINGS sample was not chosen to be uniform, but instead representative of nearby galaxies; several systems are located in dense environments such as clusters. In Figure \ref{MOHEG+all}b we show the SINGS galaxies as a function of T-type and plot galaxies that are located in a dense or interacting environment (Virgo, Fornax, Dorado clusters, compact group, pair or triple) as an open circle. 
We notice more spiral systems inhabit the red end of the mid-infrared green valley, compared to HCGs where this region is dominated by S0/a types. Intriguingly, this distribution appears somewhat similar to the HCG distribution in \Ht/7.7\micron\ PAH space. However, we cannot rule out AGN being the source of excitation in the \citet{Kan08} and SINGS MOHEG systems.

The effect of environment on the evolution of a galaxy is still contentiously debated.  Compact groups are relatively rare, but given their strong gravitational interactions, and relatively shallow potential-wells, may provide an opportunity to understand related mechanisms at work in high-density clusters, as well as loose groups. 
For example, comparisons with several different control samples led \citet{Walk10, Walk12} to conclude that the observed IRAC color bimodality in their sample of HCGs was a direct consequence of environment and appeared most similar to the distribution found for the Coma infall region.

\section{Discussion}

Recent studies have explored the connection between dense environments and the formation of early-type disk and lenticular galaxies. HCGs may be the ideal laboratory to study how galaxies are being ``pre-processed" and transformed by mechanisms within groups and clusters. 


The simulations of \citet{Bek11} indicate that the compact group environment, causing gas stripping and bulge building via tidal interactions, preferentially form S0's from spirals. They predict long tidal streams, broken rings and seemingly isolated massive clouds of neutral gas within the IGM. Deep \HI\ observations are uncovering these kind of features \citep[e.g. in HCG 44,][]{Ser12b}, as well as the strong dependence of the \HI\ morphology of early-type galaxies on environment \citep{Ser12a}. The pre-processing mechanisms occurring in groups may be key to explaining the morphology-density relation as observed in field and cluster galaxies \citep{Post84}.

Observationally, \citet{Bit11} find that S0's dominate dynamically evolved HCGs and are preferentially found in the optical green valle. If we believe that these systems are not passively evolving, but instead show accelerated evolution from the blue cloud to the red sequence, there may be a connection to the MOHEGs observed in the mid-infrared green valley. 

We have discussed the possibility that interactions with tidal debris may be a valid mechanism within compact groups, and would therefore show evidence of extended shock emission. Although we have some indication that in many cases the observed warm \Ht\ is extended on scales larger than the LL slit, without maps we lack crucial spatial information. Detailed shock diagnostic observations are currently underway with both optical IFU (Integral Field Unit) measurements and resolved {\it Herschel} spectroscopy of far-infrared cooling lines, such as high-J CO, H$_2$O, [C\,{\sc ii}] and [O\,{\sc i}].  

Tidal interactions combined with viscous stripping could produce accelerated evolution from late-type to early-type and may help explain systems with evidence of \HI-deficiency and truncated \HI\ disks that cannot be explained by simple ram pressure arguments. For example, in the Pegasus I cluster \citep{Lev07} the same correlation between sSFR and \HI-deficiency is observed in this lower density, low velocity dispersion system as found in Virgo \citep{Rose10}. They find that sSFR within \HI-depleted disks is lower, but that the sSFR of the nucleus relative to the global value increases with \HI\ deficiency. This suggests a mechanism that builds stellar mass in the bulge, but decreases \HI\ and star formation in the disk.
The evidence for suppression of star formation in low-redshift galaxy groups found by \citet{Ras12b} may be related to such a mechanism. In this study {\it GALEX} imaging indicates that, on average, the sSFR of star-forming galaxies in groups is suppressed by $\sim$40\% relative to the field.

\section{Conclusions}

In this paper we have presented the results of a mid-infrared spectral and photometric study of 74 galaxies in 23 Hickson Compact Groups using the {\it Spitzer Space Telescope} with the goal of searching for enhanced \Ht\ emission and its possible connection to evolution within HCGs. We highlight here our primary conclusions:

\begin{itemize}

\item{Searching for excited \Ht\ in the intragroup medium have led to two tentative detections, located at the edges of two disk galaxies in HCG 40 and 91, respectively. Large-scale intragroup warm \Ht\ emission as seen in Stephans' Quintet, appears to be rare.}

\item{We have detected warm \Ht\ in 32 galaxies with 14 MOHEG (Molecular Hydrogen Emission Galaxies) systems, i.e. \Ht\ emission in excess of what is expected from UV excitation associated with star formation. The observed \Ht/7.7\micron\, PAH ratios may be due to either shock-induced excitation of \Ht\ or, in a small minority of systems, possible PAH depletion due to AGN activity.}

\item{The luminosity of X-rays detected in MOHEG galaxies with archival data is insufficient to heat the observed amounts of excited \Ht. We find that in the majority of systems, AGN activity is unlikely to be responsible for the observed \Ht-enhancement. This leaves the possibility of shock-excitation through interaction with the IGM as a plausible mechanism for producing the observed emission.} 

\item{The presence of copious intragroup \HI\ suggests that collisional excitation may be due to galaxies passing through this debris, thus experiencing shock heating of their disks  -- analogous, but less energetic to what is seen in the SQ shock. The heated gas disks undergoing viscous stripping may accelerate evolution from the blue cloud across the green valley. The previously stripped material may also accrete back onto the galaxies, shock-heating the disk gas and providing fuel for ``new" star formation.}

\item{The MOHEG systems are mainly early-type disks and lie chiefly within a mid-infrared ``green valley" between the population of dusty star formers and early type galaxies. The optical colours of the majority of these systems lie in the NUV$-$r ``green valley", and their locus in specific star formation suggests they may form a transitional population moving onto the red sequence.}

\item{Interactions within a group, pair or triple environment may be producing similar \Ht-enhancement accompanied by non-passive evolution in mid-infrared or optical colour, as seen in the SINGS sample. }


\end{itemize}


\acknowledgements
We thank the referee for helpful comments and suggestions that improved the content and clarity of this paper. We are grateful to Sanch Borthakur (Johns Hopkins University) for access to unpublished \HI\ data and Michael Brown (Monash University) for the use of his galaxy template library. This work is based on observations made with the
{\it Spitzer} Space Telescope, which is operated by the Jet Propulsion Laboratory, California Institute of Technology under NASA contract 1407. MEC acknowledges support from the Australian Research Council (FS110200023). Support for this work was provided by NASA through an award issued by JPL/Caltech. UL acknowledges support  by the research projects
AYA2007-67625-C02-02 and AYA2011-24728 from the Spanish Ministerio de
Ciencia y Educaci \'on and the Junta de Andaluc \'ia (Spain) grant  
FQM108. LVM is funded by Grant AYA2008-06181-C02  and AYA2011-30491-C02-01, co-financed by
MICINN and FEDER funds, and the Junta de Andalucía (Spain) grants P08-FQM-4205 and
TIC-114. This research has made use of the NASA/IPAC Extragalactic Database (NED) which is operated by the Jet Propulsion Laboratory, California Institute of Technology, under contract with the National Aeronautics and Space Administration. This research has made use of data obtained from the Chandra Source Catalog, provided by the Chandra X-ray Center (CXC) as part of the Chandra Data Archive.

\appendix

\section{The HCG Sample and Associated Data}\label{samp_scale}

In this section we list all the galaxies that formed part of the parent HCG galaxy sample and the IRAC photometry measurements of each (as outlined in Section \ref{irac_obs}). Table \ref{tab:phot} provides the photometry for each galaxy in the sample and if a warm \Ht\ detection was made.


\begin{center}
\begin{longtable}{l l r r r r }
\caption{Measured IRAC Fluxes of HCG Sample \label{tab:phot}}\\

Galaxy    &  \Ht-detected  &  3.6\micron\ (mJy) & 4.5\micron\ (mJy) & 5.8\micron\ (mJy) & 8.0\micron\ (mJy)  \\
\hline
\\
\endfirsthead

Galaxy    &  \Ht-detected  &  3.6\micron\ (mJy) & 4.5\micron\ (mJy) & 5.8\micron\ (mJy) & 8.0\micron\ (mJy)  \\
\hline
\\
\endhead

6A  & no &    6.71  &   4.28  &   2.70  &   1.63    \\    
6B  & yes &   7.78  &   5.12   &  4.10   &  4.93  \\    
6C   & no &    5.81  &   3.72  &   2.37  &   1.64   \\  
6D  & no &  0.95  &   0.68   &  0.50  &   0.68 \\     
8A & no & 11.37  & 7.42  &  4.65  &   3.04 \\   
8C  & no & 5.99   &  3.86  &  2.73  &   2.14  \\     
8D  & no  & 5.72  &   3.72  &   2.45  &  2.25 \\    
15A & yes & 18.32  &  11.63  &  9.49  &   11.59   \\   
15C  & no &  13.79  &  8.72  &  5.87 &    3.53 \\    
15D & yes  &   8.13  &  5.49  &   4.08  &   3.01 \\    
25B  & yes  & 18.33 &  12.09  &  10.56  & 12.10  \\   
25D  & no  &  3.72  &   2.44   &  1.53  &   1.22 \\   
25F   & no &  2.86   &  1.82  &   1.25  &   0.79  \\     
31A+C & yes &  13.54  & 10.37 &  28.55 &  73.86  \\   
31B & no  &  2.46  &  1.69  &  2.23  &  4.42  \\   
40A  & no &   35.53  & 22.12 &  16.21 &  11.22 \\  
40B & yes & 10.11  & 6.41 &  4.57  &  3.50  \\    
40C  & yes & 23.77 &  16.22  & 29.04 &  73.88 \\  
40D  & yes  &  16.61 &  11.15 &  18.93  & 61.31 \\  
44A & yes  &  292.17 & 185.70 & 182.89 & 309.77 \\ 
44B  & no  &  --  & 96.29 & --  & 38.56 \\  
44D  & yes  &  13.46 &  9.11 &   20.74 &  49.71 \\   
47A$^\dagger$  & yes & 20.71 &  13.31 &  20.26 &  52.28 \\  
47B$^\dagger$  & no  & 11.84  & 7.23  &  4.05  &  2.41\\   
47D$^\dagger$  & no &  3.59  &  2.37  &  2.33  &  6.37 \\   
54A  & no &    1.14  &   0.78  &   0.82  &   1.39 \\    
54B$^\dagger$   & no &   3.70  &   2.90  &   5.50  &   11.10 \\   
54C$^\dagger$  & no & 1.25   & 0.80  &   1.80  &   4.10  \\     
55A$^\dagger$  & no & 6.53  &  3.93  &  2.66  &  1.22   \\  
55B$^\dagger$ & no &  3.44  &  2.15  &  1.33  &  0.72  \\  
55C$^\dagger$  & yes  &  5.51  &  3.80   &  3.91  &  10.91\\  
55D$^\dagger$  & no  &  6.76  &  4.36  &  6.10  &  12.97 \\ 
56B  & yes  & $^\diamond$25.36 &    $^\diamond$30.84  &   36.94  &   52.30  \\ 
56C  & yes  &  5.82  &   3.71  &   2.81  &  2.45 \\  
56D  & yes  &  4.08 &   2.90  &  5.45 &   14.53   \\  
56E  & yes  & 2.12  &  1.43  &  2.06  &   4.08 \\        
57A  & yes  & 31.11  &  20.04  &  16.82    &  20.15 \\  
57B  & no & 14.36  &  9.14   &  8.04  &   12.47  \\   
57C  & no  & 10.88  &  6.99  &   4.51   &  2.64\\   
57D & no  &  7.83  &   5.26  &   8.40   &  19.69 \\   
57E & no  &  8.98  &   5.77   &  4.43   &  5.23 \\   
62A & no & 94.42  & 58.58  & 38.77  & 54.16  \\   
62B & no &  15.16  & 9.36 &   6.27  &  4.26  \\  
62C  & no & 14.69  & 9.46  &  5.93  &  3.24  \\  
67A$^\dagger$  & no  & 54.40  &  34.10  &  20.40 &   16.10  \\   
67B$^\dagger$  & yes  & 30.70   & 20.80  &  32.30  &  72.30  \\  
67D$^\dagger$ & no &  5.30   &  3.30  &   2.30  &   1.80 \\        
68A   & yes  &  226.63  & 136.56  & 101.99 &  61.77 \\ 
68B  & yes  &  108.76 &  66.57  & 48.04  & 33.19  \\ 
68C  & yes  &  111.70 &   69.21  &   114.44  &  248.81  \\     
75A$^\dagger$  & no & 13.10  &  10.40  &  5.70  &   5.10 \\   
75C$^\dagger$  & no  &  4.00  &   2.40  &   1.60   &  1.30  \\  
75D$^\dagger$  & yes  & 3.30  &   2.10  &   3.40  &   11.80  \\ 
75E   & no  &   3.70  &   2.37  &   1.51  &   0.93 \\    
79A   & yes  &  16.09 &  10.58  &    13.91  &  26.10 \\  
79B  & yes  &  14.53 &   9.15 &  9.60 &   15.00 \\  
79C$^\dagger$ & no  &   10.22 &  5.71  &  4.03  &  2.04  \\    
82A & no  &  15.69  &   10.06 &    6.56  &   3.94  \\   
82B & yes   & 12.78  &   8.02  &   5.98  &   6.96 \\  
82C & yes  & 8.54   &  6.24  &   15.12  &  47.75 \\       
91A  & yes  &  $^\star$74.57 &   $^\diamond$59.87 &   92.65   &   608.84 \\   
91C  & yes  &  8.68   &  5.88  &   11.71   &  30.63 \\  
91D  & no  & 10.97  &  7.21  &   5.15   &  4.35  \\       
95A$^\dagger$   & no & 18.00  &    11.01 &  6.70  &   7.82  \\  
95C$^\dagger$  & yes  &   7.05  &  4.27  &  4.89  &  10.60  \\      
96A$^a$ & yes & 80.15  & 107.86 &  187.72 &  328.56 \\ 
96B & no   & 17.71  & 12.09 &  8.69 &   5.46  \\   
96C$^b$  & yes  &  8.86  &   6.21  &   11.76  &  29.78  \\     
97A$^\dagger$  & no  &  19.90  &  12.70  &  7.90  &   6.0  \\   
97C$^\dagger$  & no   &  13.20 &   7.80  &   3.50   &  2.20  \\  
97D$^\dagger$  & no  &   15.50 &   9.30  &   5.80  &   4.10  \\        
100A  & yes &   40.38  & 25.93  & 35.70  &  156.89  \\ 
100B  & no &  5.49 &   3.95 &   8.15  &  18.58  \\  
100C & no  &  3.65  &  2.56  &  3.58 &   7.14  \\  

\\
\hline
\\[0.5pt]
\multicolumn{6}{l}{$^a$ Measured using short exposure (1.2s) HDR image}\\
\multicolumn{6}{l}{$^b$ Contamination from 96A}\\
\multicolumn{6}{l}{$^\diamond$ Flux may be underestimated by $\sim$5\% due to peak pixel non-linearity}\\
\multicolumn{6}{l}{$^\star$ Flux may be underestimated by $\sim$10\% due to peak pixel non-linearity}\\
\multicolumn{6}{l}{$^\dagger$ Photometry from Bitsakis et al. (2011)}\\
\end{longtable}
\end{center}

For systems with detected excited \Ht\ emission lines, the measured fluxes are presented in Table \ref{tab:Ht}. Several systems have their SL and LL lines presented separately due to the regions sampled by the IRS not overlapping (and therefore not joined together). In addition, spectra that are not centered on the nucleus (therefore dominated by emission from the disk) are indicated. The strengths of the PAH complexes and atomic emission lines for the \Ht-detected systems are presented in Tables \ref{tab:pah} and \ref{tab:FS}.
For galaxies without SL coverage, the IRAC inferred 7.7\micron\ PAH values are:
8.77$\times 10^{-15}$\,W.m$^{-2}$, 1.33$\times 10^{-14}$\,W.m$^{-2}$ and 1.68$\times 10^{-15}$\,W.m$^{-2}$
for 40C, 44A and 75D, respectively.

\begin{sidewaystable*}[!p]
\centering
\scriptsize
{\rm
\caption{
Observed \Ht\ Fluxes in units of W\,m$^{-2}$ \label{tab:Ht}    
}
\hspace{0cm}
\begin{tabular}{l l l l l l l l l l}

\tableline
\tableline
\\[0.25pt]

Source & \multicolumn{1}{c}{H$_{2}$ 0-0 S(0)} &  \multicolumn{1}{c}{H$_{2}$ 0-0 S(1)} &  \multicolumn{1}{c}{H$_{2}$ 0-0 S(2)} &  \multicolumn{1}{c}{H$_{2}$ 0-0 S(3)} &  \multicolumn{1}{c}{H$_{2}$ 0-0 S(4)} & 
 \multicolumn{1}{c}{H$_{2}$ 0-0 S(5)} \\  
 & \multicolumn{1}{c}{$\lambda$28.21$\mu$m}   & \multicolumn{1}{c}{$\lambda$17.03$\mu$m} & \multicolumn{1}{c}{$\lambda$12.28$\mu$m}  & \multicolumn{1}{c}{$\lambda$9.66$\mu$m}  & \multicolumn{1}{c}{$\lambda$8.03$\mu$m}   & \multicolumn{1}{c}{$\lambda$6.91$\mu$m}  \\

\tableline
\\[0.25pt]

6B    &   1.710e-18   (5.093e-19)      &  8.440e-18   (8.113e-19)    &  1.251e-18   (5.748e-19)$^\Box$    &  6.185e-18   (1.739e-18)   &   $<$3.751e-18   &   $<$3.151e-18$^\dagger$   \\  
 15A    &   $<$2.605e-18        &  4.796e-18   (8.951e-19)    &  $<$2.446e-18   &  3.216e-18   (1.186e-18)   &   $<$5.385e-18   &   6.149e-18   (1.819e-18) \\  
 15D    &   2.920e-18   (1.019e-18)      &  1.107e-17   (1.864e-18)    &  3.750e-18   (1.719e-18)$^\Box$   &  8.811e-18   (2.345e-18)   &  $<$2.367e-17 &   $<$6.414e-18 $^\dagger$ \\  
 25B$^\star$    &   --      &  --    &  7.643e-19   (4.799e-20)   &  2.535e-18   (1.202e-19)   &   $<$8.727e-19   &   2.012e-18   (2.728e-19)$^\dagger$ \\  
 25B$^{\diamond\bullet}$    &   $<$1.853e-18      &  1.159e-17   (1.979e-19)    &  --   &  --   &   --  &   -- \\  
 31A+C$^{\star\bullet}$    &   --      &  --    &  1.731e-19   (3.424e-20)$^\Box$   &  1.329e-19   (5.397e-20)$^\Box$   &   $<$3.332e-19 &  $<$ 7.815e-19   \\  
 31A+C$^\diamond$    &   $<$1.423e-17      &  2.004e-17   (3.924e-20)    &  --   &  --   &   -- &   -- \\  
 40B    &   $<$2.673e-18         &  5.619e-18   (1.081e-18)    &  $<$9.519e-19   &  3.457e-18   (1.063e-18)$^\Box$   &   4.550e-18   (1.920e-18)$^\Box$ &   $<$1.631e-18$^\dagger$ \\  
 40C$^{\star\bullet}$    &   --      &  --    &  $<$1.966e-19     &  3.932e-19   (1.611e-19)$^\Box$   &   $<$6.603e-19   &  4.335e-19   (2.012e-19)$^{\Box\dagger}$ \\  
 40C$^\diamond$    &   1.857e-17   (1.602e-18)      &  7.308e-17   (1.093e-19)    &  --   &  --   &   -- &   -- \\  
 40D    &   9.173e-18   (3.047e-18)$^\Box$      &  2.697e-17   (2.675e-18)    &  1.084e-17   (1.256e-18)$^\Box$   &  2.009e-17   (1.693e-18)   &   $<$8.889e-18  &  6.978e-18   (2.348e-18)$^\Box$  \\  
 44A$^{\star\bullet}$    &   --      &  --    &  2.124e-18   (2.083e-19)$^\Box$   &  3.245e-18   (2.542e-19)$^\Box$   &    4.527e-18   ( 1.047e-18)$^\Box$ &  7.281e-18   ( 9.326e-19)$^{\Box\dagger}$\\  
 44A$^\diamond$    &   2.693e-17   (5.940e-18)      &  1.069e-16   (3.014e-20)    &  --   &  --   &  -- &  -- \\  
 44D    &   9.598e-18   (3.674e-18)      &  5.243e-17   (8.442e-18)    &  $<$1.261e-17   &  $<$3.129e-16    &   $<$7.616e-17    &   3.302e-17   (1.337e-17)$^{\Box\dagger}$\\  
 47A    &   $<$8.042e-18    &  2.556e-17   (3.100e-18)    &  7.745e-18   (3.100e-18)$^\Box$    &  $<$1.769e-17    &   1.928e-17   ( 8.152e-18)$^\Box$  &   $<$1.534e-17$^\dagger$\\  
 55C    &   1.056e-18   (1.718e-19)      &  3.927e-18   (3.630e-19)    &  1.445e-18   (3.084e-19)$^\Box$   &  3.411e-18   (5.093e-19)   &   3.035e-18   ( 6.191e-19)$^\Box$ &   $<$5.632e-18$^\dagger$  \\  
 56B    &   $<$6.768e-18        &  5.973e-18   (3.035e-18)    &  1.231e-17   (3.704e-18)   &  $<$1.575e-17      &   $<$3.544e-18 &  8.222e-18   (3.403e-18)$^\Box$ \\  
 56C    &   $<$1.143e-18      &  4.915e-18   (8.659e-19)    &  $<$4.024e-18    &  7.139e-18   (1.290e-18)   &   $<$7.657e-18   &  5.764e-18   (1.706e-18)$^{\Box\dagger}$\\  
 56D$^\star$    &   --      &  --    &  1.892e-17   (8.559e-19)$^\Box$    &  1.338e-17   (2.767e-18)   &  8.456e-18   ( 2.124e-18)$^\Box$  &   1.136e-17   (1.441e-18)$^\Box$ \\  
 56E    &   $<$3.736e-18        &  3.169e-18   (7.172e-19)    &  $<$1.122e-18   &    1.027e-18   (4.599e-19)$^\Box$   &   $<$2.200e-18   &   2.641e-18   ( 7.172e-19)$^{\Box\dagger}$\\  
 57A$^\bullet$    &   3.521e-18   (4.759e-19)      &  1.311e-17   (5.390e-19)    &  2.622e-18   (5.269e-19)   &  5.754e-18   (4.856e-19)   &   1.343e-18   ( 6.313e-19)$^\Box$ &  2.792e-18   (6.507e-19)$^{\Box\dagger}$\\  
 57A$^\diamond$    &   7.728e-18   (1.699e-18)      &  7.612e-17   (2.456e-19)    &  --   &  --   &  -- &  -- \\  
 67B    &   5.306e-18   (1.044e-18)      &  1.800e-17   (2.793e-18)    &  1.036e-17   (1.655e-18)$^\Box$   &  1.528e-17   (3.829e-18)$^\Box$   &   2.852e-17   ( 4.397e-18)$^\Box$  &   $<$2.102e-17    \\  
 68A    &   $<$1.642e-18      &  1.212e-17   (9.984e-19)    &  1.037e-17   (9.887e-19)   &  2.026e-17   (1.115e-18)   &  $<$1.193e-17 &  $<$1.605e-17 $^\dagger$\\  
 68B    &   $<$1.709e-18        &  4.415e-18   (9.170e-19)    &  $<$2.312e-18      &  3.685e-18   (8.024e-19)   &   $<$3.436e-18 &   $<$7.999e-18$^\dagger$   \\  
 68C    &   7.839e-18   (8.386e-19)      &  3.919e-17   (1.540e-18)    &  2.625e-17   (2.388e-18)   &  5.779e-17   (3.792e-18)   &   $<$1.566e-17   &   $<$2.433e-17   \\  
 75D$^\diamond$    &   2.163e-18   (6.665e-19)      &  7.513e-18   (1.533e-18)    &  --   &  --   &  -- &  -- \\  
 79A    &   3.830e-18   (4.552e-19)      &  1.248e-17   (1.053e-18)    &  5.138e-18   (3.508e-19)$^\Box$   &  1.062e-17   (7.780e-19)   &   $<$5.298e-18 &  $<$ 3.660e-18    \\  
 79B    &   1.028e-18   (2.694e-19)      &  6.854e-18   (5.797e-19)    &  1.867e-18   (5.214e-19)$^\Box$   &  2.563e-18   (7.620e-19)   &   $<$5.196e-18   &  1.783e-17   (2.512e-18)$^\Box$ \\  
 82B    &   1.684e-18   (3.707e-19)      &  4.619e-18   (8.734e-19)    &  $<$2.300e-18     &  2.644e-18   (1.024e-18)   &   $<$4.643e-18  &   $<$1.333e-18$^\dagger$\\  
 82C    &   1.026e-17   (8.441e-19)      &   3.333e-17   (1.375e-18)   &  9.751e-18   (7.786e-19)$^\Box$   &  1.208e-17   (1.113e-18)   &   $<$6.353e-18    &  1.528e-17   (2.511e-18)$^\Box$ \\  
 91A    &   1.346e-17 (1.010e-18)     &   4.585e-17 (1.844e-18)    &   1.078e-17 (1.468e-18)$^\Box$   & 1.831e-17 (1.662e-18)$^\Box$    &  $<$1.186e-17 &  2.959e-17 (5.604e-18)$^{\Box\dagger}$\\   
 91C    &   $<$2.228e-18       &  1.028e-17   (1.750e-18)    &  2.327e-18   (7.695e-19)$^\Box$   &  7.423e-18   (1.079e-18)   &   $<$7.423e-18    &   $<$7.430e-18 \\  
 95C    &   2.249e-18   (7.984e-19)      &  8.838e-18   (1.317e-18)    &  5.388e-18   (7.923e-19)   &  1.707e-17   (1.975e-18)   &   $<$5.144e-18  &   7.497e-18   (2.633e-18) \\  
 96A    &   3.337e-17   (2.346e-18)      &  5.041e-17   (3.424e-18)    &  5.085e-17   (1.418e-18)$^\Box$   &  $<$3.897e-18   &  5.609e-17   (3.380e-18)$^\Box$ &  1.472e-16   (4.225e-18)$^{\Box\dagger}$\\  
 96C    &   2.954e-18   (9.037e-19)      &  1.795e-17   (1.498e-18)    &  7.339e-18   (8.066e-19)   &  1.104e-17   (1.407e-18)   &   $<$5.604e-18 &  1.462e-17   ( 3.051e-18)$^\Box$ \\  
 100A    &   5.569e-18   (9.621e-19)      &  3.240e-17   (1.723e-18)    &  6.564e-18   (5.994e-19)   &  1.662e-17   (9.197e-19)   &   $<$9.418e-18 &   2.584e-17   (2.536e-18)$^\Box$ \\

\\
\tableline
\\
\multicolumn{7}{l}{Uncertainties are listed in parentheses. Marginal detections are low signal to noise as determined by visual inspection.}\\
\multicolumn{5}{l}{$^\dagger$ Blended with [Ar{\sc ii}]\,6.98\micron}\\
\multicolumn{5}{l}{$^\Box$ Marginal Detection}\\
\multicolumn{5}{l}{$^\star$ SL coverage}\\
\multicolumn{5}{l}{$^\diamond$ LL coverage}\\
\multicolumn{5}{l}{$^\bullet$ Off nuclear extraction}
\end{tabular}

}

\end{sidewaystable*}

\begin{sidewaystable*}[!p]
\scriptsize
{\rm
\caption{
Observed PAH Fluxes in units of W m$^{-2}$ \label{tab:pah}    
}
\hspace{0cm}
\begin{tabular}{l l l l l l l l l l}

\tableline
\tableline
\\[0.25pt]

Source & \multicolumn{1}{c}{6.2\micron } &  \multicolumn{1}{c}{7.7\micron} &  \multicolumn{1}{c}{8.6\micron} &  \multicolumn{1}{c}{11.3\micron} &  \multicolumn{1}{c}{12.0\micron} & \multicolumn{1}{c}{12.6\micron} & \multicolumn{1}{c}{17\micron} \\  

\tableline
\\[0.25pt]

6B  &  4.887e-17   (4.680e-18)    &  8.243e-17     (1.297e-17)    &   2.040e-17    (4.442e-18)   &   5.249e-17   (3.756e-18)   &   1.523e-17   (2.321e-18)   &   2.342e-17   (2.536e-18)   &    3.676e-17    (2.710e-18) \\  
 15A  &  3.540e-17   (7.305e-18)$^\Box$    &  1.801e-16     (8.771e-18)    &   4.074e-17    (5.875e-18)   &   6.837e-17   (2.427e-18)   &   $<$1.070e-17   &   2.597e-17   (3.200e-18)   &    7.251e-17    (4.287e-18) \\  
 15D  &  $<$6.174e-17     &  2.032e-16     (2.877e-17)    &   $<$2.337e-17   &   6.964e-17   (5.563e-18)   &   1.485e-17   (6.701e-18)   &   $<$8.418e-18  &    2.192e-17    (7.626e-18) \\  
 25B$^\star$  &  7.689e-18   (7.358e-19)    &  3.357e-17     (1.538e-18)    &   6.939e-18    (5.788e-19)   &   1.848e-17   (1.940e-19)   &   7.228e-18   (2.091e-19)   &   9.607e-18   (1.871e-19)   &    -- \\  
 25B$^{\diamond\bullet}$  &  --   &  --    &   --   &   --   &   --  &   --   &    6.379e-16    (1.166e-18) \\  
 31A+C$^{\star\bullet}$  &  1.365e-17   (3.994e-19)    &  3.780e-17     (8.320e-19)    &   7.066e-18    (3.023e-19)   &   1.278e-17   (1.103e-19)   &   4.095e-18   (1.370e-19)   &   7.486e-18   (1.484e-19)   &    -- \\  
 31A+C$^\diamond$  &  --   &  --    &   --   &   --   &   --   &  --   &    2.376e-16    (4.624e-19) \\  
 40B  &  7.170e-17   (5.749e-18)$^\Box$    &  6.738e-17     (7.512e-18)     &   1.622e-17    (4.595e-18)   &   4.554e-17   (3.003e-18)   &   $<$3.919e-18   &   1.491e-17 (4.326e-18)   &    $<$1.469e-17  \\  
 40C$^{\star\bullet}$  &  1.319e-17   (5.119e-19)    &  3.887e-17     (1.208e-18)    &   9.010e-18    (5.650e-19)   &   1.555e-17   (3.918e-19)   &   3.898e-18   (2.845e-19)   &   9.877e-18   (2.920e-19)   &    -- \\  
 40C$^\diamond$  &  --   &  --   &   --   &   --   &   --   &   --   &    1.524e-15    (2.062e-18) \\  
 40D  &  1.033e-15   (1.337e-17)    &  3.561e-15     (2.860e-17)    &   6.633e-16    (9.244e-18)   &   9.058e-16   (1.484e-17)   &   2.417e-16   (5.517e-18)   &   4.741e-16   (6.437e-18)   &    4.988e-16    (1.937e-17) \\  
 44A$^{\star\bullet}$  &  $<$1.051e-17    &  1.037e-16     (5.614e-18)$^\Box$    &   2.189e-17    (1.916e-18)   &   3.914e-17   (4.805e-19)   &   1.815e-17   (6.773e-19)   &   9.427e-18   (3.950e-19)   &    -- \\  
 44A$^\diamond$  &  --    &  --   &   --   &  --   &   --   &   --   &    1.857e-15    (4.175e-19) \\  
 44D  &  7.656e-16   (5.555e-17)    &  2.879e-15     (2.036e-16)    &   1.021e-15    (1.427e-16)   &   1.255e-15   (1.397e-16)   &   $<$1.529e-16   &   2.680e-16   (4.162e-17)   &    7.096e-16    (7.267e-17) \\  
 47A  &  3.529e-16   (2.302e-17)    &  7.118e-16     (6.860e-17)    &   9.348e-17    (2.120e-17)   &   3.640e-16   (1.234e-17)   &   1.126e-16   (1.318e-17)   &   1.650e-16   (1.282e-17)   &    1.179e-16    (1.423e-17) \\  
 55C  &  1.302e-16   (3.253e-18)    &  4.601e-16     (8.403e-18)    &   1.027e-16    (2.327e-18)   &   1.293e-16   (2.369e-18)   &   2.390e-17   (1.269e-18)   &   5.445e-17   (1.442e-18)   &    7.819e-17    (2.511e-18) \\  
 56B  &  $<$7.599e-17    &  3.457e-16     (6.633e-17)    &   $<$1.439e-17  &   2.538e-16   (1.399e-17)   &   1.972e-16   (1.630e-17)   &   1.330e-16   (1.568e-17)   &    2.918e-16    (1.856e-17) \\  
 56C  &  4.972e-17   (9.903e-18)    &  3.228e-16     (1.748e-17)    &   6.813e-17    (5.966e-18)   &   1.218e-16   (1.496e-18)   &   5.649e-17   (2.109e-18)   &   2.935e-17   (1.229e-18)   &    $<$1.666e-17 \\  
 56D$^\star$  &  3.337e-16   (1.170e-17)    &  7.553e-16     (2.350e-17)    &   1.618e-16    (1.319e-17)   &   1.914e-16   (1.259e-17)   &   $<$9.407e-18   &   3.028e-17   (1.515e-18)   &    -- \\  
 56E  &  6.294e-17   (2.653e-18)    &  2.540e-16     (6.618e-18)    &   4.730e-17    (2.157e-18)   &   5.496e-17   (1.055e-18)   &   1.294e-17   (1.323e-18)   &   2.286e-17   (1.447e-18)   &    2.499e-17    (3.501e-18) \\  
 57A$^\bullet$  &  3.918e-17   (1.651e-18)    &  1.433e-16     (5.967e-18)    &   2.508e-17    (2.045e-18)   &   5.307e-17   (1.507e-18)   &   1.823e-17   (2.041e-18)   &   2.444e-17   (2.253e-18)   &    5.068e-17    (2.757e-18) \\  
 57A$^\diamond$  &  --    &  --   &  --    &   --   &  --   &   --  &    4.564e-16    (1.887e-18) \\  
 67B  &  1.104e-15   (1.766e-17)    &  4.201e-15     (6.808e-17)    &   7.128e-16    (1.455e-17)   &   9.034e-16   (1.341e-17)   &   2.917e-16   (7.759e-18)   &   5.261e-16   (8.408e-18)   &    2.416e-16    (9.284e-18) \\  
 68A  &  1.168e-16   (5.306e-18)    &  $<$8.205e-17    &   $<$3.644e-17   &   1.442e-16   (2.168e-18)   &   3.786e-17   (3.059e-18)   &   3.908e-17   (3.000e-18)   &    9.871e-17    (4.280e-18) \\  
 68B  &  5.253e-17   (6.416e-18)    &  1.117e-16     (6.815e-18)    &   1.542e-17    (3.073e-18)   &   1.028e-16   (1.665e-18)   &   2.346e-17   (2.352e-18)   &   4.577e-17   (3.457e-18)   &    1.060e-16    (4.077e-18) \\  
 68C  &  1.045e-15   (2.108e-17)    &  2.964e-15     (7.289e-17)    &   4.906e-16    (1.552e-17)   &   1.046e-15   (7.681e-18)   &   3.649e-16   (8.636e-18)   &   5.015e-16   (7.960e-18)   &    7.284e-16    (9.992e-18) \\  
 75D$^\diamond$  &  --    &  --  &  --  &  --   &  --   &   --   &    1.844e-16    (2.583e-17) \\  
 79A  &  6.546e-16   (5.593e-18)    &  2.483e-15     (1.268e-17)    &   3.424e-16    (3.394e-18)   &   4.291e-16   (2.668e-18)   &   1.281e-16   (2.033e-18)   &   2.652e-16   (1.990e-18)   &    2.608e-16    (6.256e-18) \\  
 79B  &  2.760e-16   (6.515e-18)    &  9.979e-16     (2.698e-17)    &   1.606e-16    (3.098e-18)   &   2.040e-16   (4.496e-18)   &   7.498e-17   (2.311e-18)   &   1.327e-16   (2.878e-18)   &    9.559e-17    (4.119e-18) \\  
 82B  &  4.659e-17   (5.080e-18)    &  1.977e-16     (1.941e-17)    &   3.197e-17    (3.731e-18)   &   5.797e-17   (2.708e-18)   &   9.219e-18   (2.504e-18)   &   2.647e-17   (2.714e-18)   &    4.386e-17    (5.425e-18) \\  
 82C  &  1.157e-15   (9.503e-18)    &  4.116e-15     (2.664e-17)    &   6.894e-16    (7.531e-18)   &   7.082e-16   (7.277e-18)   &   2.260e-16   (3.150e-18)   &   4.615e-16   (3.427e-18)   &    4.445e-16    (1.119e-17) \\  
91A  &  7.837e-16   (1.712e-17)    &  1.996e-15     (5.703e-17)    &   2.009e-16    (7.835e-18)   &   6.046e-16   (4.341e-18)   &   1.983e-16   (5.965e-18)   &   2.840e-16   (6.344e-18)   &    6.052e-16    (8.812e-18) \\  
91C  &  3.142e-16   (7.112e-18)    &  1.098e-15     (2.869e-17)    &   2.135e-16    (5.177e-18)   &   2.488e-16   (3.136e-18)   &   6.280e-17   (3.899e-18)   &   1.510e-16   (4.302e-18)   &    1.586e-16    (9.371e-18) \\  
 95C  &  1.081e-16   (8.582e-18)    &  4.714e-16     (2.685e-17)    &   9.415e-17    (4.675e-18)   &   1.364e-16   (4.887e-18)   &   4.481e-17   (3.553e-18)   &   8.102e-17   (3.839e-18)   &    1.477e-16    (1.020e-17) \\  
 96A  &  2.593e-15   (1.179e-17)    &  1.197e-14     (4.933e-17)    &   7.595e-16    (8.759e-18)   &   1.736e-15   (4.418e-18)   &   1.007e-15   (6.902e-18)   &   1.410e-15   (6.369e-18)   &    2.030e-15    (2.042e-17) \\  
 96C  &  7.005e-16   (1.056e-17)    &  2.209e-15     (3.928e-17)    &   4.341e-16    (7.732e-18)   &   5.684e-16   (1.190e-17)   &   1.558e-16   (4.062e-18)   &   2.825e-16   (4.642e-18)   &    2.439e-16    (1.070e-17) \\  
 100A  &  1.226e-15   (7.659e-18)    &  4.220e-15     (2.289e-17)    &   6.740e-16    (3.410e-18)   &   1.209e-15   (2.044e-18)   &   2.752e-16   (2.881e-18)   &   6.563e-16   (3.496e-18)   &    6.432e-16    (7.450e-18) \\

\\
\tableline
\\
\multicolumn{6}{l}{Uncertainties are listed in parentheses. Marginal detections are low signal to noise as determined by visual inspection.}\\
\multicolumn{6}{l}{$^\Box$ Marginal Detection}\\
\multicolumn{6}{l}{$^\star$ SL coverage}\\
\multicolumn{6}{l}{$^\diamond$ LL coverage}\\
\multicolumn{6}{l}{$^\bullet$ Off nuclear extraction}
\end{tabular}
}

\end{sidewaystable*}

\begin{sidewaystable*}[!p]
\scriptsize

{\rm
\caption{
Observed Forbidden Line Fluxes in units of W m$^{-2}$ \label{tab:FS}    
}
\hspace{0cm}
\begin{tabular}{l l l l l l  l l l  l}

\tableline
\tableline
\\[0.25pt]

Source & \multicolumn{1}{c}{[Ar{\sc ii}]\,6.98\micron} &  \multicolumn{1}{c}{[Ne{\sc ii}]\,12.81\micron} &  \multicolumn{1}{c}{[Ne{\sc iii}]\,15.56\micron} &  \multicolumn{1}{c}{[S{\sc iii}]\,18.71\micron} &  \multicolumn{1}{c}{[O{\sc iv}]\,25.89\micron/} & \multicolumn{1}{c}{[S{\sc iii}]\,33.48\micron} & \multicolumn{1}{c}{[Si{\sc ii}]\,34.82\micron} \\  
 & & &  &  &  \multicolumn{1}{c}{[Fe{\sc ii}]\,25.98\micron} &  &  \\
\tableline
\\[0.25pt]

6B    &   $<$3.400e-18$^{\dagger}$      &  1.364e-18   (4.693e-19)$^\Box$   &  2.568e-18   (4.584e-19)$^\Box$   &  $<$1.306e-18   &   1.575e-18   (4.075e-19)  &   $<$3.423e-18  &   3.674e-18   (1.066e-18) \\  
 15A    &   5.290e-18   (1.470e-18)      &  2.452e-18   (6.782e-19)    &  6.353e-18   (7.423e-19)   &  4.737e-18   (7.139e-19)   &  $<$2.554e-18 &   $<$4.0479e-18 &   1.033e-17   (2.001e-18) \\  
 15D    &   $<$1.072e-17$^{\dagger}$      &  4.369e-18   (1.704e-18)$^\Box$    &  $<$4.799e-18   &  $<$3.546e-18   &   $<$6.896e-18 &  $<$3.141e-18  &   4.005e-18   (1.893e-18)$^\Box$ \\  
 25B$^\star$     &   9.752e-19   (2.419e-19)$^{\Box\dagger}$      &  5.534e-19   (4.702e-20)    &  --      &   --  &  -- &  -- &   -- \\  
 25B$^{\diamond\bullet}$    &   --      &  --    &  1.772e-18   (1.979e-20)   &  $<$4.689e-18   &   $<$2.629e-18 &  $<$4.047e-18 &   $<$4.604e-18 \\  
 31A+C$^\diamond$    &   --      &  --    &  3.737e-16   (6.930e-20)   &  2.718e-16   (1.148e-19)   &   $<$1.690e-17 &   2.786e-16   (1.446e-17) &   1.269e-16   (1.366e-17) \\  
40B    &   $<$3.457e-18      &  3.542e-18   (1.002e-18)$^\Box$     &  3.353e-18   (1.045e-18)$^\Box$    &  $<$2.196e-18  &   3.748e-18   (6.926e-19)$^\Box$  &   $<$3.748e-18 &   $<$7.405e-18  \\  
 40C$^{\star\bullet}$    &   4.938e-19   (1.332e-19)$^{\Box\dagger}$      &  5.706e-19   (7.023e-20)    &  --   & --   &   -- &   -- &  -- \\  
 40C$^\diamond$     &   --      & --   &  2.452e-17   (2.294e-20)   &  8.971e-17   (1.469e-19)   &    3.861e-18   (1.208e-18) &   6.325e-17   (1.651e-20) &   1.602e-16   (2.331e-20) \\  
 40D    &   5.886e-17   (2.446e-18)      &  1.114e-16   (1.431e-18)    &  8.048e-18   (1.824e-18)   &  3.385e-17   (2.402e-18)   &   1.365e-17   (1.376e-18) &   4.870e-17   (5.056e-18) &   1.190e-16   (4.805e-18) \\  
 44A$^{\star\bullet}$    &   2.621e-18   (8.208e-19)$^\Box$      &  1.849e-18   (2.263e-19)    &  --   &  --    &   -- &  -- &   -- \\  
 44A$^\diamond$    &   --      &  --    &  5.692e-17   (1.718e-20)   &  3.348e-17   (4.615e-20)   &   2.096e-17   (9.390e-18)  &   3.348e-17   (7.745e-18) &   1.863e-16   (6.930e-18)$^\Box$  \\  
 44D    &   4.933e-17   (1.245e-17)$^{\Box\dagger}$       &  4.149e-17   (8.400e-18)    &  5.717e-17   (7.080e-18)   &  1.540e-16   (1.346e-17)   &   7.843e-18   (3.406e-18)  &   9.845e-17   (5.490e-18) &   8.545e-17   (5.903e-18) \\  
 47A    &   3.295e-17   (5.367e-18)$^{\Box\dagger}$      &  3.745e-17   (3.049e-18)    &  5.792e-18   (2.624e-18)   &  1.928e-17   (2.403e-18)   &      $<$9.792e-18  &   4.433e-17   (2.369e-18) &   4.356e-17   (2.998e-18) \\  
 55C    &   $<$5.808e-18$^\dagger$      &  1.002e-17   (2.574e-19)    &  2.021e-18   (4.941e-19)   &  3.873e-18   (7.648e-19)   &   $<$1.045e-18 &   6.434e-18   (4.134e-19) &   1.487e-17   (7.770e-19) \\  
 56B    &   2.725e-17   (4.189e-18)      &  3.636e-17   (4.169e-18)    &  6.748e-17   (3.365e-18)   &  1.813e-17   (2.695e-18)   &     8.377e-17   (2.220e-18) &   1.377e-17   (2.569e-18) &   1.610e-17   (3.423e-18) \\  
 56C    &   8.005e-18   (1.469e-18)$^{\Box\dagger}$      &  4.830e-18   (1.316e-18)$^\Box$    &  2.555e-18   (9.593e-19)$^\Box$   &  2.581e-18   (5.985e-19)$^\Box$   &   $<$4.609e-18 &   6.053e-18   (1.783e-18)$^\Box$ &   $<$7.828e-18 \\  
 56D$^\star$     &   3.089e-17   (1.583e-18)      &  2.381e-17   (8.803e-19)    &  --   &  --   &    -- &   -- &   -- \\  
 56E    &   3.111e-18   (7.657e-19)$^{\Box\dagger}$      &  5.185e-18   (3.979e-19)    &  1.682e-18   (4.357e-19)   &  4.846e-18   (5.428e-19)   &   $<$2.224e-18 &   5.767e-18   (1.502e-18) &   2.559e-18   (1.575e-18) \\  
 57A$^\bullet$    &   1.216e-18   (5.026e-19)$^{\Box\dagger}$      &  4.006e-18   (5.876e-19)    &  3.084e-18   (5.026e-19)   &  1.585e-18   (3.229e-19)   &   1.148e-18   (5.099e-19)  &   2.719e-18   (7.600e-19) &   4.637e-18   (8.110e-19) \\  
 57A$^\diamond$    &  --      &  --    &  1.184e-17   (2.592e-20)   &  3.971e-18   (7.379e-20)   &  $<$1.875e-18 &   1.010e-17   (2.515e-18) &   2.845e-17   (2.864e-18) \\  
 67B    &   4.092e-17   (7.869e-18)      &  7.530e-17   (1.715e-18)    &  3.404e-18   (1.672e-18)$^\Box$   &  8.065e-18   (1.426e-18)   &  $<$5.662e-18 &   1.825e-17   (2.334e-18) &   5.212e-17   (3.192e-18) \\  
 68A    &   1.706e-17   (2.472e-18)$^{\Box\dagger}$      &  1.270e-17   (7.531e-19)    &  1.435e-17   (7.590e-19)   &  6.979e-18   (9.325e-19)   &    4.381e-18   (5.709e-19) &   6.465e-18   (1.202e-18) &   1.774e-17   (1.338e-18) \\  
 68B    &   4.644e-18   (2.089e-18)$^{\Box\dagger}$      &  3.880e-18   (6.767e-19)    &  1.206e-17   (6.631e-19)   &  5.383e-18   (7.310e-19)   &   $<$3.986e-18 &   2.921e-18   (8.830e-19)$^\Box$ &   8.915e-18   (1.155e-18)$^\Box$ \\  
 68C    &   9.516e-17   (7.711e-18)      &  1.207e-16   (2.188e-18)    &  2.078e-17   (1.134e-18)   &  3.974e-17   (1.464e-18)   &      8.149e-18   (9.425e-19) &   3.573e-17   (1.721e-18) &   7.328e-17   (1.969e-18) \\  
 75D$^\diamond$    &   --      &  --    &  $<$1.730e-18   &  1.666e-17   (2.527e-18)   &    1.763e-18   (7.331e-19) &   1.309e-17   (1.569e-18) &   2.460e-17   (1.939e-18) \\  
 79A    &   2.106e-17   (1.427e-18)      &  3.949e-17   (4.816e-19)    &  7.949e-18   (6.939e-19)   &  1.639e-17   (9.172e-19)   &   2.293e-18   (4.442e-19) &  3.040e-17   (1.113e-18) &   5.215e-17   (1.384e-18) \\  
 79B    &   9.151e-18   (2.436e-18)      &  2.363e-17   (5.177e-19)    &  4.156e-18   (3.445e-19)   &  9.698e-18   (6.162e-19)   &      3.719e-18   (3.252e-19) &   1.371e-17   (6.089e-19) &   2.767e-17   (8.313e-19) \\  
 82B    &   $<$4.450e-18$^\dagger$     &  4.770e-18   (6.114e-19)    &  3.503e-18   (6.453e-19)   &  5.095e-18   (8.831e-19)   &   $<$1.618e-18 &   2.829e-18   (1.058e-18)$^\Box$ &   4.949e-18   (1.271e-18)$^\Box$ \\  
 82C    &   5.974e-17   (2.780e-18)      &  1.179e-16   (9.169e-19)    &  1.921e-17   (1.077e-18)   &  3.471e-17   (2.088e-18)   &   8.369e-18   (2.263e-18)  &   6.899e-17   (1.761e-18) &   9.751e-17   (3.246e-18) \\  
 91A    &   3.796e-17   (5.058e-18)$^{\Box\dagger}$      &  5.676e-17   (1.286e-18)    &  2.693e-17   (1.504e-18)   &  4.185e-17  (1.468e-18)   &   2.474e-17   (3.857e-18) &   5.955e-17   (2.402e-18) &   1.261e-16   (2.511e-18) \\  
 91C    &   1.376e-17   (3.007e-18)      &  2.140e-17   (9.767e-19)    &  3.949e-18   (1.580e-18)   &  1.469e-17   (2.013e-18)   &  $<$7.169e-18 &   1.418e-17   (2.166e-18) &   4.348e-17   (3.185e-18) \\  
 95C    &   6.705e-18   (3.206e-18)      &  1.481e-17   (7.070e-19)    &  6.089e-18   (1.085e-18)   &  9.996e-18   (1.201e-18)   &   1.792e-18   (7.802e-19)  &   1.195e-17   (1.268e-18) &   1.664e-17   (1.810e-18) \\  
 96A    &   7.912e-17   (4.590e-18)$^{\Box\dagger}$      &  2.506e-16   (1.260e-18)    &  2.448e-16   (2.681e-18)   &  1.544e-16   (2.346e-18)   &   3.031e-16   (1.923e-18)  &   2.506e-16   (3.744e-18) &   1.661e-16   (4.866e-18) \\  
 96C    &   2.650e-17   (3.245e-18)      &  5.489e-17   (1.007e-18)    &  1.274e-17   (1.225e-18)   &  1.528e-17   (1.322e-18)   &   4.755e-18   (1.134e-18)  &   2.638e-17   (2.117e-18) &   4.719e-17   (2.942e-18) \\  
 100A    &   6.115e-17   (2.560e-18)      &  1.080e-16   (7.013e-19)    &  2.548e-17   (1.359e-18)   &  9.609e-17   (1.274e-18)   &   1.003e-17   (1.614e-18)  &   1.164e-16   (2.293e-18) &   1.359e-16   (2.269e-18) \\

\\
\tableline
\\

\multicolumn{7}{l}{Uncertainties are listed in parentheses. Marginal detections are low signal to noise lines as determined by visual inspection.}\\
\multicolumn{6}{l}{$^\dagger$ Blended with \Ht\ S(5)\,6.91\micron}\\
\multicolumn{6}{l}{$^\Box$ Marginal Detection}\\
\multicolumn{6}{l}{$^\star$ SL coverage}\\
\multicolumn{6}{l}{$^\diamond$ LL coverage}\\
\multicolumn{6}{l}{$^\bullet$ Off-nuclear extraction}
\end{tabular}
}

\end{sidewaystable*}

\begin{table}[!htp]
\begin{minipage}[c]{17cm}
{\scriptsize
\caption{\Ht-detected Extraction Areas \label{tab:extract}}
\begin{center}
\begin{tabular}{l r r r }
\hline
\hline
\\[0.2pt]
Galaxy    &  SL Area &  LL Area & SF$^a$   \\
     &  (arcsec$^2$)  & (arcsec$^2$)    &    \\
\hline
\\[0.2pt]
6B          &    27.3       &      154.8      &      1 \\
 15A        &     75.3      &		 309.6	&  1 \\                                    
 15D        &     47.9      &	309.8	&  2.50\\                                      
 25B           &    82.3        &	--	&  \\                                  
25B$^{\bullet}$   --         &   516.5    	& \\                             
 31A+C          &   --        &	722.7	&   \\                                 
 31A+C$^{\bullet}$ & 82.3       &    --       &   \\                           
 40B           &   47.8        &	258.5	&   3.37\\                                   
 40C           &   --          &	 516.5	&   \\                              
40C$^{\bullet}$ & 77.8          &   --    	&   \\                       
 40D           &   95.6        &	464.6	&  1.96\\                                   
 44A           &    --       &	619.4	&   \\                                 
44A$^{\bullet}$  &  116.0       &   --      	&   \\                       
 44D           &    93.5       &	878.1	&  5.90\\                                   
 47A           &    75.0       &	361.3	&  3.79\\                                   
 55C           &    68.4       &	258.2	&   1\\                                   
 56B           &    61.9       &	412.5	&  1.59\\                                   
 56C           &    54.7       &	361.2	&  4.09\\                                   
 56D		  &    54.8        &	--	&  \\                                         
 56E           &     41.0      &	206.2	&  1\\                                    
 57A            &    --       &	 413.1	&  \\                                
 57A$^{\bullet}$  &  67.9        &   103.3    	&  1\\                        
 67B           &    170.4       &	361.2	&  1\\                                    
 68A           &    61.7       &	412.4	& 5.33\\                                    
 68B           &   75.4        &	 361.2	& 1\\                                    
 68C           &    308.6       &	775.6	& 2.24\\                                    
75D              &    --       &	 257.8	&  \\                               
 79A           &    61.6       &	361.3	&  1.58\\                                   
 79B           &     68.6      &	155.1	&  1.24\\                                   
 82B           &    34.3       &	206.4	&  1\\                                    
 82C           &   54.7        &	309.6	& 1.60\\                                    
 91A           &   225.1        &	516.0	&  1.53\\                                   
 91C           &    47.8       &	361.3	&  2.36\\                                   
 95C           &  82.7         &	259.3	&  1\\                                    
 96A           &   130.9        &	619.9	&  1.64\\                                   
 96C           &   41.4        &	258.0	&  1.61\\                                   
 100A          &   68.4        &	516.2	&  1\\                                    
\hline
\\[0.5pt]
\multicolumn{4}{l}{$^a$ Scale Factor}\\
\multicolumn{4}{l}{$^\bullet$ Off-nuclear extraction}\\

\\[0.5pt]
\end{tabular}
\end{center}
}
\end{minipage}
\end{table}

\begin{figure*}[!thp]
\begin{center}

\subfigure[Comparison of 7.7\micron\ PAH and 24\micron\ Luminosity]{\includegraphics[width=8cm]{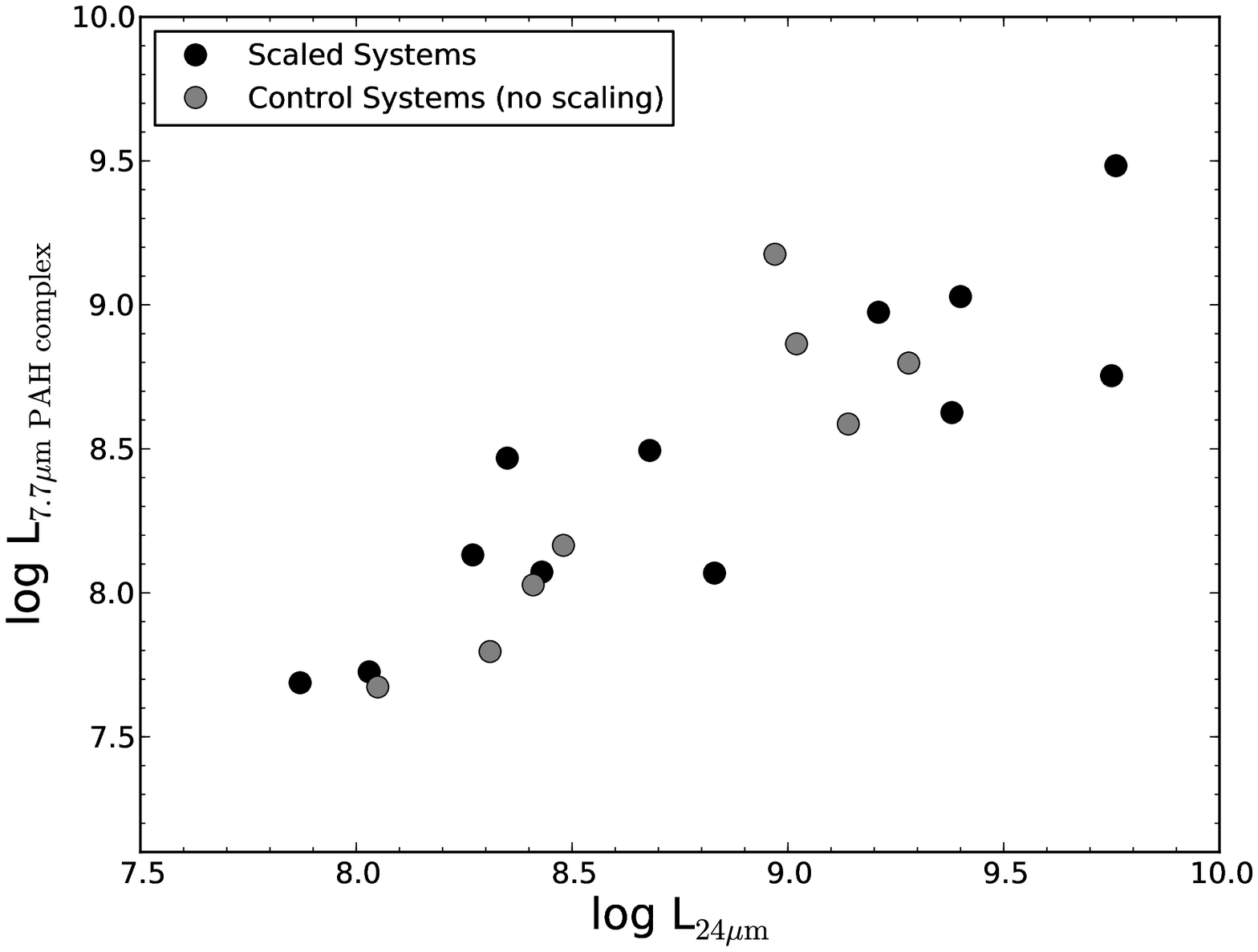}}
\hfill
\subfigure[Comparison of 11.3\micron\ PAH and 24\micron\ Luminosity]{\includegraphics[width=8cm]{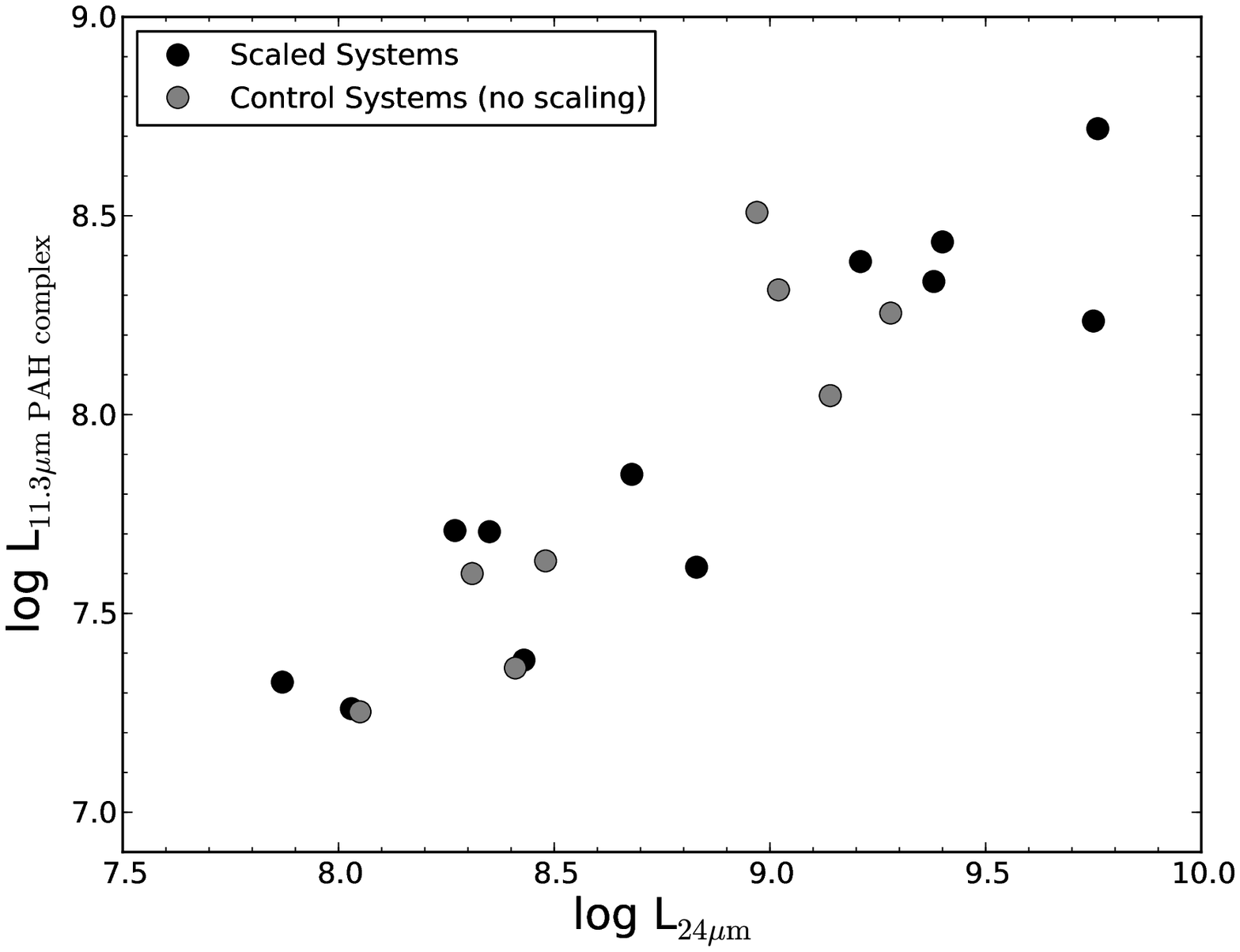}}
\caption[]
{Comparison of scaled and unscaled continuum and emission features}
\label{fig:scale}
\end{center}
\end{figure*}

The extracted areas of the \Ht-detected galaxies are listed in Table \ref{tab:extract}, as well as any scaling factors employed to match the LL and SL spectral orders. The scaling between modules makes the assumption that the emission lines scale with the continuum i.e. the SL spectrum is scaled to bring it in line with the LL spectrum continuum. As an independent check we use the 7.7\micron\ PAH and 11.3\micron\ PAH emission features compared to the matched 24\micron\ photometry. For systems not dominated by AGN heating, these measures should scale similarly with star formation. We plot these comparisons in Figure \ref{fig:scale} for systems where the respective PAH emission lines are well-determined and the spectra (and global mid-infrared colors) are not AGN-dominated. Since systems that have been scaled show the same behaviour as those that have not, the scaling does not appear to have produced inconsistent fluxes.
Uncertainties in the SL scaling factor (due to noise of the continuum) are estimated to be $<$0.02, corresponding to a flux uncertainty of $<$5\%.

\section{Spectra of Non-MOHEG \Ht\ Detections}\label{non_spec}

\begin{figure*}[!thp]
\begin{center}

\subfigure[HCG 31A+C -- SL (Off-Nuclear)]{\includegraphics[width=5.3cm]{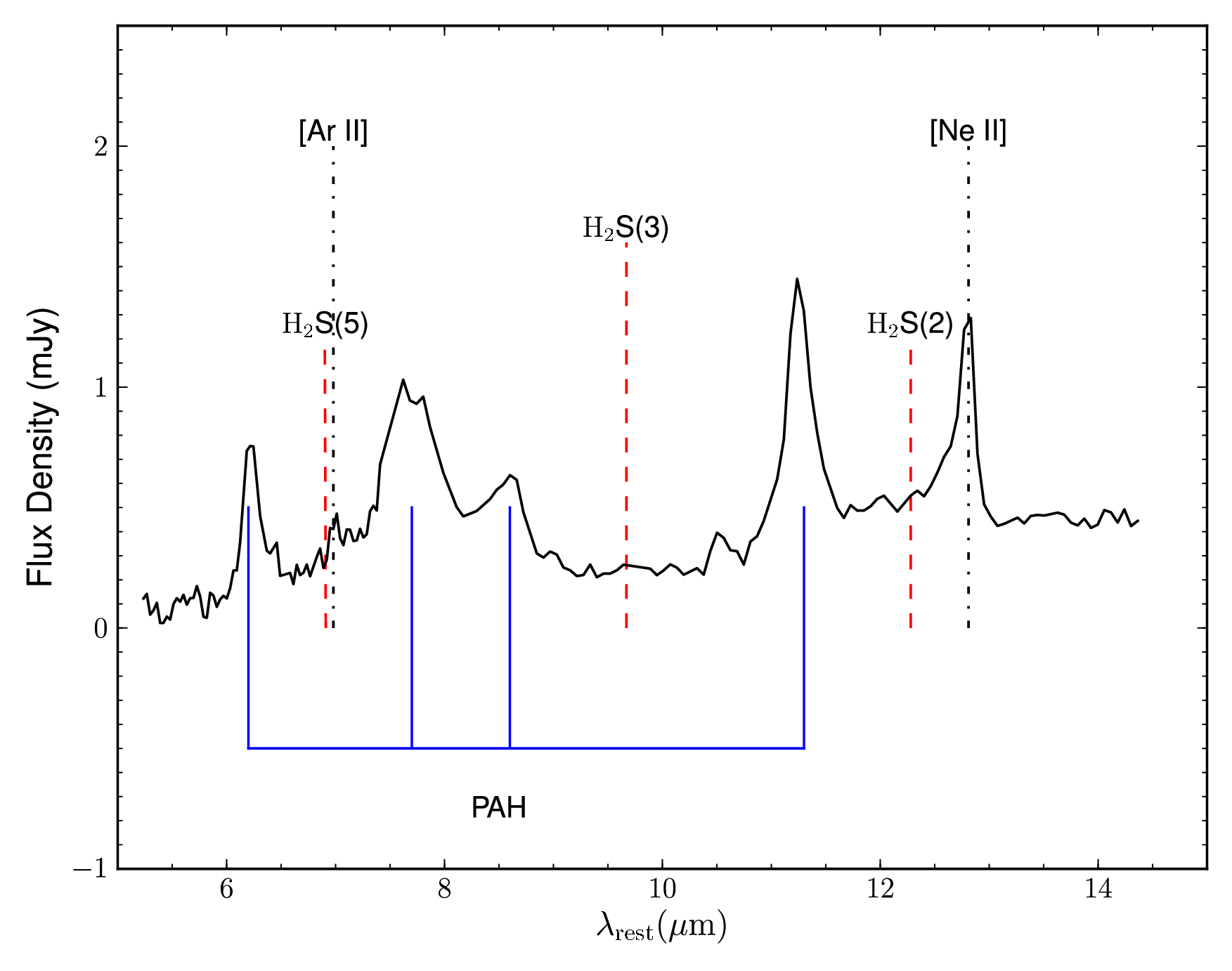}}
\hfill
\subfigure[HCG 31A+C -- LL (Nuclear)]{\includegraphics[width=5.3cm]{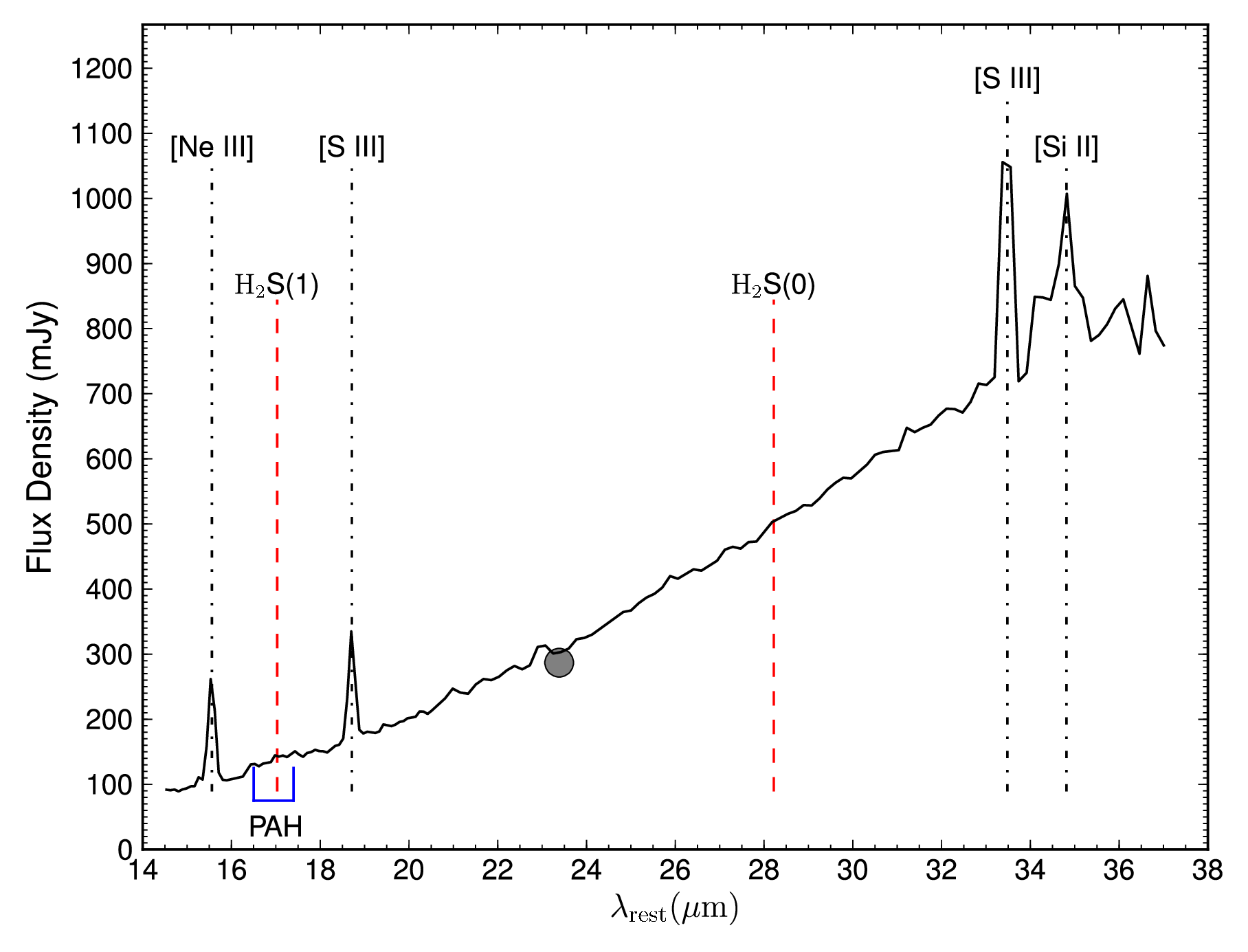}}
\hfill
\subfigure[HCG 40D (Nuclear)]{\includegraphics[width=5.3cm]{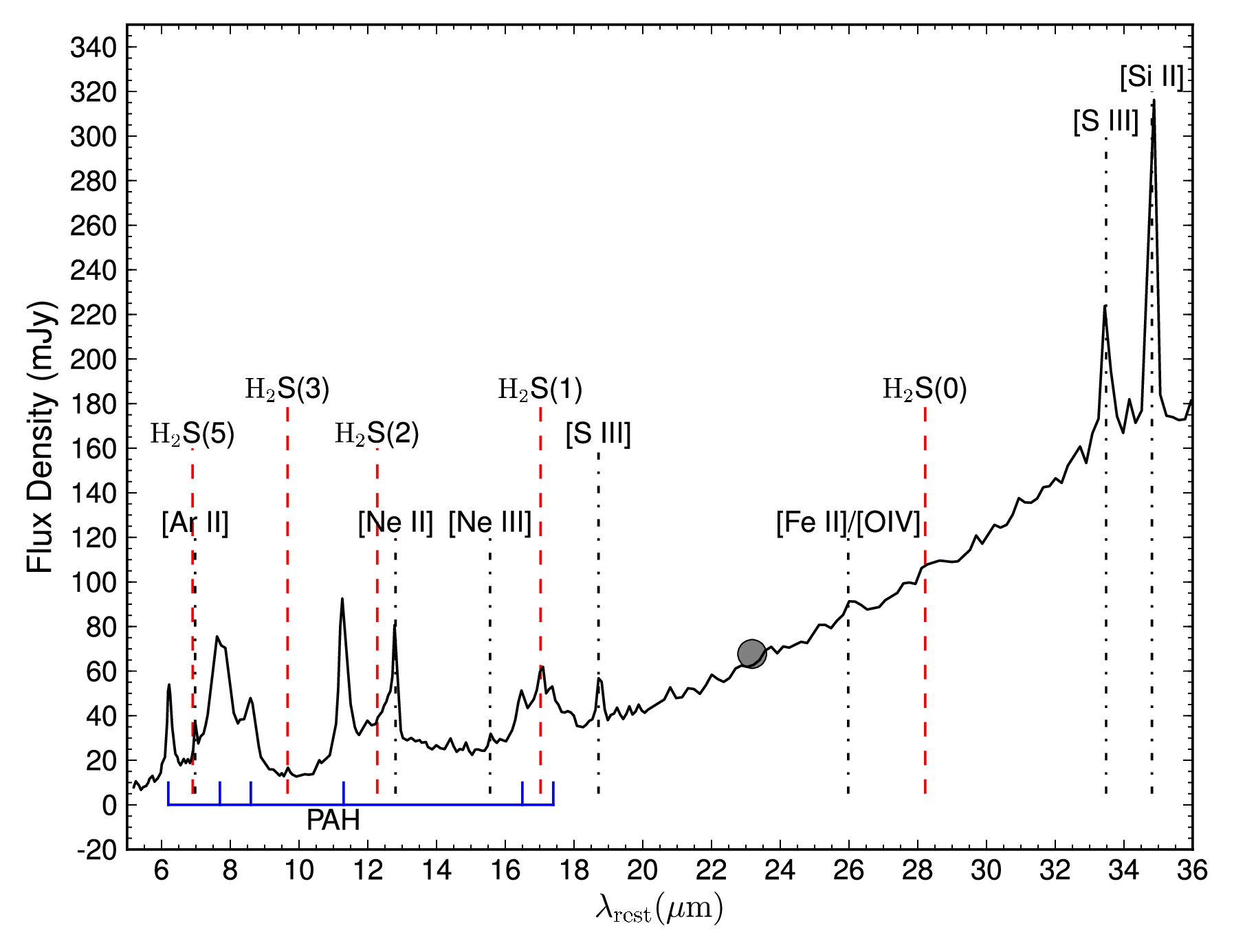}}
\vfill
\subfigure[HCG 40C -- SL (Off-Nuclear)]{\includegraphics[width=5.3cm]{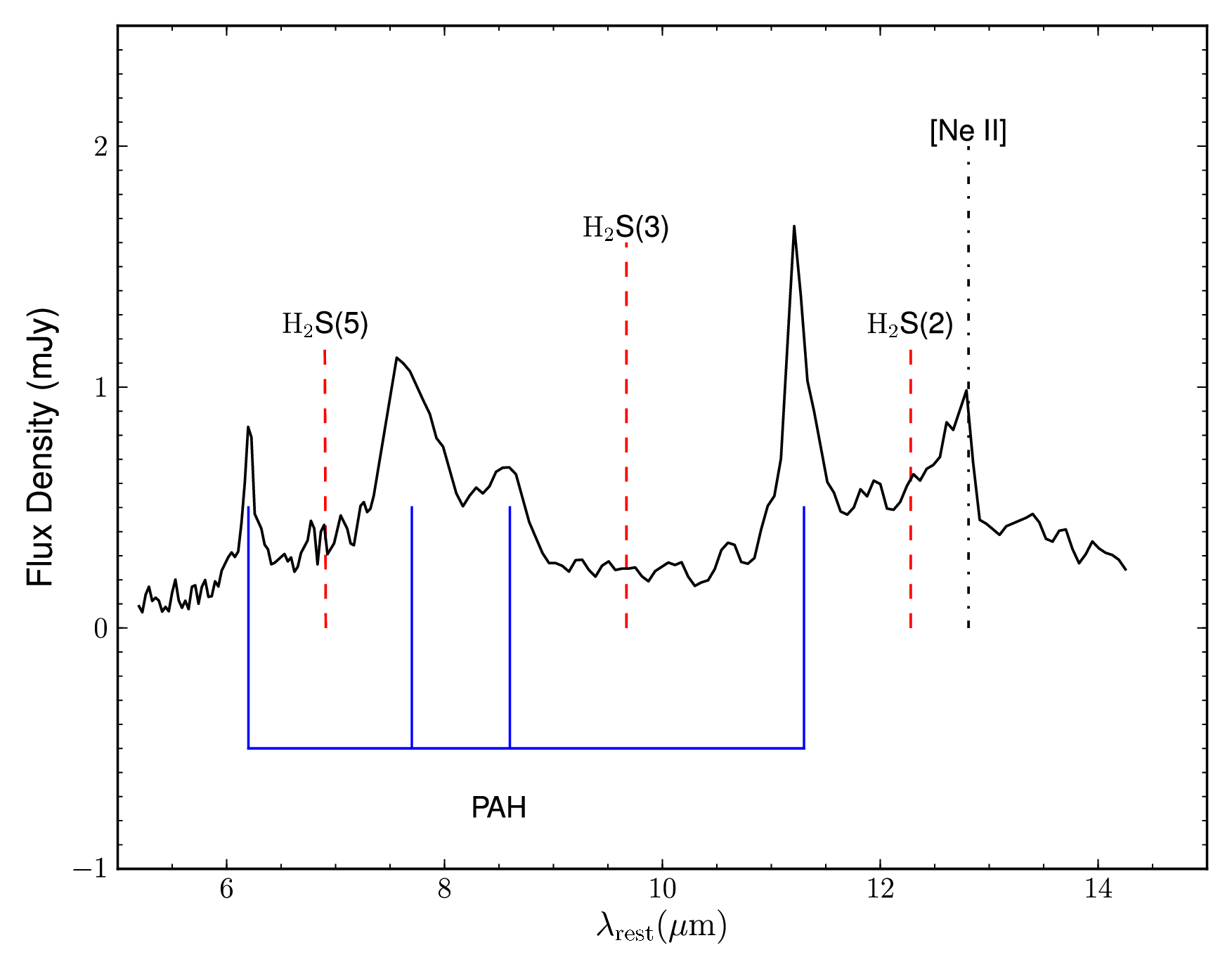}}
\hfill
\subfigure[HCG 40C -- LL (Nuclear)]{\includegraphics[width=5.3cm]{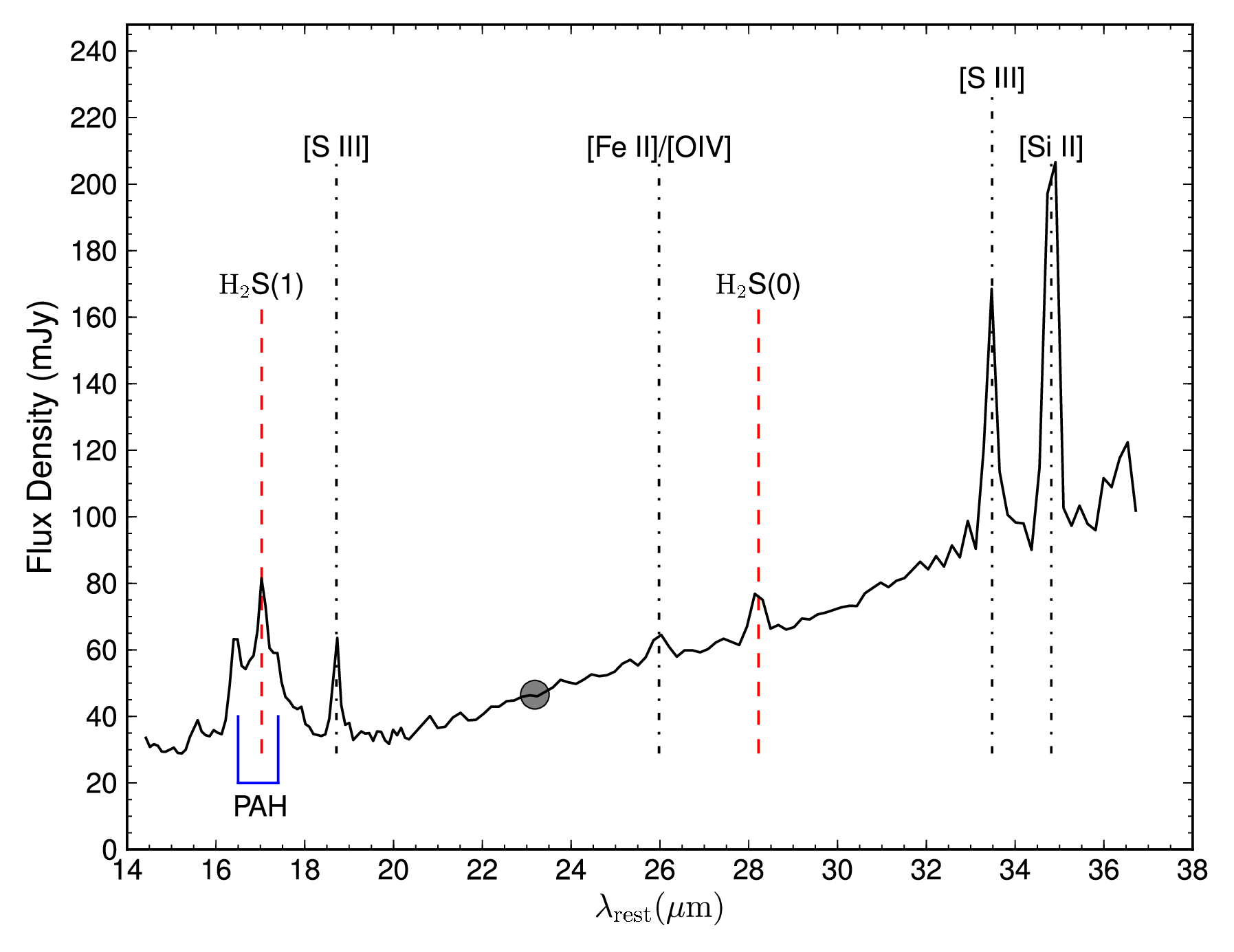}}
\hfill
\subfigure[HCG 44D (Nuclear)]{\includegraphics[width=5.3cm]{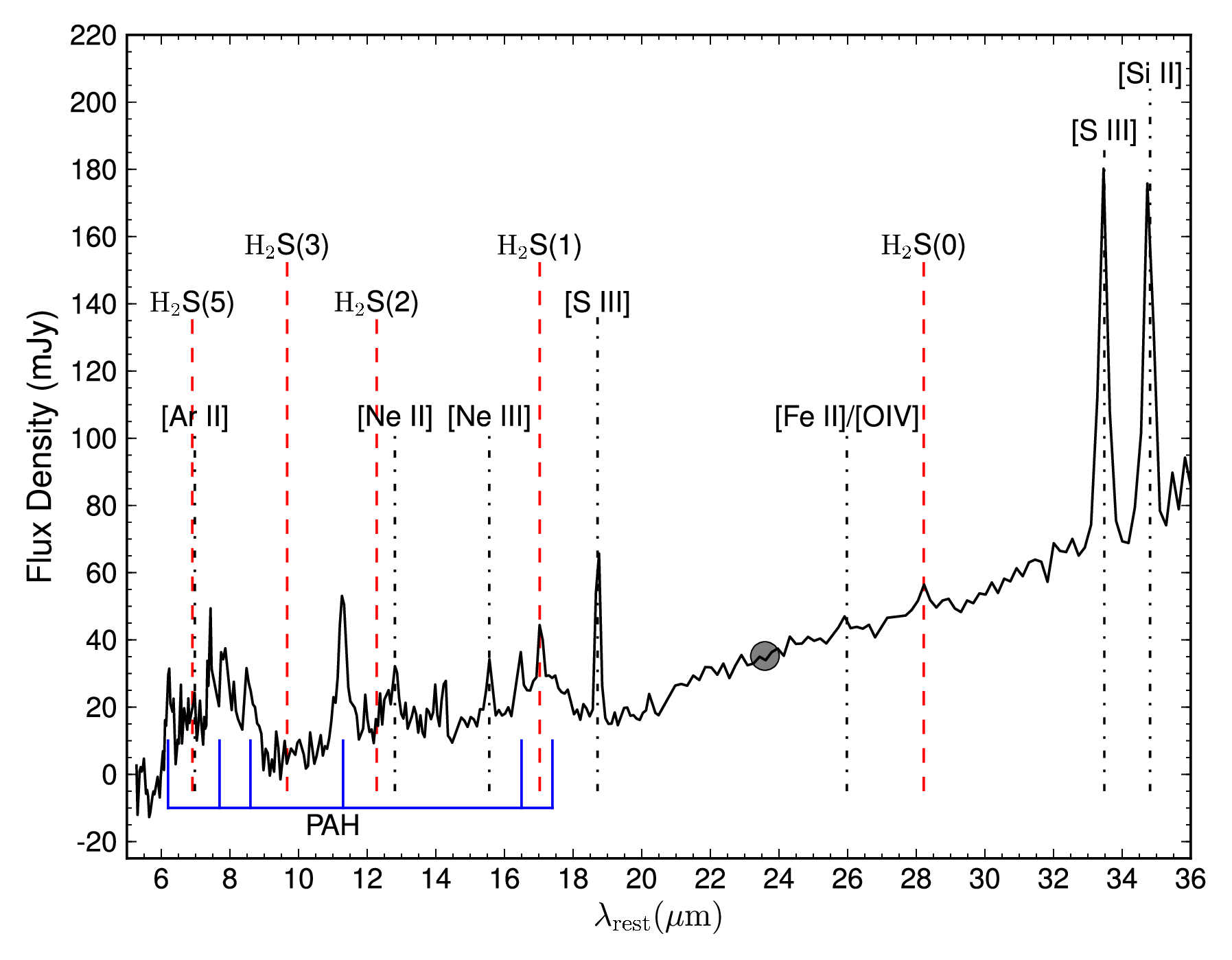}}
\vfill
\subfigure[HCG 47A (Nuclear)]{\includegraphics[width=5.3cm]{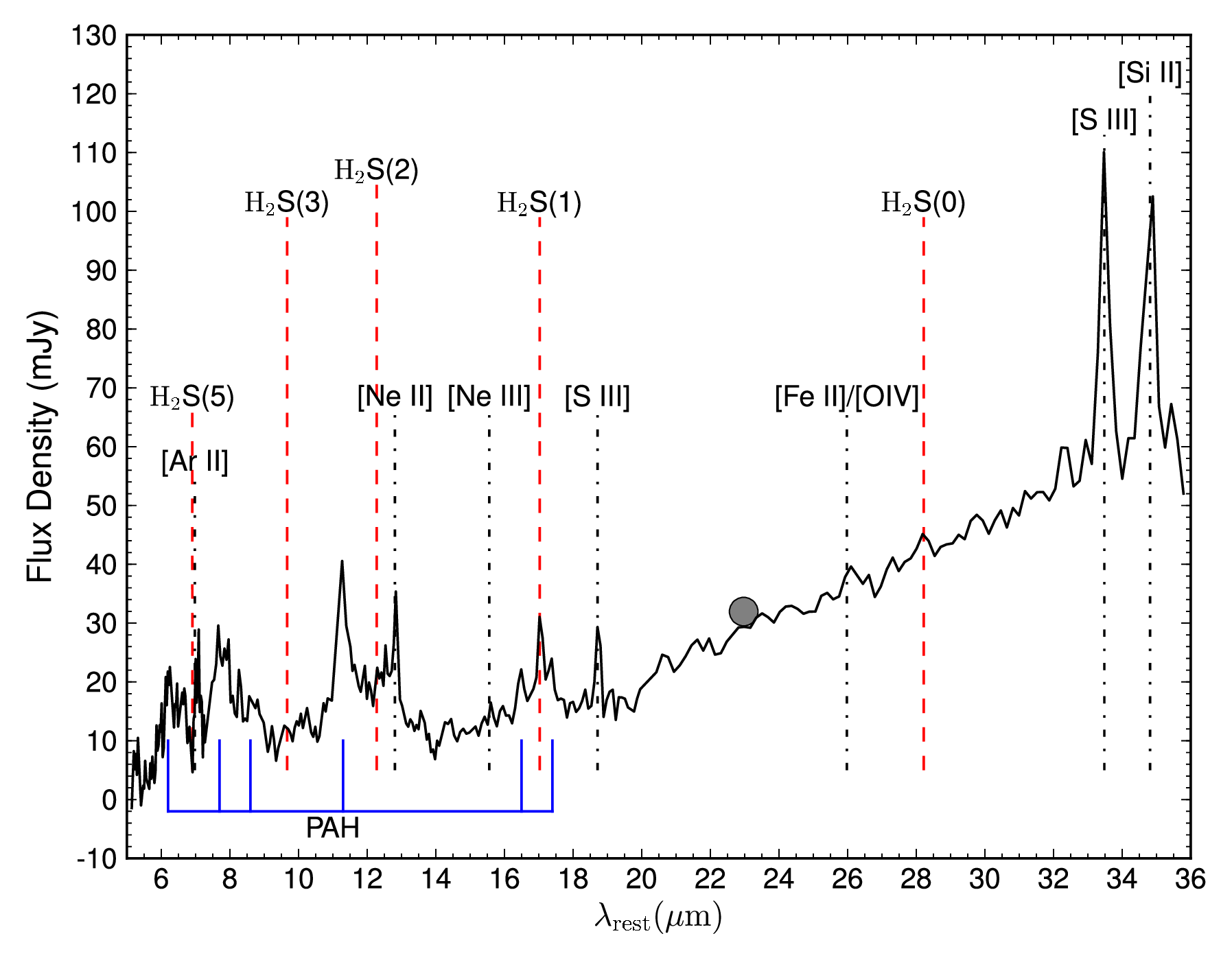}}
\hfill
\subfigure[HCG 55C (Nuclear)]{\includegraphics[width=5.3cm]{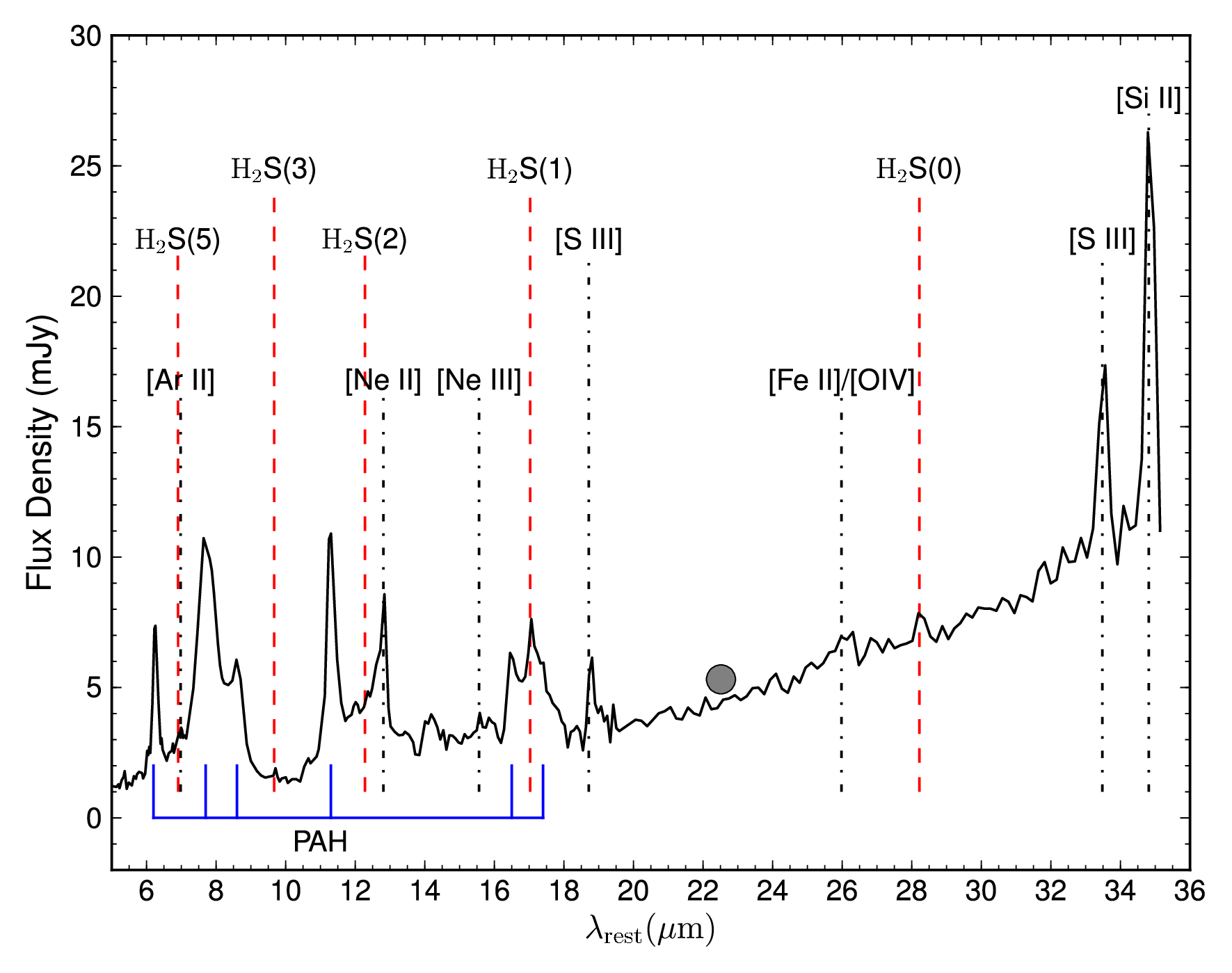}}
\hfill
\subfigure[HCG 56D -- SL (Nuclear)]{\includegraphics[width=5.3cm]{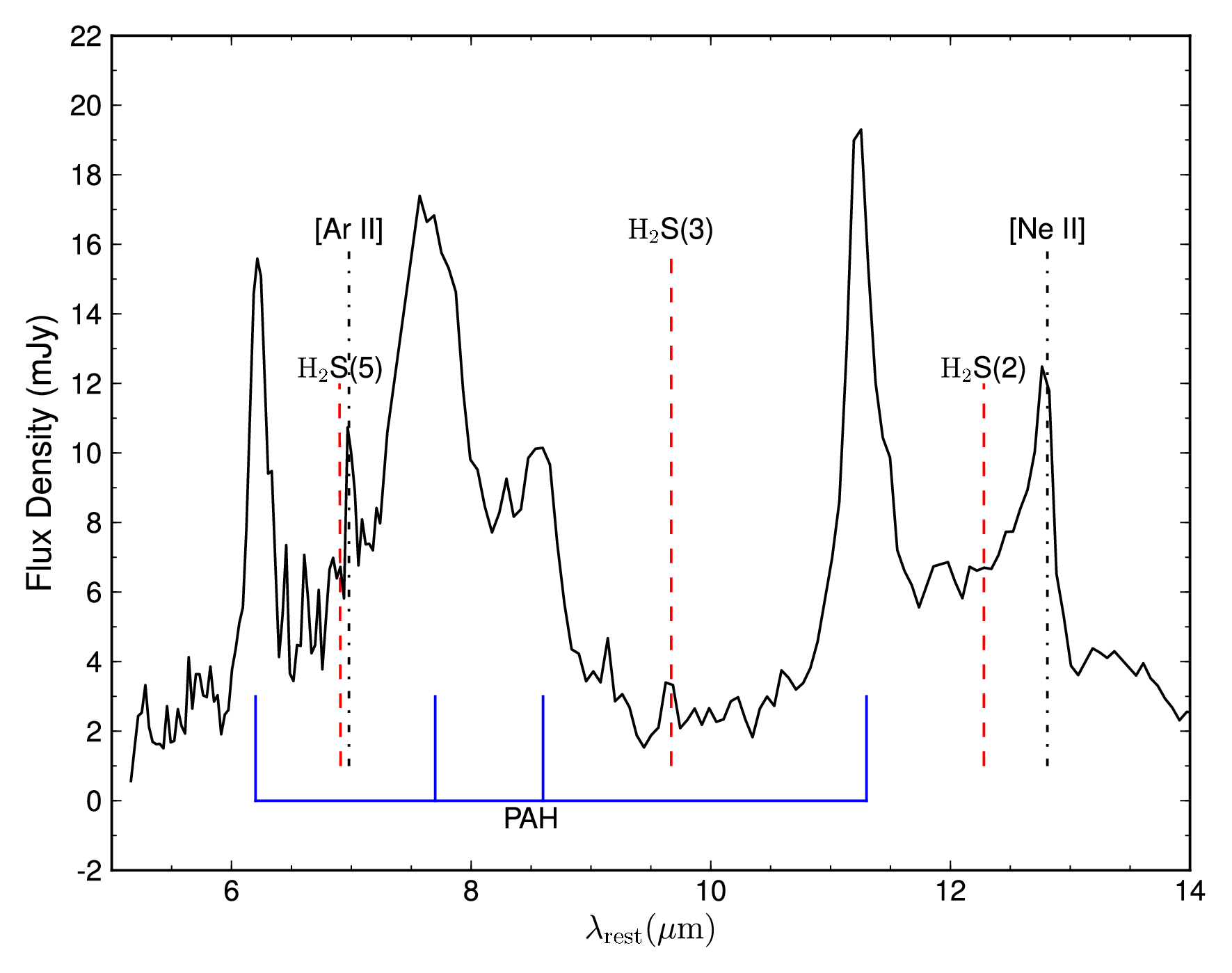}}
\vfill
\subfigure[HCG 56E (Nuclear)]{\includegraphics[width=5.3cm]{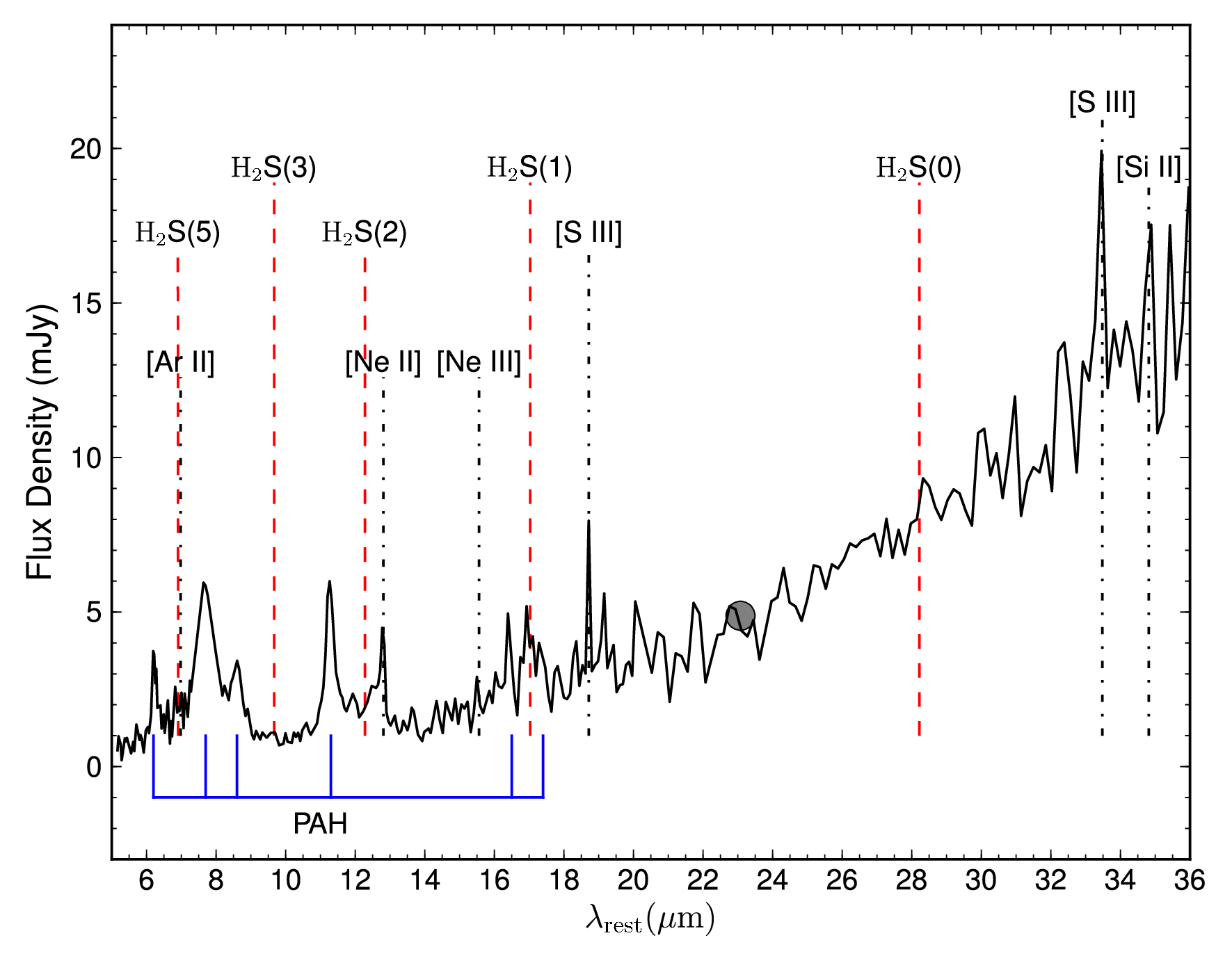}}
\hfill
\subfigure[HCG 67B (Nuclear)]{\includegraphics[width=5.3cm]{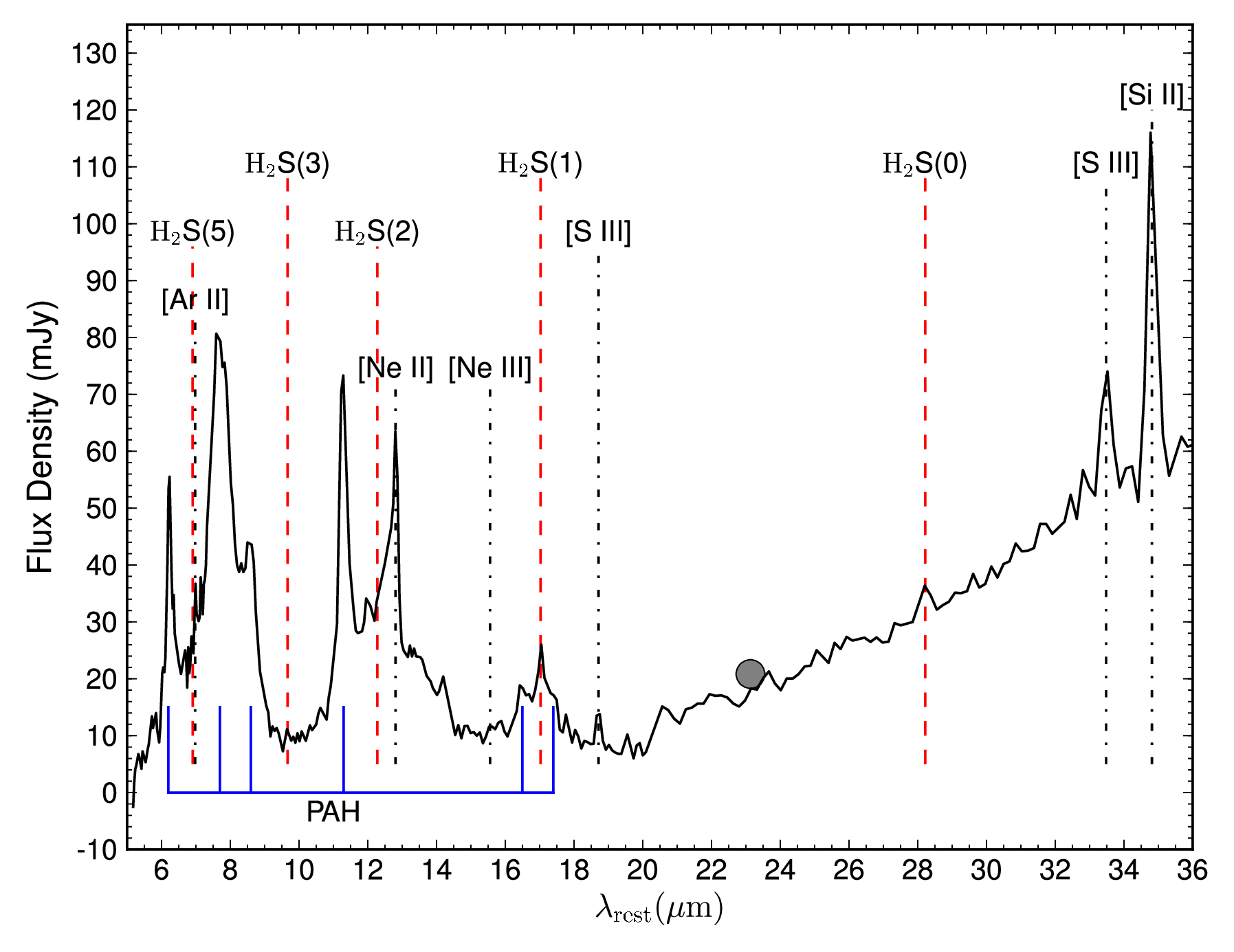}}
\hfill
\subfigure[HCG 75D -- LL (Nuclear)]{\includegraphics[width=5.3cm]{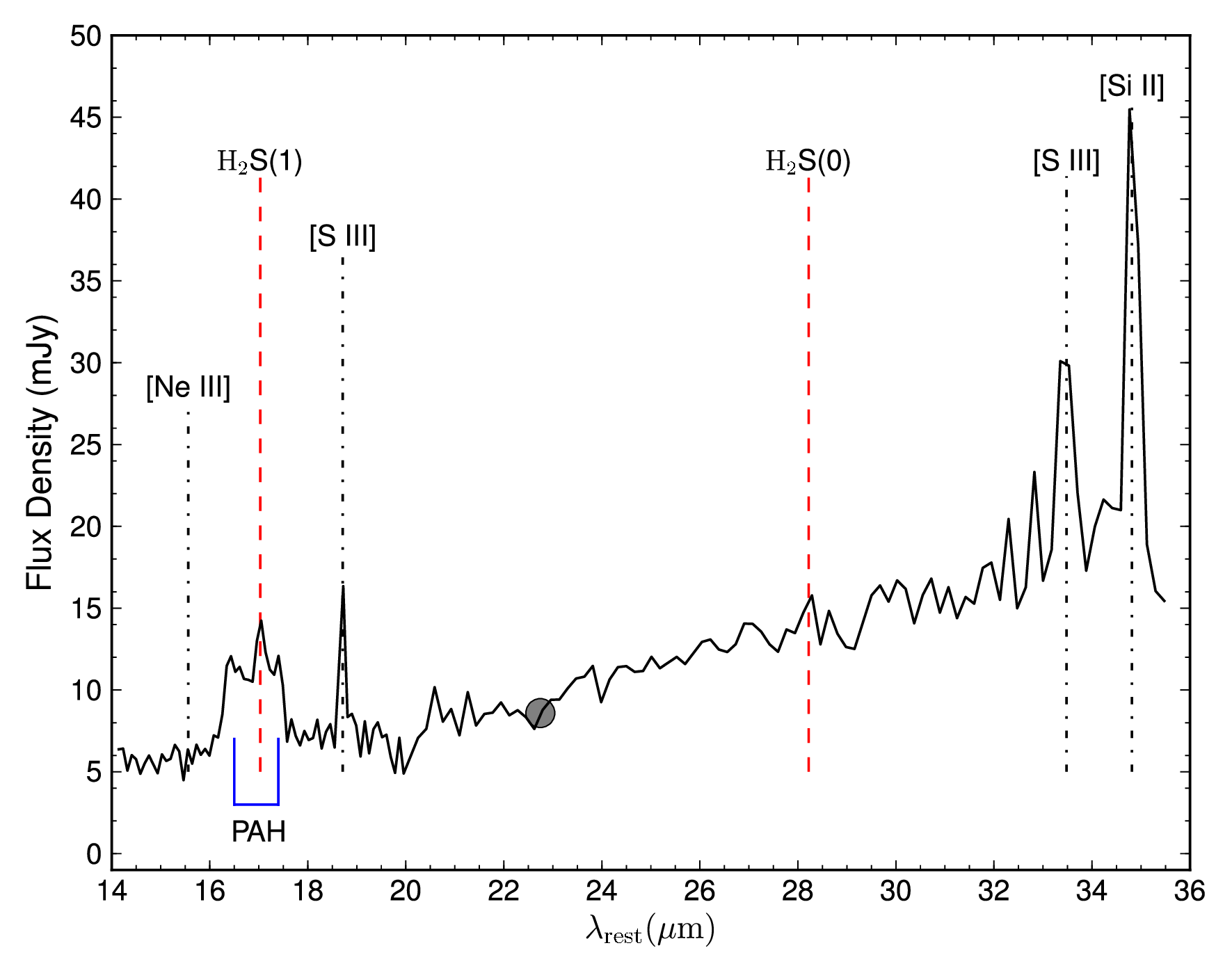}}
\hfill

\caption[]
{Non-MOHEG, \Ht-detected HCG Galaxies. The matched MIPS 24\micron\ photometry is shown as a grey point.}
\label{fig:spectra_gen}
\end{center}
\end{figure*}

\begin{figure*}[!thp]
\begin{center}
\subfigure[HCG 79A (Nuclear)]{\includegraphics[width=5.3cm]{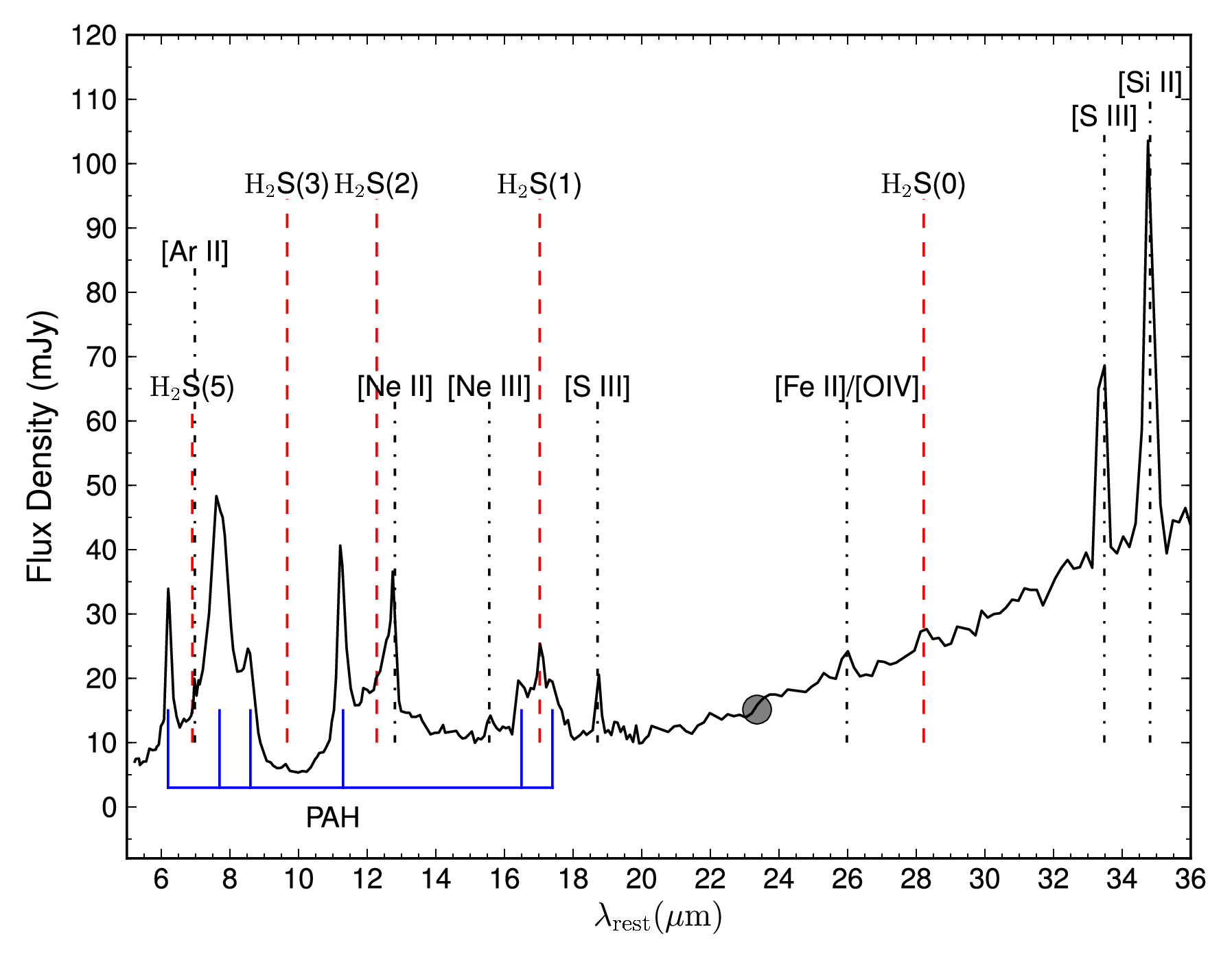}}
\hfill
\subfigure[HCG 79B (Nuclear)]{\includegraphics[width=5.3cm]{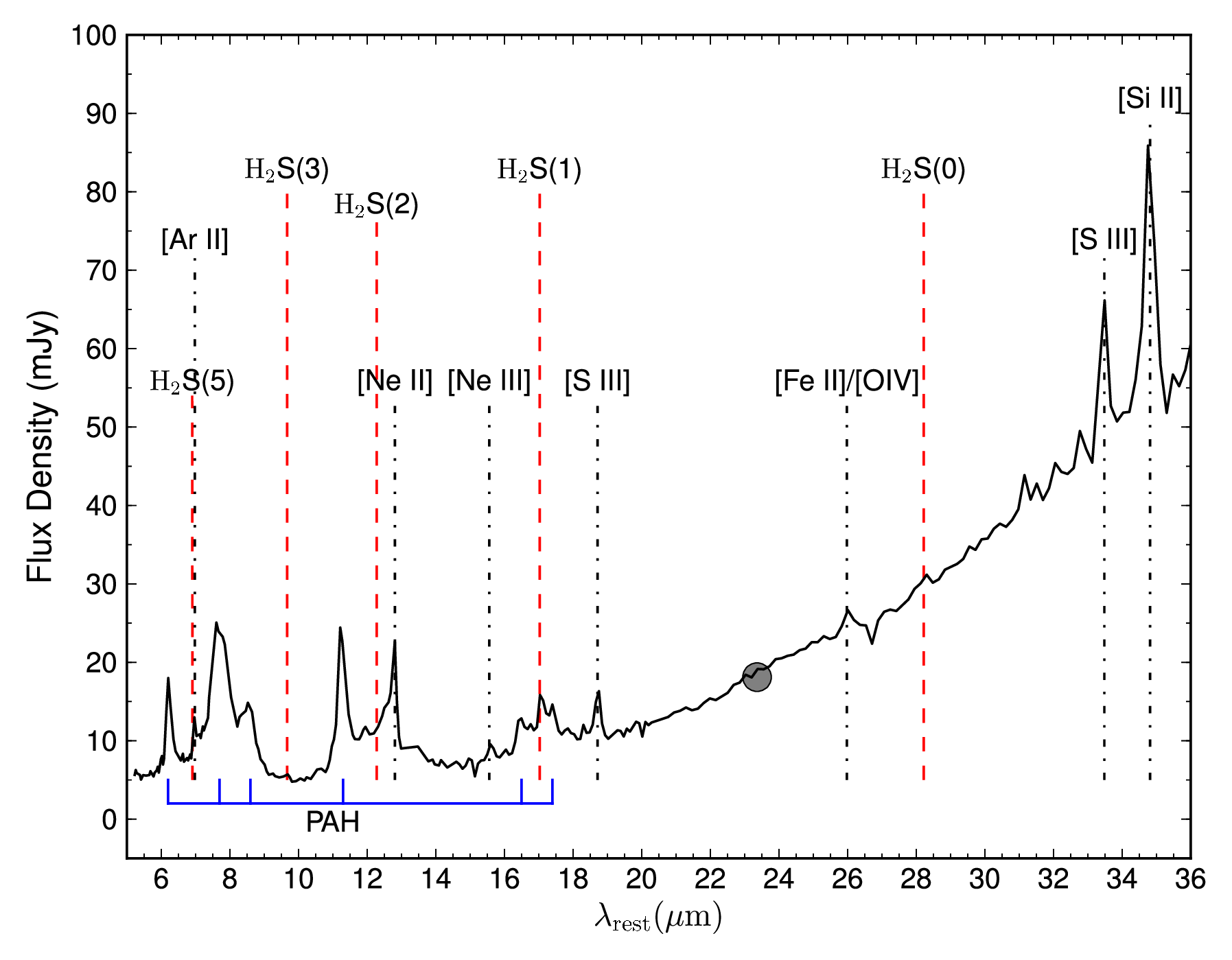}}
\hfill
\subfigure[HCG 82C (Nuclear)]{\includegraphics[width=5.3cm]{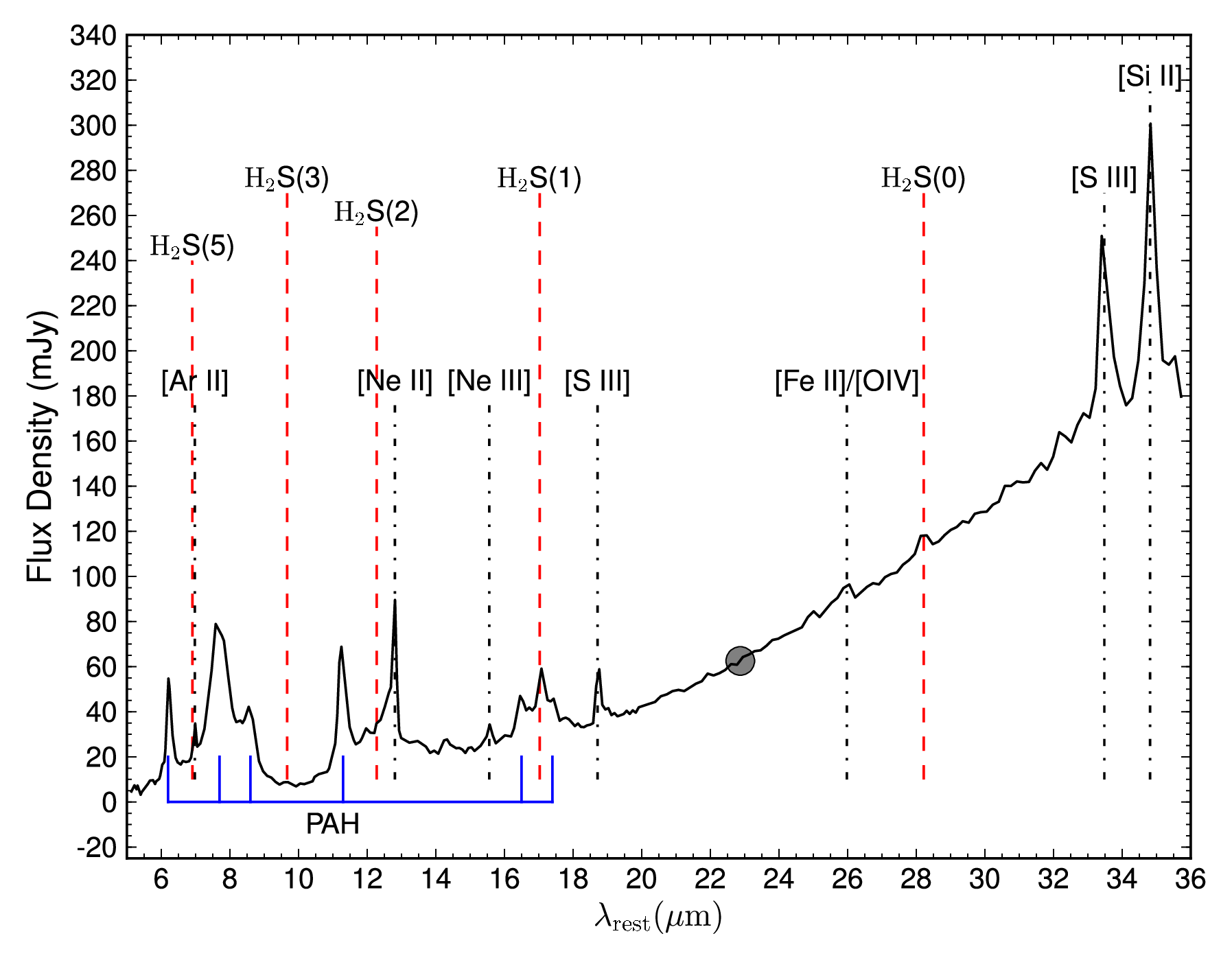}}
\vfill
\subfigure[HCG 91A (Nuclear)]{\includegraphics[width=5.3cm]{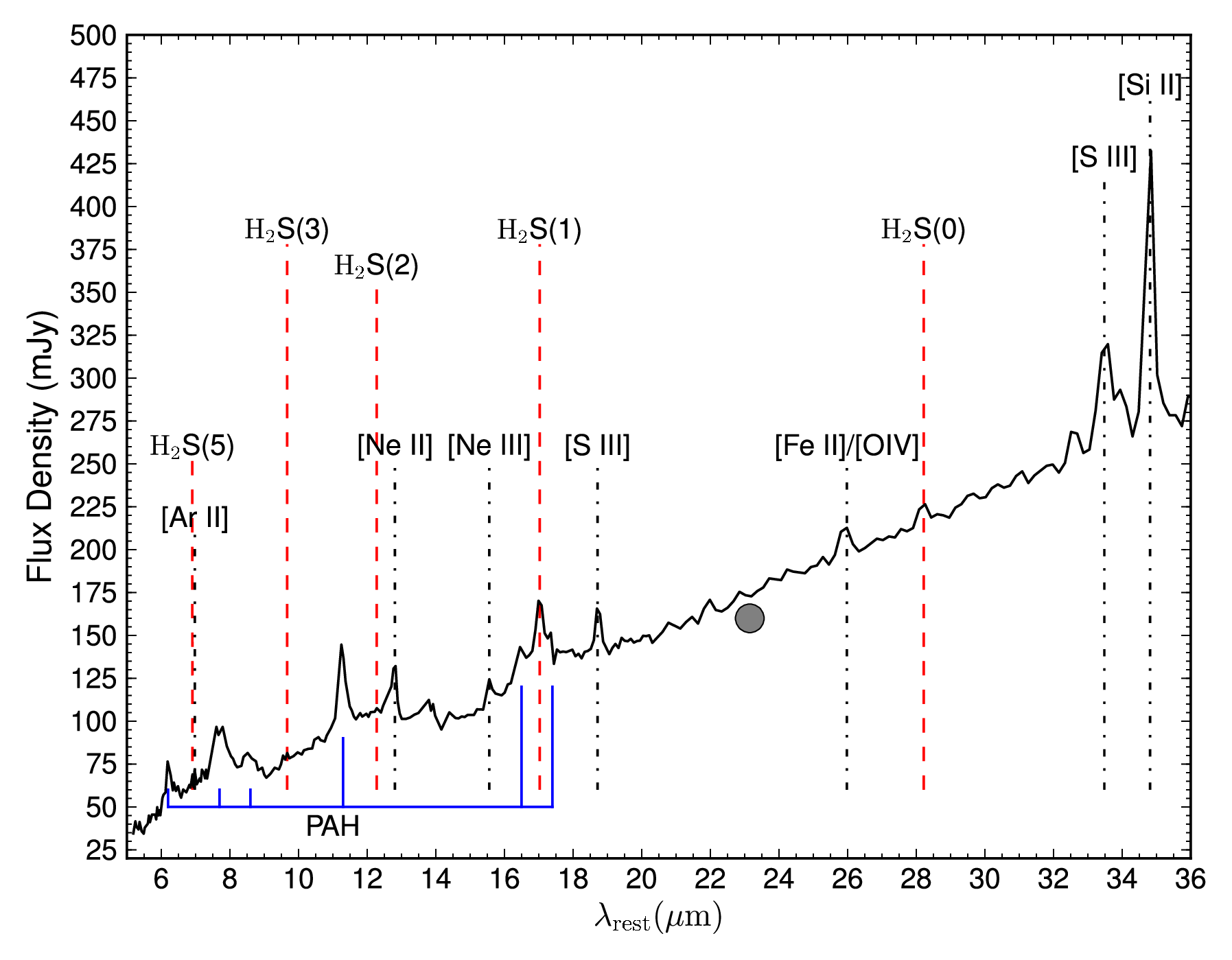}}
\hfill
\subfigure[HCG 91C (Nuclear)]{\includegraphics[width=5.3cm]{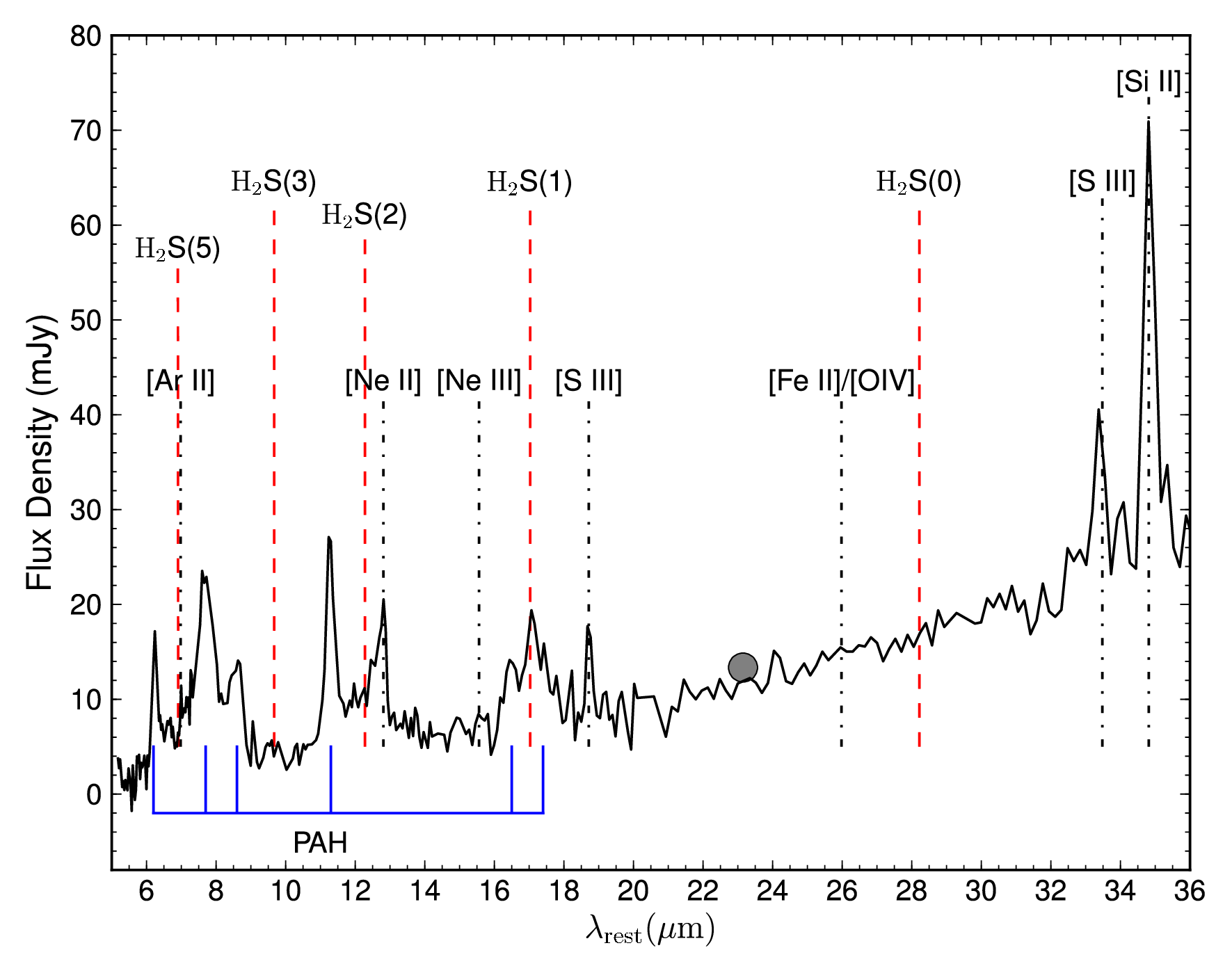}}
\hfill
\subfigure[HCG 96A (Nuclear)]{\includegraphics[width=5.3cm]{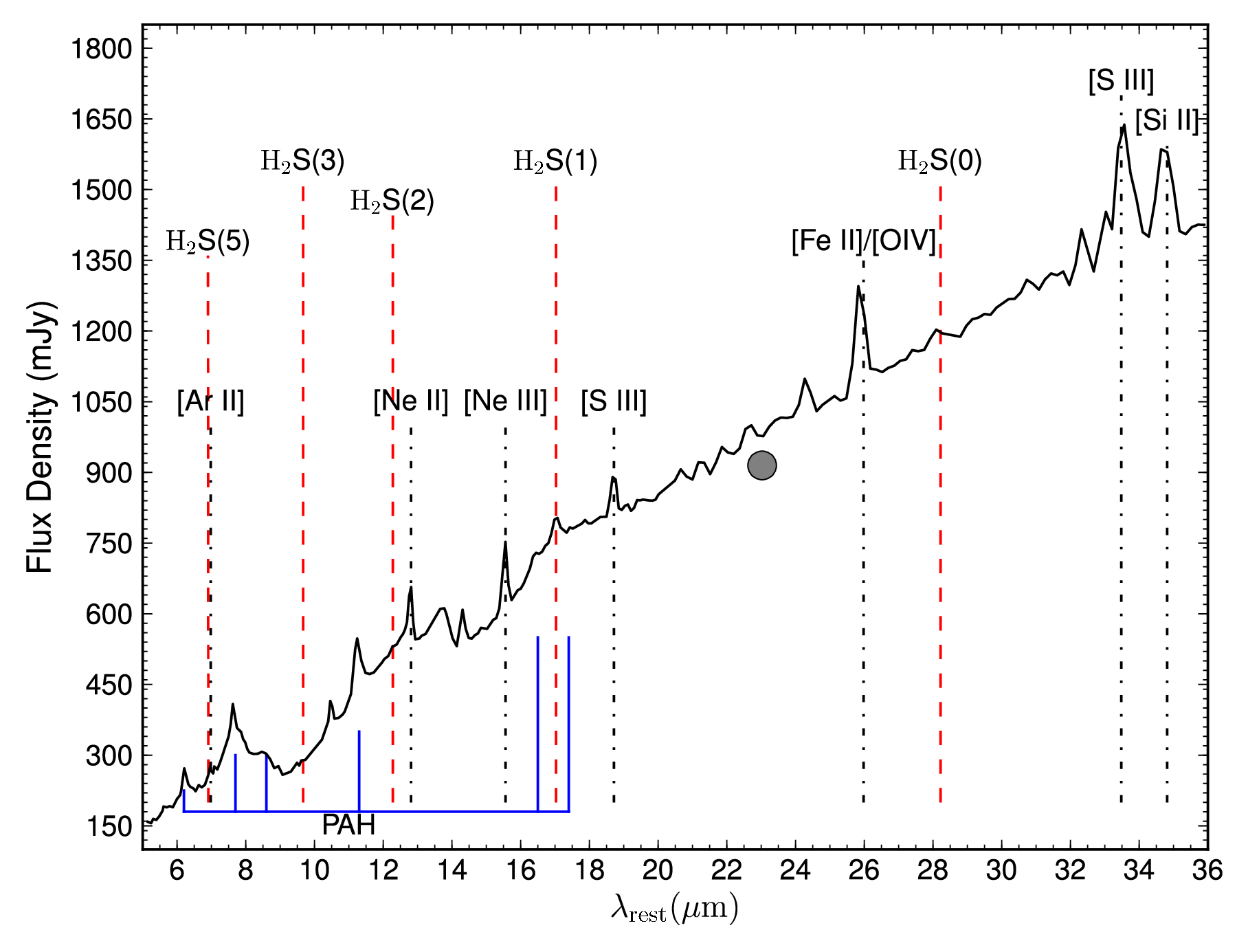}}
\vfill
\subfigure[HCG 96C (Nuclear)]{\includegraphics[width=5.3cm]{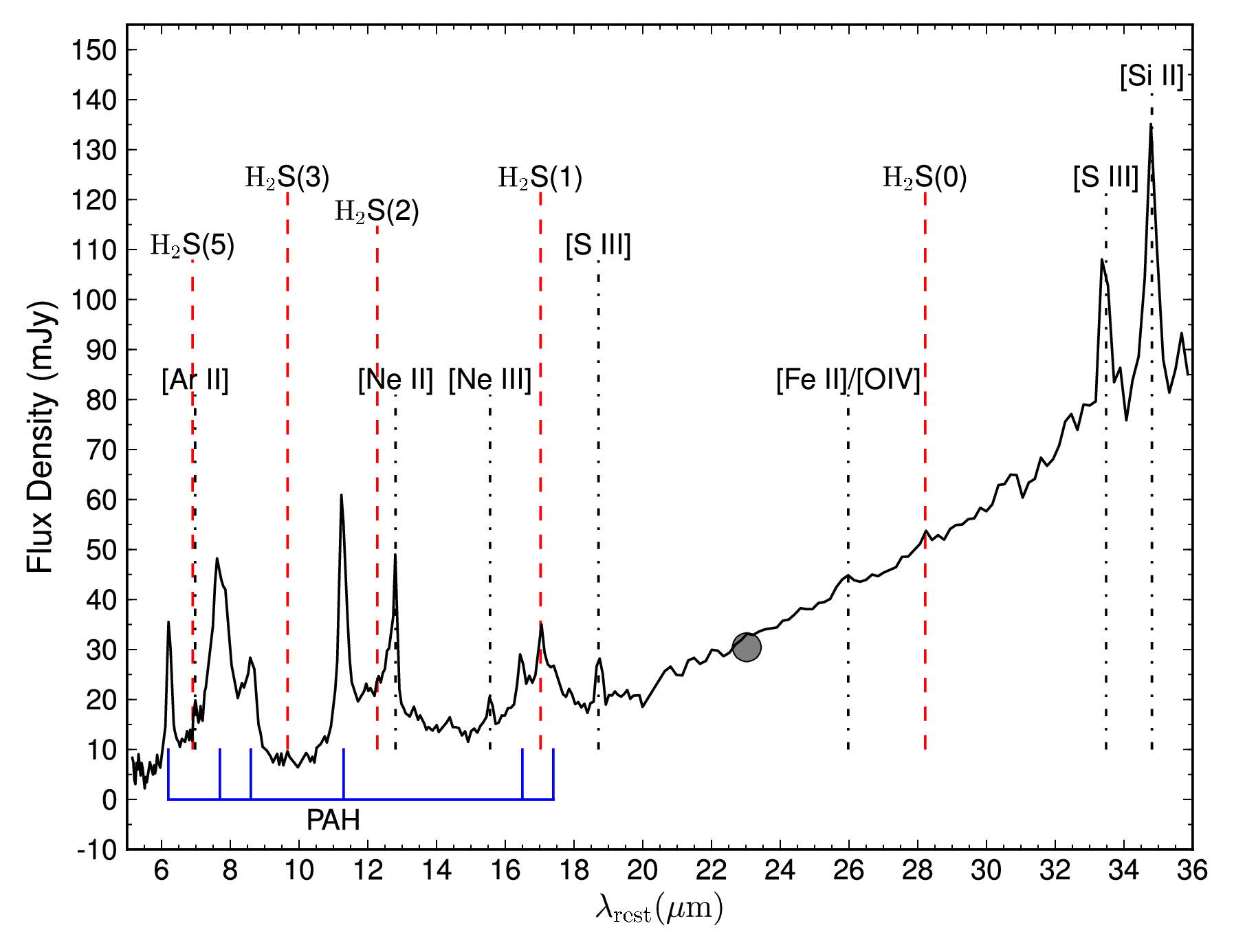}}
\hfill
\subfigure[HCG 100A (Nuclear)]{\includegraphics[width=5.3cm]{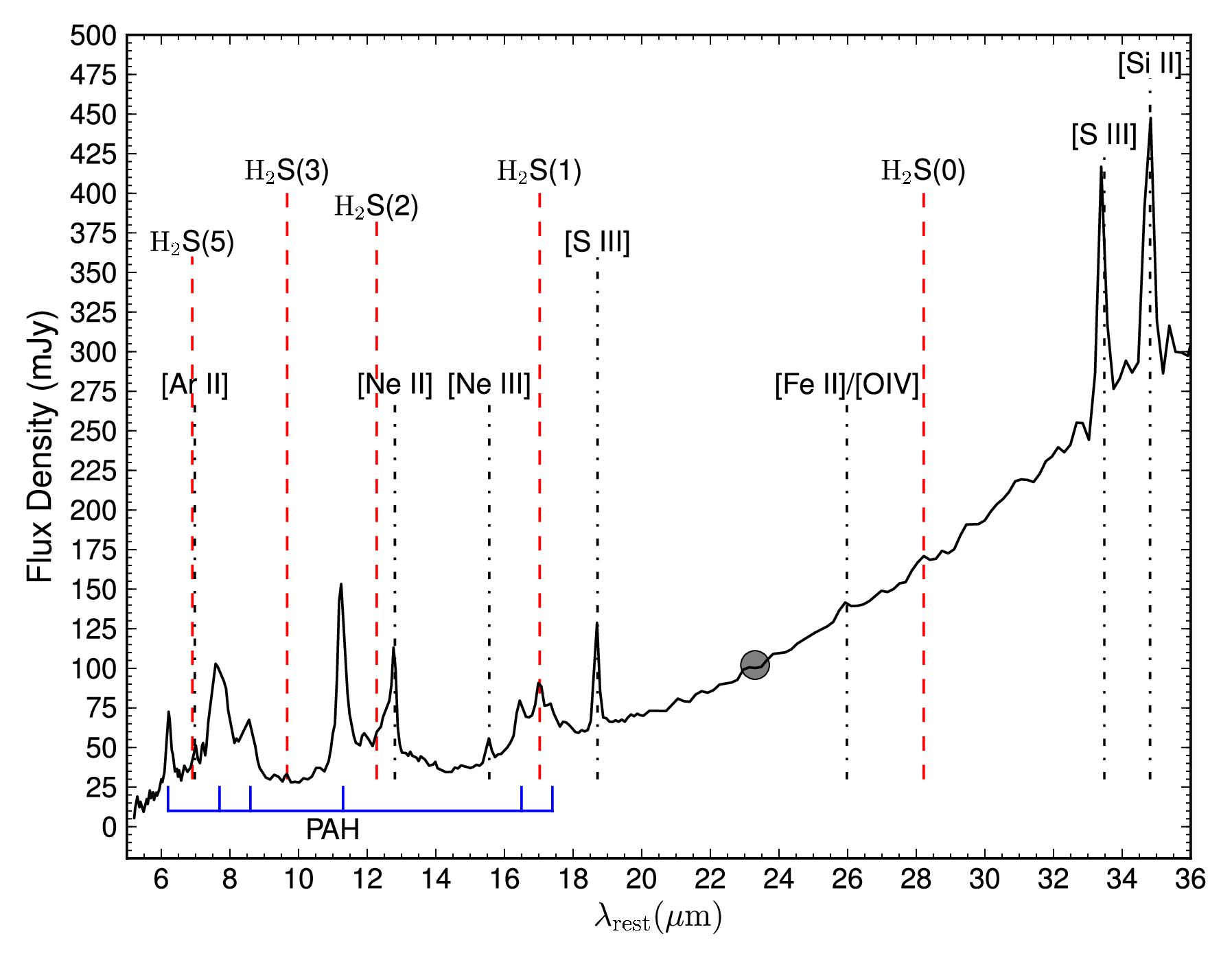}}

\caption[]
{Non-MOHEG, \Ht-detected HCG Galaxies. The matched MIPS 24\micron\ photometry is shown as a grey point.}
\label{fig:spectra_gen_more}
\end{center}
\end{figure*}

The spectra of the non-MOHEG \Ht-detected HCG galaxies are shown in Figure  \ref{fig:spectra_gen} and \ref{fig:spectra_gen_more}. As expected, the majority of spectra appear consistent with star-forming systems, having excited \Ht\ associated with PDRs and UV excitation. The spectrum of HCG 96A indicates the presence of a powerful AGN given the prominent hot and warm dust and strong high excitation lines (Fig. \ref{fig:spectra_gen_more}f), also reflected in its mid-infrared colors. This galaxy is a known LIRG \citep{San03} and Seyfert 2 galaxy \citep{Ost93}. 

The HCG 31A+C complex (Fig. \ref{fig:spectra_gen}a and b) has an irregular morphology and high levels of star formation; it is likely experiencing a triggered starburst due to the mutual interaction between the two parent systems \citep{Gal10}. HCG 91A appears strongly star forming (Fig. \ref{fig:spectra_gen_more}d), but has an optically-identified active nucleus \citep[Sy1.2, ][]{Rad98}; we note our  extraction is optimised to include the disk.

If we consider the morphological types of the non-MOHEG \Ht\ galaxies listed in Table \ref{tab:morph}, many are classified as peculiar early type galaxies (Sa or S0). However, these galaxies have spectra consistent with current star formation (strong PAH emission and a sharply rising mid-infrared continuum). \citet{Bit11} have suggested that several systems are misclassified (due to orientation and dust obscuration). For the galaxies 55C, 56D and 56E (see Fig. \ref{fig:spectra_gen}) our spectra don't rule out this possibility, as proposed by \citet{Bit11}, based on their SEDs and IRAC mid-infrared morphology. In addition, their global mid-infrared colors are consistent with late type systems (log[f$_{5.8\mu \rm m}$/f$_{3.6\mu \rm m}$]$>0$) and their nuclear classifications suggest they are star-forming systems \citep{Coz04}.

\begin{figure*}[!thp]
\begin{center}


\subfigure[79A and 79B]{\includegraphics[height=7.5cm]{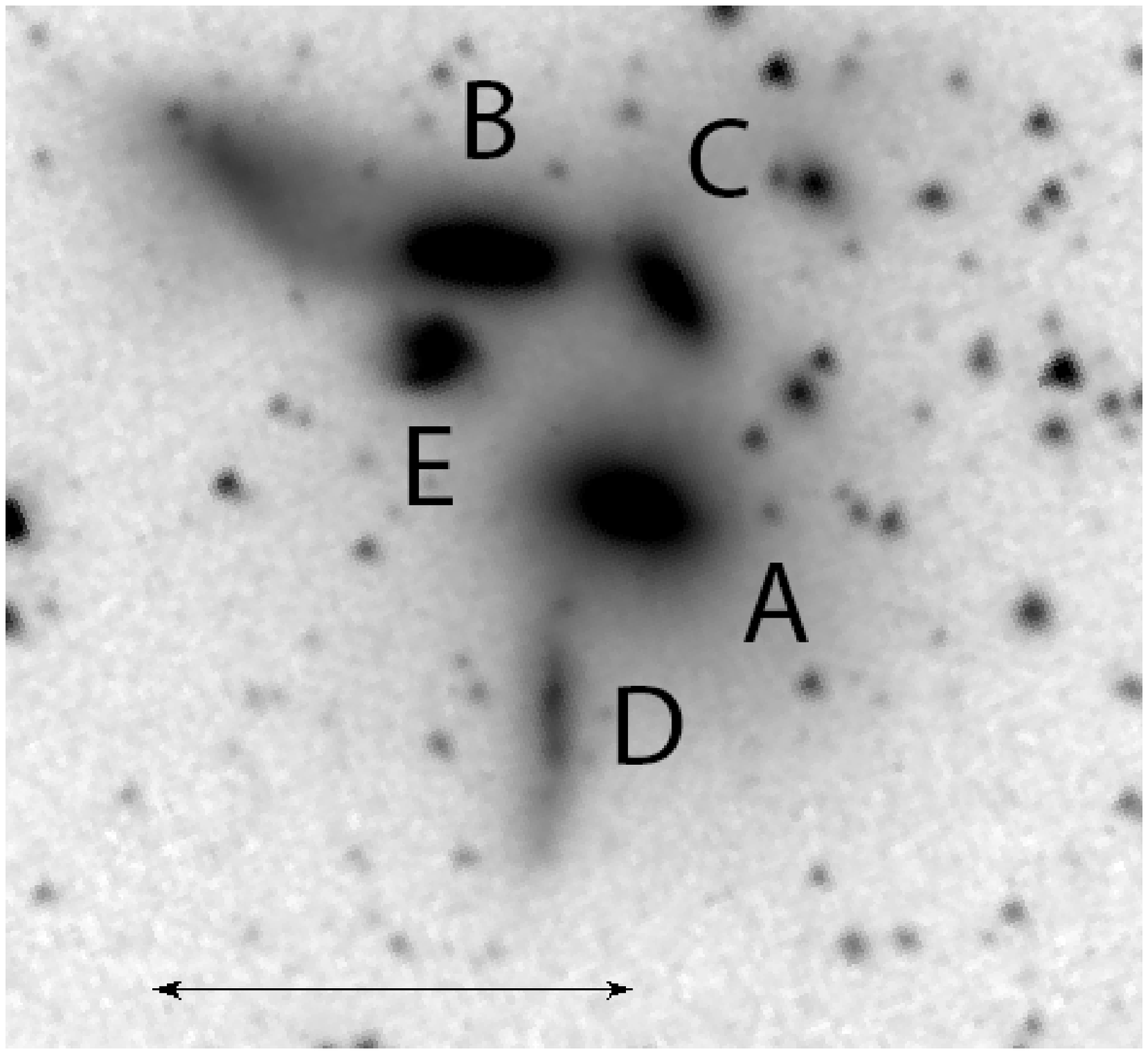}}
\hfill
\subfigure[100A]{\includegraphics[height=7.5cm]{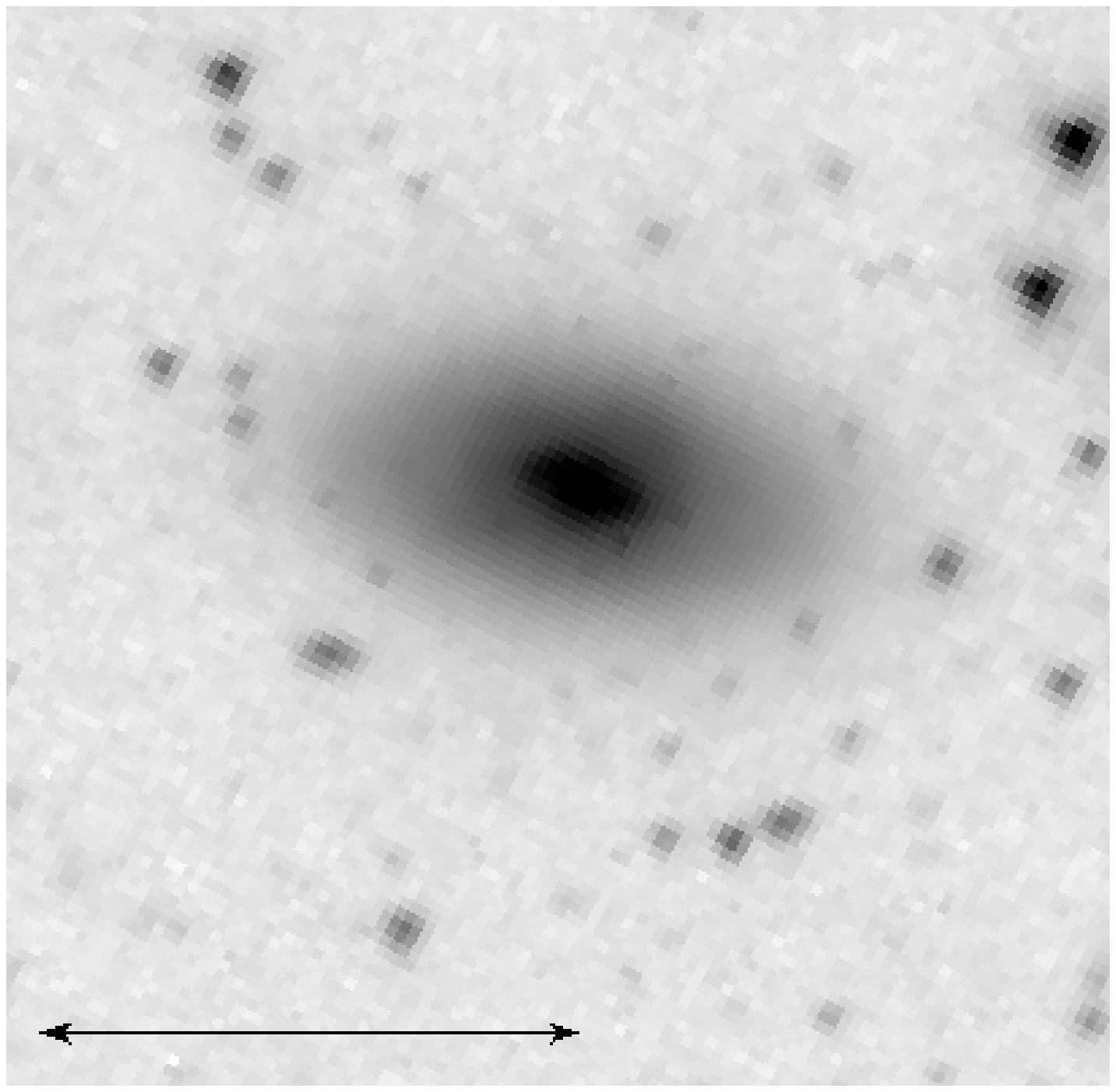}}

\caption{IRAC 3.6\micron\ images of the non-MOHEG \Ht-detected galaxies which lie within the green valley; the arrow is 1\arcmin\ in length. These systems all have spectra that indicate active star formation. }
\label{hybrids}
\end{center}
\end{figure*}

HCG 79A has a star-forming spectrum (Fig. \ref{fig:spectra_gen_more}a), but a mid-infrared color of log[f$_{5.8\mu \rm m}$/f$_{3.6\mu \rm m}$] $=$ -0.06, i.e. at the blue end of the mid-infrared green valley. Inspection of its IRAC image (included as Figure \ref{hybrids}a) suggests that 79A is an S0/a type. The star formation seen in its spectrum could be the result of a minor merger or gas accretion that has introduced fuel for star formation and is expected to further build the lenticular bulge \citep{Shap10}. The H$\alpha$ velocity field determined by \citet{Dur08} suggests a cross-fueling (discrete infall or fueling event rather than a continuous flow) from HCG 79D. Evidence for a minor merger in HCG 79A was also seen by \citet{Coz07} based on the asymmetry observed in their near-infrared images.

Within our sample, \citet{Bit11} have also suggested 79B and 100A are late type galaxies misclassified as early types. Inspection of their spectra indicates star formation is occurring (Figures \ref{fig:spectra_gen_more}b and \ref{fig:spectra_gen_more}h, respectively). We note that their SEDs \citep{Bit11} indicate the presence of a warm dust continuum superimposed on a substantial stellar component.  However, we draw attention to these galaxies since they have IRAC colors (log[f$_{5.8\mu \rm m}$/f$_{3.6\mu \rm m}$]) of  -0.18 and -0.05, for HCG 79B and 100A, respectively. This places them at intermediate mid-infrared color. We show these galaxies in Figure \ref{hybrids}a and \ref{hybrids}b, respectively, and draw attention to their disturbed morphologies. HCG 79B displays a magnificent tidal feature and there is evidence that it has accreted a dwarf \citep{Dur08}. These galaxies are all classified as early type disk systems (S0/a), but we may be seeing the effect of minor mergers or gas accretion producing star formation and giving rise to the combination of relatively dusty mid-infrared colours with mid-infrared spectra indicative of star formation. Alternatively star formation is being actively shut off or ``starved" by some mechanism. It is therefore not clear that these galaxies are misclassifications.

\section{\Ht\ Upper Limits for non-detections in the rest of the sample}\label{non-detect}

In this section we give the \Ht\ upper limits for the S(0), S(1), S(2) and S(3) lines (Table \ref{tab:h2_upp}) and measured PAH values and upper limits for galaxies in the sample with an extracted spectrum (Table \ref{tab:pah_upp}.

\begin{table}[!htp]
{\scriptsize
\caption{\Ht\ Upper Limits for the Remainder of Sample in units of W\,m$^{-2}$ \label{tab:h2_upp}}
\begin{center}
\begin{tabular}{l l l l l}
\hline

\\[0.5pt]
Galaxy & \multicolumn{1}{c}{H$_{2}$ 0-0 S(0)} &  \multicolumn{1}{c}{H$_{2}$ 0-0 S(1)} &  \multicolumn{1}{c}{H$_{2}$ 0-0 S(2)} &  \multicolumn{1}{c}{H$_{2}$ 0-0 S(3)}  \\  
 & \multicolumn{1}{c}{$\lambda$28.21$\mu$m}   & \multicolumn{1}{c}{$\lambda$17.03$\mu$m} & \multicolumn{1}{c}{$\lambda$12.28$\mu$m}  & \multicolumn{1}{c}{$\lambda$9.66$\mu$m}    \\

\\
\hline
\\[0.5pt]

8A  &  $<$1.513e-18 &  $<$2.286e-18  &  $<$5.699e-19 &  $<$1.142e-18\\
8C  &  $<$1.897e-18 &  $<$2.509e-18  &  $<$7.261e-19 &  $<$1.566e-18\\
8D  &  $<$1.752e-18 &  $<$3.618e-18  &  $<$1.298e-18 &  $<$1.592e-18\\
15C & $<$1.506e-18 &  $<$3.299e-18  &  $<$1.196e-18  & $<$2.198e-18\\
25D &  $<$1.650e-18 &  $<$2.958e-18 &   $<$1.161e-18  & $<$1.252e-18\\
25F & $<$1.864e-18 &  $<$3.186e-18  &  $<$9.264e-19 &  $<$1.825e-18 \\
40A & $<$1.415e-18 &  $<$2.303e-18  &  $<$7.692e-19 &  $<$1.173e-18\\
31B &  $<$ 2.864e-18  & $<$ 2.383e-18   & $<$ 3.292e-18 &  $<$ 4.508e-18 \\
44B & $<$1.976e-18 &  $<$3.267e-18  &  $<$1.883e-18  & $<$6.087e-18\\
47B & $<$1.284e-18 &  $<$1.454e-18  &  $<$1.342e-18  & $<$2.156e-18\\
47D & $<$4.018e-18 &   $<$5.080e-18 &  $<$3.122e-18  & $<$3.052e-18\\
54A & $<$1.415e-18  & $<$4.105e-18  &  $<$1.108e-18  & $<$2.264e-18\\
54B  & $<$2.152e-18  & $<$2.184e-18  &  $<$2.537e-18  & $<$6.094e-18\\
54C & $<$1.564e-18  & $<$2.643e-18  &  $<$8.198e-19 &  $<$1.662e-18\\
55A & $<$7.940e-19 &  $<$9.644e-19  &  $<$3.727e-19 &  $<$5.379e-19\\
55B$^\star$ & --      &           --     &             $<$6.945e-19 &  $<$5.913e-19\\
55D$^\star$  & --     &            --    &              $<$2.814e-19 &  $<$4.005e-19\\
57B$^\diamond$ &  $<$1.108e-18 &  $<$2.865e-18 &  --  &        --\\
57B$^\star$ & --    &     --            &         $<$1.709e-19  & $<$1.994e-19\\
57C  & $<$1.945e-18 &  $<$2.728e-18  &     --    &    --  \\
57D & $<$2.085e-18  & --      &   --         &   -- \\
57E & $<$1.361e-18  & $<$2.108e-18  &   $<$6.653e-19 &  $<$9.469e-19\\
62A$^\diamond$ & $<$9.610e-19 &  $<$1.881e-18  &  --    &  --           \\
62A$^\star$  &   --         &      --       &        $<$1.776e-18  & $<$3.169e-18\\
62B &  $<$1.082e-18 &  $<$1.588e-18   &  $<$1.578e-18 &  $<$3.723e-18\\
62C &  $<$9.707e-19 &  $<$1.422e-18  &   $<$9.435e-19 &  $<$2.819e-18\\
67A & $<$2.099e-18  & $<$3.888e-18  &   $<$4.141e-18  & $<$7.609e-18\\
67D & $<$1.704e-18 &  $<$2.371e-18  &   --      &           --\\
75A & $<$2.375e-18  & $<$2.323e-18  &   $<$2.861e-18 &  $<$6.098e-18\\
75C & $<$9.559e-19  & $<$1.815e-18  &   $<$1.804e-18  & $<$2.296e-18\\
75E & $<$1.333e-18 &  $<$1.768e-18  &   $<$6.877e-19  & $<$1.205e-18\\
82A  & $<$1.331e-18 &  $<$2.363e-18  &   $<$3.547e-19  & $<$7.162e-19\\
91D & $<$2.280e-18  & $<$5.068e-18   &  $<$5.111e-19  & $<$1.894e-18\\
95A & $<$1.506e-18 &  $<$2.906e-18  &   $<$1.561e-18 &  $<$3.768e-18\\
96B & $<$1.241e-18  & $<$3.459e-18   &  $<$1.041e-18  & $<$2.242e-18\\
97A & $<$1.798e-18  & $<$2.312e-18  &   $<$1.042e-18  & $<$1.490e-18\\
97C & $<$1.510e-18 &  $<$3.099e-18   &  $<$1.357e-18  & $<$2.160e-18\\
97D & $<$1.988e-18 &  $<$2.146e-18   &  $<$1.450e-18  & $<$3.818e-18\\
100B & $<$2.716e-18 & $<$5.954e-18 &   $<$1.867e-18  & $<$4.176e-18\\
100C & $<$1.965e-18 & $<$3.052e-18  &  $<$1.643e-18  & $<$2.809e-18\\

\\
\hline
\\[0.5pt]
\multicolumn{4}{l}{$^\star$ SL coverage only}\\
\multicolumn{4}{l}{$^\diamond$ LL coverage only}\\

\\
\end{tabular}
\end{center}
}
\end{table}

\begin{figure*}[!thp]
\begin{center}
\subfigure[31B]{\includegraphics[width=8cm]{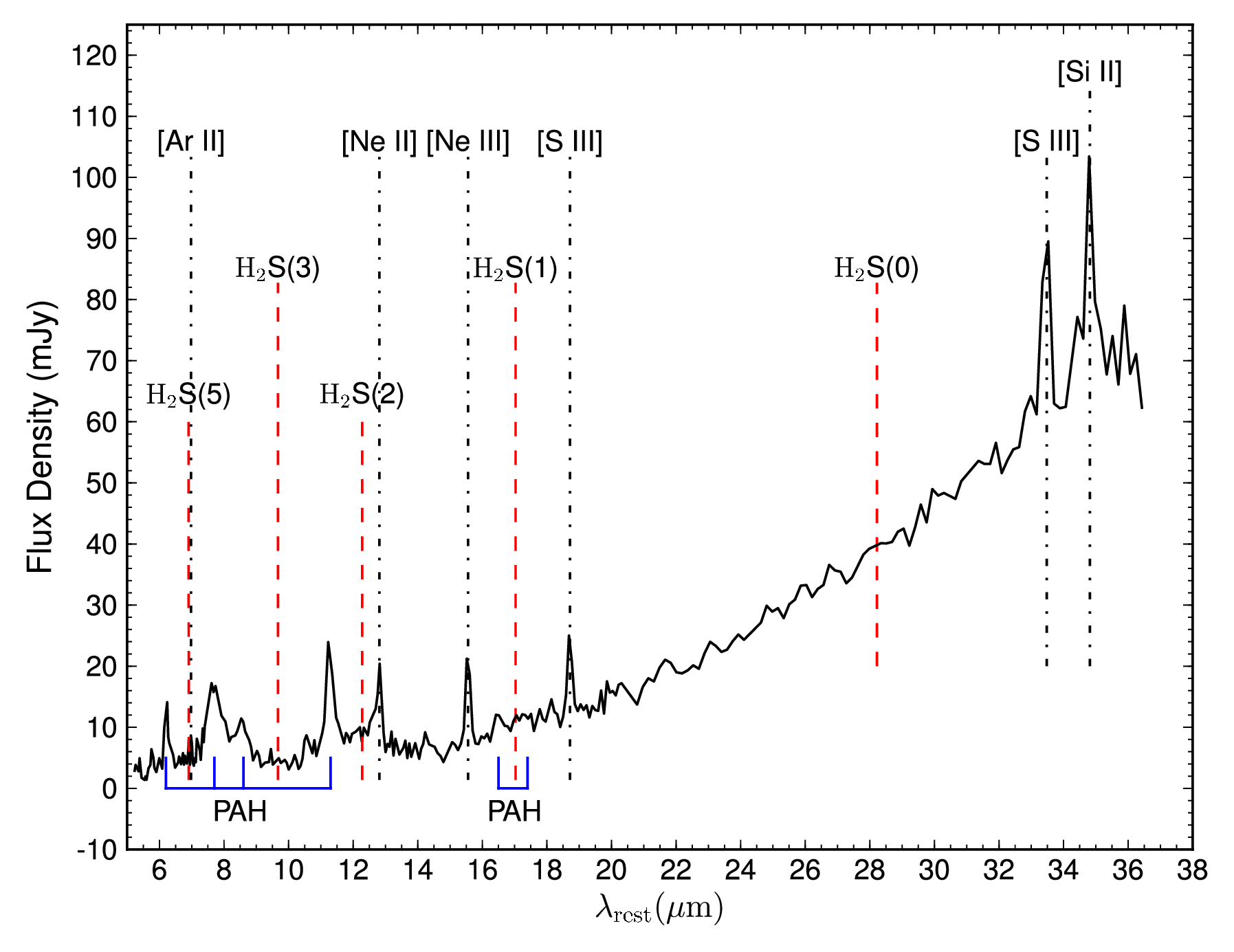}}
\hfill
\subfigure[LL slit on an IRAC 3.6\micron\ image (2\arcmin$\times$2\arcmin)]{\includegraphics[width=8cm]{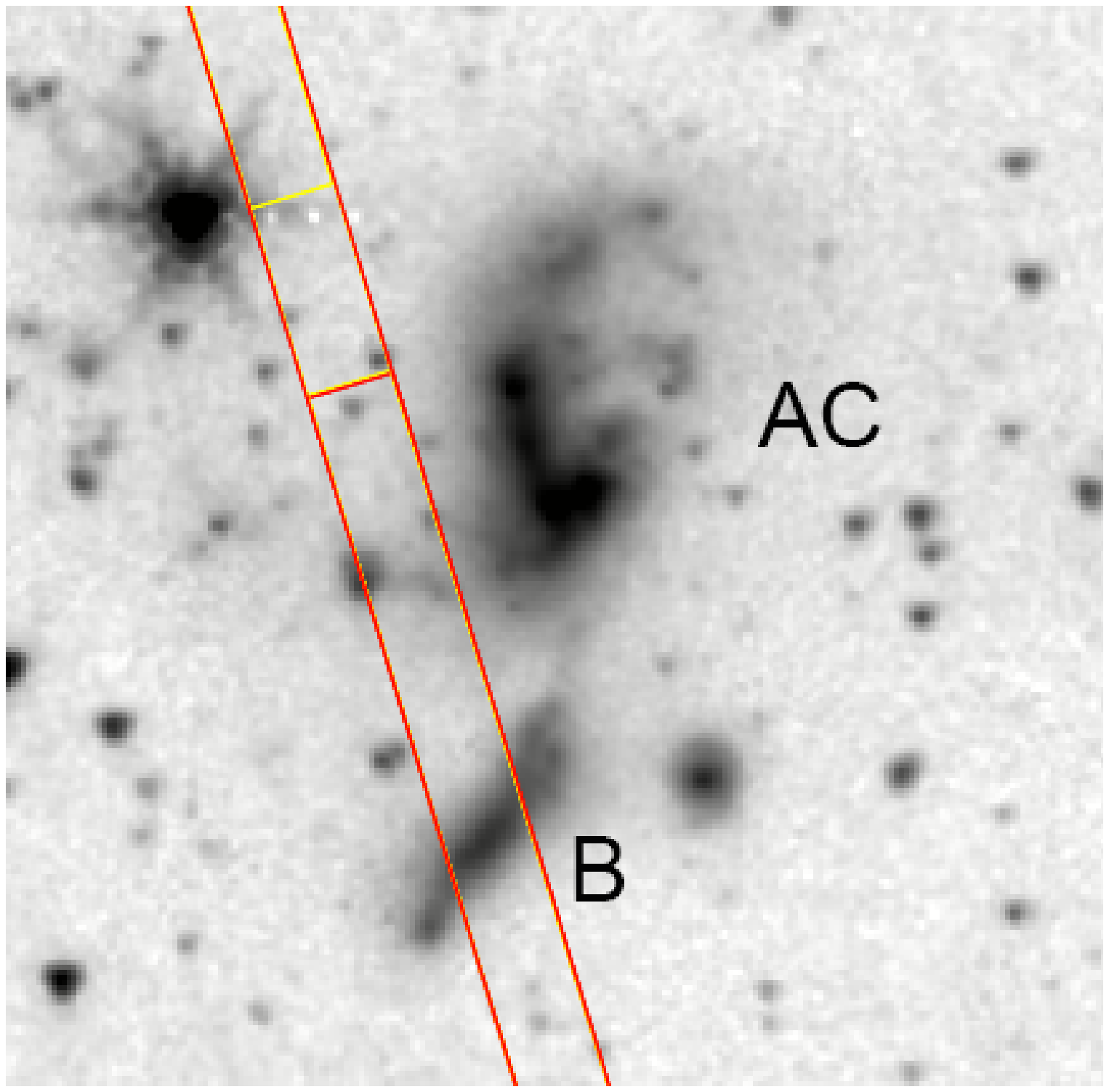}}
\caption[]
{a) The spectrum of HCG 31B is shown as typical of galaxies with star-forming mid-infrared colours, but no \Ht\ detections. b) Here the LL slit overlay shows that the coverage for the galaxy includes the nuclear and much of the disk region, thus the spectrum in (a) is broadly representative of the galaxy's global properties.}
\label{ht_nodetect}
\end{center}
\end{figure*}

As noted in the text, several galaxies with star-forming, mid-infrared colours do not have detected \Ht\ emission. In Figure \ref{ht_nodetect}a we use HCG 31B to illustrate the typical spectrum of these systems. Excited \Ht\ emission is associated with PDRs in star-forming galaxies \citep{HT}. However, given the large aperture (the LL slit coverage is shown in Figure \ref{ht_nodetect}b) of our extractions, which include the nuclear and disk regions, and depending on the geometry of dust and star forming regions, the \Ht\ line contrast can be effected by the warm dust continuum which may dominate entirely. This is less of an effect when targeting the nuclei and \HII\ regions of nearby galaxies \citep[see, for example,][]{Rous07}, but a feature of our study since our observations are tailored to finding IGM detections of excited \Ht, where continuum levels are low, and shock-driven \Ht, which have been shown to be a powerful cooling channel \citep[see, for example,][]{Clu10}.

\begin{table}[!hbtp]
{\scriptsize
\caption{Measurements for Main PAH features for Remainder of Sample in units of W\,m$^{-2}$ \label{tab:pah_upp}}
\begin{center}
\begin{tabular}{l l l l }
\hline

\\[0.5pt]
Galaxy & \multicolumn{1}{c}{6.2\micron } &  \multicolumn{1}{c}{7.7\micron} &    \multicolumn{1}{c}{11.3\micron}  \\  
\\
\hline
\\[0.5pt]

8A  &   $<$6.014e-18      &    $<$7.630e-18   &   9.170e-18  (1.826e-18)\\
8C  &   $<$7.029e-18      &    $<$7.961e-17   &   1.754e-17  (2.363e-18)\\
8D  &   $<$7.180e-18      &    5.434e-17  (4.884e-18)  &   2.076e-17  (8.858e-19)\\
15C &  $<$1.788e-17      &    $<$5.850e-17   &   $<$5.884e-18\\
25D &  $<$7.194e-18      &    $<$2.707e-17   &   $<$4.555e-18\\
25F  &  $<$4.319e-18     &    $<$3.667e-17   &    $<$2.530e-18\\
31B  &  2.269e-16 (1.306e-17)    &  6.022e-16 (5.166e-17)  &  2.119e-16 (6.787e-18)  \\
40A &  $<$7.947e-18      &    1.058e-17 (2.645e-18)  &  1.242e-17 (5.437e-19)\\
44B  & $<$1.912e-17      &    $<$1.127e-16     &   $<$9.197e-18\\
47B &  $<$1.214e-17      &    $<$6.908e-17     &   $<$6.273e-18\\
47D & $<$4.196e-17       &   $<$4.482e-17 &  2.309e-17   (3.805e-18)\\
54A  & $<$1.180e-17      &  6.653e-17 (6.314e-18)   &  1.318e-17  (1.314e-18)\\
54B  &  9.182e-17 (9.197e-18)   &   1.764e-16 (1.756e-17)  &  6.839e-17  (7.564e-18)\\
54C &  $<$8.366e-18       &   6.307e-17  (6.260e-18)  &     1.568e-17  (2.409e-18)\\
55A & $<$9.556e-18       &    $<$7.376e-18   &  1.248e-18 (3.539e-19)\\
55B &  $<$4.214e-18      &    $<$1.768e-17    &   $<$1.422e-18\\
55D &  $<$2.492e-18      &    $<$4.115e-18       &  2.180e-18  (6.428e-19)\\
57B  & $<$3.270e-18   &  4.168e-18  (1.015e-18)  &   2.027e-18 (3.227e-19)\\
57C$^\diamond$ &  --   &    --  &    --\\
57D$^\diamond$ &  --  &   --   &   --\\
57E &  $<$5.498e-18  &  $<$7.503e-17    &     1.541e-17  (3.929e-18)\\
62A  & 3.606e-17  (7.201e-18) &    $<$ 6.775e-17  &    8.293e-18  (2.755e-18)\\
62B  &  $<$1.723e-17  &  $<$6.527e-17    &  $<$1.132e-17\\
62C  &  $<$1.735e-17  &   $<$  9.991e-17 &  $<$4.573e-18\\
67A  &  $<$5.361e-17  &   $<$  1.850e-16  &  $<$2.205e-17\\
67D$^\diamond$ &  --  &  --  & -- \\
75A  &  $<$2.737e-17   & 3.330e-17  (6.426e-18)  &    2.916e-17  (1.506e-18)\\
75C  & 2.154e-17 (6.270e-18)  &  8.114e-17  (2.569e-17)   &  3.254e-17  (2.010e-18)\\
75E &  $<$2.183e-17  &   $<$1.652e-17  &  $<$6.830e-18\\
82A  & $<$6.455e-18  &   $<$4.775e-17  &  $<$2.735e-18\\
91D &  $<$1.781e-17  &  3.659e-17  (7.704e-18)  &  7.828e-18  (2.189e-18)\\
95A  &  $<$2.470e-17   &  $<$2.100e-17  &     $<$6.839e-18\\
96B &  $<$1.568e-17   &   $<$7.431e-17  &     $<$4.826e-18\\
97A  &  $<$1.236e-17  &   $<$8.951e-17  &     $<$5.459e-18\\
97C  &  $<$1.510e-17  &   $<$2.646e-17  &    $<$2.602e-18\\
97D  & $<$2.322e-17  &    $<$1.220e-16  &    $<$6.714e-18\\
100B &  2.164e-16 (9.738e-18) &  5.821e-16  (4.674e-17)  &  1.361e-16  (7.608e-18)\\
100C &  1.999e-17  (7.049e-18)  &  5.320e-17  (8.919e-18)  &   2.929e-17  (2.515e-18)\\
\\
\hline
\\[0.5pt]
\multicolumn{4}{l}{$^\diamond$ LL coverage only}\\

\\
\end{tabular}
\end{center}
}
\end{table}

\clearpage

\section{X-ray and Radio Properties of the HCG Sample}\label{xray}

The X-ray distribution and luminosity in our HCG MOHEGs is an important consideration since XDRs (X-ray dominated regions) associated with AGN may excite \Ht. It is therefore necessary to rule out photo-heating of \Ht\ from within the galaxies.

\subsection{Diffuse X-ray Emission from Intragroup Gas}\label{diffuse}

First we consider the amount and distribution of X-ray emission (including diffuse plasma) in the MOHEG groups to explore any connections to the \Ht-enhanced systems. We have X-rays detected in 5 groups and we list the diffuse X-ray measurements obtained from the literature in Table \ref{tab:xray}. 
The MOHEGs HCG 15A and D appear to have some X-ray structure associated with them, in particular HCG 15D appears connected to 15F \citep{Ras08}. In HCG 40, B is \Ht-enhanced, but the strongest X-ray sources are C and D, which are only \Ht-detected systems and not MOHEGs \citep{Ras08}. However, HCG 40B and C appear connected in X-ray emission, particularly interesting since we detect \Ht\ outside 40C (see section \ref{igm}). In HCG 68, the map of \citet{For06} indicates emission around MOHEGs HCG 68A and C show emission and it appears that the emission from A connects to B (also a MOHEG).

\begin{table}[!th]
{\scriptsize
\caption{Diffuse X-ray emission in MOHEG Compact Groups \label{tab:xray}}
\begin{center}
\begin{tabular}{l c l l}
\hline
\hline
\\[0.5pt]
Group     & MOHEGs  & \multicolumn{2}{c}{X-ray Luminosity}    \\
  &   &  L$_{X}$ (erg\,s$^{-1}$) &   L$_\odot$  \\
\\
\hline
\\
HCG 6    & 1  & $<$2.7$\times 10^{41}$  (3$\sigma$, a)   & $<$7.0$\times 10^{7}$ \\
HCG 15   & 2  &  3.2$\pm$0.2$\times 10^{41}$  (b)    &  8.3$\times 10^{7}$  \\
HCG 25    & 1   &  --   &  -- \\
HCG 40   & 1 &    3.1$\pm$0.5$\times 10^{40}$ (b)  & 8.1$\times 10^{6}$   \\
HCG 56   &  1  &  $<$1.7$\times 10^{42}$   (3$\sigma$, a) & $<$4.4$\times 10^{8}$  \\
HCG 57   &  1  &  8.3$+$3.2$-$3.5$\times 10^{41}$   (c)   & 2.2$\times 10^{8}$   \\
HCG 68 & 3 &   3.3$\pm$0.3$\times 10^{41}$   (d) & 8.6$\times 10^{7}$  \\
HCG 82  & 1  &  1.9$\pm$0.8$\times 10^{42}$   (a) & 4.9$\times 10^{8}$  \\
HCG 95   & 1   & $<$2.7$\times 10^{42}$   (3$\sigma$, a) & $<$7.0$\times 10^{8}$  \\

\\
\hline
\\[0.5pt]
\multicolumn{3}{l}{(a) From \citet{Pon96}}\\
\multicolumn{3}{l}{(b) From \citet{Ras08}}\\
\multicolumn{3}{l}{(c) From \citet{Mulch03}}\\
\multicolumn{3}{l}{(d) From \citet{Os04}}\\
\\
\end{tabular}
\end{center}
}
\end{table}


Unlike SQ, X-ray emission from a shock-front is not clearly distinguishable from the diffuse emission in the groups. SQ is, of course, notable due to the high relative velocity ($\sim$\,1000\,km/s) of the collision of the intruder with the intragroup \HI. Collisions of the group galaxies with tidal material would be occurring at far lower relative velocities ($100-1000$\,km/s), thus the kinetic energy would be far less compared to SQ, resulting in weaker X-ray emission. 

\subsection{X-ray Emission within Group Members}\label{xray_gal}

Several galaxies appear to have nuclear enhancements in their X-ray data. Searching the archives for suitable coverage of the HCG MOHEGs, we have determined that 
HCG 6B, 25B, 82B and 95C have not been covered by {\it Chandra/XMM} and were not detected by {\it ROSAT}.  All other available measurements (from the literature and data, as specified) are listed in Table \ref{tab:xray_nuc}. We include the radius of the measurement and the luminosity of the \Ht\ in relation to the X-ray. In addition, archival radio data is presented. The galaxies are listed in order of decreasing \Ht/7.7\micron\ PAH.

\begin{table}[!thb]
{\scriptsize
\caption{MOHEG Nuclear X-ray and Radio Data \label{tab:xray_nuc}}
\begin{center}
\begin{tabular}{l l l r r | c  r  r    r}
\hline
\hline
\\[0.5pt]
 & \multicolumn{3}{c}{X-ray Luminosity}  &   & \multicolumn{4}{c}{Radio Data}   \\
    &  L$_{\rm X}$ &  Radius &  L$_{\rm X}$  &  L$_{\rm H_2}$/L$_{\rm X}\,^\dagger$   &  & $F_{\rm 1.4 GHz}$  & L$_{\rm 1.4 GHz}$  &  log\,L$_{\rm 1.4 GHz}$  \\
    & (erg\,s$^{-1}$)  &  & (L$_\odot$)  &  &  &  (mJy)    &   (erg\,s$^{-1}$) & [W\,Hz$^{-1}$]\\
\\
\hline
\\
68A    & 1.6$\times 10^{40}$ $^{a,b}$ & 4.7\arcs & 4.1$\times 10^6$ & $>$0.53 & NVSS$^\diamond$    &  40.5 (1.3)  & 8.7$\times 10^{37}$  & 21.6  \\
6B    & -- & -- & -- & -- & --   & \\
57A    & 4.1$\times 10^{39}$ $^{c,d}$ & 15\arcs & 1.1$\times 10^6$ & $>$28.9 & --\\

40B   &   1.8$\times 10^{39}$ $^{c,e}$ & 2.1\arcs & 4.7$\times 10^5$ &  $>$2.42 & --\\
25B    & -- & -- &-- & -- & --\\
15D   &   6.7$\times 10^{41}$($\star$) $^{c,e}$ & 25\arcs  & 1.7$\times 10^8$  & $>$0.03  &  NVSS  &  4.6 (0.5)  & 6.6$\times 10^{37}$ & 21.7\\
68B    & 9.0$\times 10^{39}$ $^{a,b}$ & 2.3\arcs & 2.3$\times 10^6$ & $>$0.18 & NVSS &  8.0 (0.5)  & 1.7$\times 10^{37}$  & 21.1  \\
95C & -- & -- & -- & -- & --\\
15A   &   2.6$\times 10^{40}$($\star$) $^{c,e}$ & 25\arcs  & 6.8$\times 10^6$ & $>$0.22   & NVSS &  3.6 (0.6)  & 5.1$\times 10^{37}$  &  21.6 \\
56B &$<$1.5$\times 10^{40}$ $^{c,f}$  & 3\arcs & $<$3.8$\times 10^6$ & $>$1.42 & NVSS  &  26.7 (1.2)  & 6.1$\times 10^{38}$ & 22.6 \\

68C & 6.9$\times 10^{39}$ $^{a,b}$  & 1.2\arcs & 1.8$\times 10^6$ & $>$3.70 & NVSS &  16.8 (1.7)  & 3.6$\times 10^{37}$  & 21.4 \\
82B     & -- & -- & --  & --& --\\
56C    & $<$1.5$\times 10^{40}$ $^{c,f}$  &  3\arcs & $<$3.8$\times 10^6$&  $>$1.69 & --\\

\\
\hline
\\[0.5pt]
\multicolumn{8}{l}{$^\dagger$ $L$(\Ht\,0-0 S(0)-S(3))/$L_{X}$(2-10 keV)}\\
\multicolumn{8}{l}{$^\diamond$ NRAO VLA Sky Survey; \citet{Con98}}\\
\multicolumn{8}{l}{($\star$) {\it XMM} -- Can't discern point-like from diffuse emission}\\
\multicolumn{8}{l}{$^a$ 0.5-7 keV}\\
\multicolumn{8}{l}{$^b$ From Chandra Source Catalogue; \citet{Ev10}}\\
\multicolumn{8}{l}{$^c$ 0.3-2 keV}\\
\multicolumn{8}{l}{$^d$ Derived from {\it XMM} pn data}\\
\multicolumn{8}{l}{$^e$ From \citet{Ras08}}\\
\multicolumn{8}{l}{$^f$ Upper limit: not detected in Chandra Source Catalogue}\\

\end{tabular}
\end{center}
}
\end{table}

\Ht\ heating by X-rays via photoelectrons in XDR models leads to $\sim$2\% of the total gas cooling via the pure rotational lines of \Ht\ \citep{Mal96}. Assuming all of the X-ray flux is absorbed by an XDR, the maximum $L$(\Ht\,0-0 S(0)-S(3))/$L_{\rm X}$(2-10 keV) ratio is 0.01 \citep{Og10}. Assuming a power-law spectrum typical of an AGN \citep[as in ][]{Og10}, we scale the fluxes in the energy bands listed in Table \ref{tab:xray_nuc} to the $L_{\rm X}$(2-10 keV) band. Given that our IRS spectra have only sampled a fraction of the HCG galaxies, we do not have a measure of the total amount of warm \Ht\ and hence the ${\rm L_{\rm H_2}}/{\rm L_{\rm X}}$ values are lower limits, but still illustrative. We note that all the HCG MOHEGS have ${\rm L_{\rm H_2}}/{\rm L_{\rm X}}> 0.01$, usually having \Ht\ luminosities more than an order of magnitude greater than what can be generated by X-rays (except HCG 15D, although this measurement is contaminated by diffuse emission). We can thus rule out XDRs as being responsible for the enhanced \Ht\ seen in the HCG MOHEGs. This is not unexpected as even though HCG galaxies have increased AGN activity (likely due to tidal interactions), they are dominated by low luminosity AGN due to low accretion rates.

Archival radio data is also included in Table \ref{tab:xray_nuc}. The highest 1.4\,GHz luminosity is associated with the Seyfert 2 galaxy HCG 56B, a candidate MOHEG for jet-ISM interactions producing the excited \Ht\ \citep[see][]{Og10}.

\end{document}